\documentclass[a4paper,11pt]{article}
\usepackage{jcappub} 
\usepackage{comment}
\DeclareMathOperator{\icov}{\boldsymbol{\Psi}}
\usepackage{hanging}
\usepackage{orcidlink}

\arxivnumber{2404.07283} 
\title{A comparison between ShapeFit compression and Full-Modelling method with PyBird for DESI 2024 and beyond}

\author[1]{{Y.~Lai}\orcidlink{0000-0001-9054-4324},}

\author[1]{{C.~Howlett}\orcidlink{0000-0002-1081-9410},}
\author[2]{{M.~Maus}\orcidlink{0000-0002-9020-911X},}
\author[3]{{H.~Gil-Mar\'in}\orcidlink{0000-0003-0265-6217},}
\author[3, 4, 5]{{H.~E. Noriega}\orcidlink{0000-0002-3397-3998},}
\author[5]{S.~Ram\'irez-Solano,}
\author[6]{{P.~Zarrouk}\orcidlink{0000-0002-7305-9578},}
\author[7]{{J.~Aguilar}\orcidlink{0000-0001-5998-3986},}
\author[8]{{S.~Ahlen}\orcidlink{0000-0001-6098-7247},}
\author[9]{O.~Alves,}
\author[4, 10]{{A.~Aviles}\orcidlink{0000-0001-5998-3986},}
\author[11]{D.~Brooks,}
\author[12]{{S.~Chen}\orcidlink{0000-0002-5762-6405},}
\author[7]{T.~Claybaugh,}
\author[1]{{T.~M. Davis}\orcidlink{0000-0002-4213-8783},}
\author[13]{K.~Dawson,}
\author[5]{{A.~de la Macorra}\orcidlink{0000-0002-1769-1640},}
\author[11]{P.~Doel,}
\author[14, 15]{{J.~E. Forero-Romero}\orcidlink{0000-0002-2890-3725},}
\author[16, 17, 18]{E.~Gaztañaga,}
\author[7]{{S.~Gontcho A Gontcho}\orcidlink{0000-0003-3142-233X},}
\author[19, 20, 21]{K.~Honscheid,}
\author[22]{S.~Juneau,}
\author[7]{{M.~Landriau}\orcidlink{0000-0003-1838-8528},}
\author[23, 24]{{M.~Manera}\orcidlink{0000-0003-4962-8934},}
\author[24. 25]{R.~Miquel,}
\author[26]{E.~Mueller,}
\author[17]{{S.~Nadathur}\orcidlink{0000-0001-9070-3102},}
\author[27, 28]{{G.~Niz}\orcidlink{0000-0002-1544-8946},}
\author[7, 29]{{N.~Palanque-Delabrouille}\orcidlink{0000-0003-3188-784X},}
\author[30, 31, 32]{{W.~Percival}\orcidlink{0000-0002-0644-5727},}
\author[2, 7, 33]{C.~Poppett,}
\author[34]{{M.~Rezaie}\orcidlink{0000-0001-5589-7116},}
\author[35]{G.~Rossi,}
\author[36]{{E.~Sanchez}\orcidlink{0000-0002-9646-8198},}
\author[9, 37]{M.~Schubnell,}
\author[22]{D.~Sprayberry,}
\author[9]{{G.~Tarl\'{e}}\orcidlink{0000-0003-1704-0781},}
\author[5]{{M.~Vargas-Magaña}\orcidlink{0000-0003-3841-1836},}
\author[3, 25]{{L.~Verde}\orcidlink{0000-0003-2601-8770},}
\author[38]{{S.~Yuan}\orcidlink{0000-0002-5992-7586},}
\author[7]{{R.~Zhou}\orcidlink{0000-0001-5381-4372},}
\author[39]{{H.~Zou}\orcidlink{0000-0002-6684-3997}}

\affiliation{Affiliations are in Appendix \ref{sec:affiliations}}

\emailAdd{y.lai1@uqconnect.edu.au}

\abstract{DESI aims to provide one of the tightest constraints on cosmological parameters by analysing the clustering of more than thirty million galaxies. However, obtaining such constraints requires special care in validating the methodology and efforts to reduce the computational time required through data compression and emulation techniques. In this work, we perform a rigorous validation of the \textsc{PyBird} power spectrum modelling code with both a traditional emulated \textit{Full-Modelling} approach and the model-independent \textit{ShapeFit} compression approach. By using cubic box simulations that accurately reproduce the clustering and precision of the DESI survey, we find that the cosmological constraints from \textit{ShapeFit} and \textit{Full-Modelling} are consistent with each other at the \(\sim0.5\sigma\) level for the \(\Lambda\)CDM model. Both \textit{ShapeFit} and \textit{Full-Modelling} are also consistent with the true \(\Lambda\)CDM simulation cosmology down to a scale of \(k_{\mathrm{max}} = 0.20 h\mathrm{Mpc}^{-1}\) even after including the hexadecapole. For extended models such as the \(w\)CDM and the \(o\)CDM models, we find that including the hexadecapole can significantly improve the constraints and reduce the modelling errors with the same \(k_{\mathrm{max}}\). While their discrepancies between the constraints from \textit{ShapeFit} and \textit{Full-Modelling} are more significant than \(\Lambda\)CDM, they remain consistent within \(0.7\sigma\). Lastly, we also show that the constraints on cosmological parameters with the correlation function evaluated from \textsc{PyBird} down to \(s_{\mathrm{min}} = 30 h^{-1} \mathrm{Mpc}\) are unbiased and consistent with the constraints from the power spectrum.}

\begin{document}
\maketitle
\flushbottom

\section{Introduction}
\label{sec:introduction}
The universe's large-scale structure (LSS) encoded in galaxies' positions contains valuable information on cosmic growth, which helps to constrain cosmological parameters. Unlike the Cosmic Microwave Background (CMB), an effectively 2-dimensional surface at the last scattering, galaxy surveys can explore the full 3-dimensional spectrum of modes and perturbations in the universe over a much larger volume. This extra dimension means galaxy surveys can potentially provide tighter constraints on the cosmological parameters than the CMB \citep{Carrasco_2012, Baumann_2012, Carrasco_2014}. However, this is made challenging due to the nonlinear growth of structure between the epoch of recombination and now. Recent developments in nonlinear modelling, particularly in Effective Field Theory models of Large-scale Structure (EFTofLSS; \citep{Carrasco_2014, Porto_2014}), have enabled us to extract more information from the nonlinear regimes. This improvement allows current and future surveys to reach their scientific potential.

The Dark Energy Spectroscopic Instrument (DESI) is one such example. DESI \citep{Snowmass2013.Levi, DESI2016a.Science, DESI2016b.Instr, DESI2022.KP1.Instr, DESI2023a.KP1.SV, DESI2023b.KP1.EDR, DESI2024.V.KP5} is forecasted to obtain more than an order of magnitude more galaxies than any previous survey. Although the huge amount of data will provide us with the strongest constraints on cosmological parameters with galaxy surveys to date, it also means that 1) our methodologies require more precise validation than what was necessary for previous surveys and 2) a much longer computing time is required to analyse the data. The aim of this work, along with a series of other papers to be released coincidentally \citep{KP5s1-Maus, KP5s2-Maus, KP5s3-Noriega, KP5s5-Ramirez}, is to perform this validation. We validate methods designed to extract cosmological information by modelling the clustering of galaxies measured with DESI. We also investigate how to reduce the information in the two-point statistics to a smaller set of compressed parameters. We can then use these parameters to fit for wider ranges of cosmological models.

In the brute force approach to constrain cosmological parameters, one can consider computing the two-point clustering of galaxies (i.e., the power spectrum) within a specific cosmological model. We then use the Markov Chain Monte Carlo (MCMC) algorithm to find the range of model power spectra that best fit a dataset. This process is referred to in this work as ``\textit{Full-Modelling}'' fitting. There are now many codes on the market to achieve this; within the DESI collaboration, we have been testing four different pipelines: \textsc{PyBird} \citep{Colas_2020,d_Amico_2020, d_Amico_2021}, \textsc{Velocileptors} \citep{Chen_2020, Chen_2021, KP5s2-Maus}, \textsc{FOLPS$\nu$} \citep{Noriega_2022, KP5s3-Noriega}, and \textsc{EFT-GSM} \citep{Ramirez_2023, KP5s5-Ramirez}. However, one limitation of \textit{Full-Modelling} methods (without emulators) is that they can be time-consuming because the Boltzmann code can take up to a few seconds to compute the linear power spectrum for a specific cosmology. In \textsc{PyBird}, the IR resummation terms (see section \ref{sec:theory}), which also take a few seconds to compute, further hinder the computation speed.  

To speed up the \textit{Full-Modelling} fitting, we can either shrink the size of the data vector \citep{Heavens_2000, lai_2023} or use emulators \citep{Colas_2020, Donald_McCann_2022}. However, \textit{Full-Modelling} fitting is also model-dependent --- we have to re-fit the entire set of power spectra when we change the underlying cosmological model. Another widely used compression technique is to fit for parameterized, model-independent, growth and Baryon Acoustic Oscillation (BAO) parameters first and then convert these parameters to cosmological parameters \citep{Beutler_2016, Grieb_2017, Satpathy_2017, Gil_Mar_n_2020, Maus_2023, Yu_2023}. This compression technique is the standard ``\textit{Template}'' fitting method. It does not require us to re-fit the underlying clustering when testing multiple cosmologies, so long as the model of interest remains valid for the template \citep{Vargas_Maga_a_2018, Carter_2018}.

However, previous research \citep{Philcox_2021, Brieden_2021} has found that \textit{Template} Fitting has less constraining power than the \textit{Full-Modelling} method when using constraints from the LSS alone. Recent works \cite{Brieden_2021, Brieden_2023} demonstrated that we can improve the \textit{Template} method by introducing an additional shape parameter \(m\), which measures the tilt of the power spectrum along with the BAO and growth parameters. We can then convert the shape parameter to cosmological parameters like the \textit{Template} parameters. Calling this method ``\textit{ShapeFit}'', Ref.~\citep{Brieden_2021b} found this additional degree of freedom helped to break the degeneracy between the scalar amplitude and the matter density and gave similar constraints to the \textit{Full-Modelling} methodology with a fixed scalar index \(n_s\) and a BBN (Big Bang Nucleosynthesis) prior on the baryon density \(\Omega_b\). Crucially, they also found \textit{ShapeFit} retains its model independence so that we do not need to re-fit the data power spectrum when we change the cosmological model. Consequently, if the conversion from \textit{ShapeFit} parameters to the cosmological parameters is fast, \textit{ShapeFit} is potentially faster than the traditional \textit{Full-Modelling} when fitting multiple cosmologies.  

Within this context, the aims of this paper are two-fold. Firstly, we validate that the \textsc{PyBird} EFTofLSS pipeline can reliably produce cosmological constraints within DESI, with sufficient control of systematic errors in the theoretical modelling of the clustering to ensure the robustness of upcoming constraints from the Year 1 data release \citep{DESI2024.I.DR1}. Secondly, we test whether \textit{ShapeFit} can still get similar constraints to the \textit{Full-Modelling} fit within DESI, such that the main output product of the collaboration can be a simple set of compressed BAO, growth, and shape compressed parameters with which the community can perform their cosmological inference. This paper forms one in a set of four, each focused on similar tests with the other pipelines mentioned above (\textsc{Velocileptors} \citep{KP5s2-Maus}, \textsc{FOLPS$\nu$} \citep{KP5s3-Noriega}, and EFT-GSM \citep{KP5s5-Ramirez}). A fifth paper in the series \citep{KP5s1-Maus} compares and contrasts the results across the different pipelines, demonstrating consistency in the precision of DESI cosmological parameter constraints even in the extreme case of a \(200h^{-3} \mathrm{Gpc}^3\) volume, despite different intrinsic modelling assumptions. This result shows without reasonable doubt that different EFT setups agree. 

We organize this paper as follows: Section \ref{sec:mock} presents the simulations we used to test our pipeline. Section \ref{sec:theory} briefly explains the theory behind the \textsc{PyBird} model and how we adopted the \textit{ShapeFit} methodology into this ecosystem. Section \ref{sec:result} shows the configurations we used for our tests and the main results of this work. From there, for the interested reader, we dig deeper into the details of the \textit{ShapeFit} and \textit{Full-Modelling} methods (in Sections \ref{sec:Shapefit} and \ref{sec:Full-Modelling} respectively), testing various systematic effects and extended cosmological models. In Section \ref{sec:corr_func}, we show a brief but representative set of results from the correlation function and compare them to the constraints with the power spectrum. Lastly, we conclude in Section \ref{sec:conclusion}. 

\section{Simulated Samples}
\label{sec:mock}
We use different methods to generate the power spectra and covariance matrices for our DESI-based mock galaxy catalogues. For the power spectrum, we use the more accurate N-body simulations from the \textsc{AbacusSummit} simulation suite \citep{Maksimova_2021, Garrison_2021, Hadzhiyska_2021}. For the covariance matrix, we use the less accurate but much faster Extended Zel'dovich (\textsc{EZ}) \citep{Chuang_2014} mocks. We calculate the covariance matrix from 1000 \textsc{EZ} mocks for each tracer.

\subsection{Mock Samples}

Mock catalogues for the Luminous Red Galaxy (LRG), the Emission Line Galaxy (ELG), and the Quasi Steller Objects (QSO) samples in DESI are produced from the \textsc{AbacusSummit} Simulations with \(6912^3\) particles within a cubic box of \(2000h^{-1} \mathrm{Mpc}\). These mocks assume a Planck2018 \citep{Planck_2020} cosmology with \(A_s = 2.083 \times 10^{-9}\), \(h = 0.6736\), \(\omega_{cdm} = 0.12\), \(\omega_b = 0.02237\), \(w = -1.0,\) and \(\Omega_k = 0.0\).\footnote{The baseline \(\Lambda\)CDM model in the \textsc{AbacusSummit} also set a single scalar index \(n_s = 0.9649\). Additionally, we also have two massless neutrino species (\(N_{\mathrm{ur}} = 2.0328\)) and one massive neutrino species (\(N_{\mathrm{ncdm}} = 1\)) with a total mass of \(\Omega_{\mathrm{ncdm}} = 0.00064420\)eV.} Throughout this work, we will test our method with these cubic box mocks. These mocks do not contain the survey geometry of DESI, but this is appropriate for our model comparisons and tests, as we expect the window function to apply equally regardless of the underlying theoretical model. 

There are 25 different realizations for the LRG, ELG, and QSO samples. The snapshot redshift is 0.8 for the LRG sample \citep{Zhou_2020, LRG.TS.Zhou.2023}, 1.1 for the ELG sample \citep{ELGPrelim.Raichoor.2020, Raichoor_2023, rocher2023desi}, and 1.4 for the QSO sample \citep{QSOPrelim.Yeche.2020, QSO.TS.Chaussidon.2023, yuan2023desi}. Each tracer is reproduced in the simulation snapshot by applying a Halo Occupation Distribution (HOD) model calibrated to match the small-scale clustering and the large-scale bias evolution of the DESI sample. These mocks are then further randomly subsampled to approximately match the mean galaxy number density distribution \citep{Moon_2023} of galaxies that DESI will observe. In this work, we will use two different fitting configurations. If we fit a single tracer, we fit the mean of the 25 mocks (hereon called the first fitting configuration), resulting in an effectively noise-free measurement of the clustering. If we combine three different tracers, we fit the mean of LRG with the first eight mocks, ELG with the mean of the 9th to the 16th mock, and QSO with the mean of the 17th to the 24th mock (hereon called the second fitting configuration). We use this configuration to avoid the cross-correlation between two different tracers built from the same realizations (albeit at different redshifts). We leave out the 25th mock because we want the mock mean from each tracer to be the average over the same number of mocks such that we remove a similar amount of noise. The power spectrum is produced with the estimator from Ref.~\cite{Bianchi_2015, Scoccimarro_2015} and a triangular-shaped cloud mass assignment scheme \citep{Grove_2022}, while the correlation function is generated with the Landy-Szalay estimator \citep{Landy_1993}. In both cases, we use these implementations in \textsc{Nbodykit} \citep{Hand_2018}.

\subsection{Covariance matrix}
We calculate the covariance matrices from the \textsc{EZ} mocks. The \textsc{EZ} mocks extend on the Zel'dovich approximation \citep{Zel_1970} by including the stochastic scale-dependent, non-local, and nonlinear biasing contributions \citep{Chuang_2014}. Ref.~\citep{Moon_2023} shows the \textsc{EZ} mocks have very similar average clustering properties as the \textsc{AbacusSummit} mocks in terms of the one-point, two-point, and three-point statistics and can be produced at much cheaper computational cost than the full N-body  \textsc{AbacusSummit} mocks. 1000 \textsc{EZ} mock relizations per tracer are produced with a \(2000h^{-1} \mathrm{Mpc}\) cubic box, matching the configurations of the \textsc{AbacusSummit} mocks. We then calculate the covariance matrices from these 1000 mocks. To obtain the unbiased inverse covariance matrix \(\hat{\icov}\), we apply both the Hartlap factor \citep{Hartlap_2006}
\begin{equation}
    \hat{\icov} = \frac{N_S - N_P - 2}{N_S - 1} \icov
    \label{eq:Hartlap}
\end{equation}
and the Percival factor \citep{Percival_2014} 
\begin{equation}
m_{1} = \frac{1 + B(N_{D} - N_{P})}{1+A+B(N_{P}+1)}
\label{eq:m1}
\end{equation}
to the inverse covariance matrix \(\icov\). Here,  
\begin{align}
A &= \frac{2}{(N_{S}-N_{D}-1)(N_{S}-N_{D}-4)} \notag \\
B &= \frac{N_{S}-N_{D}-2}{(N_{S}-N_{D}-1)(N_{S}-N_{D}-4)}
\end{align}
and \(N_s\) is the number of realizations employed to estimate the covariance matrix, \(N_D\) is the length of the data vector, and \(N_P\) is the number of free parameters in the model. Ref.~\citep{Sellentin_2015} points out that the Hartlap factor is only an approximation. However, we have 1000 simulations, so the Hartlap factor is a good approximation. Furthermore, the modified likelihood in \citep{Sellentin_2015} is non-Gaussian, so we cannot analytically marginalize over the nuisance parameters (see section \ref{sec:theory}). Therefore, we decided to use the Hartlap factor in our analysis. We use single box covariance matrices (i.e., corresponding to a volume of \(8 h^{-3} \mathrm{Gpc^{3}}\)) in this work, which is sufficient to explore the impact on DESI-like volumes for each tracer, and the three tracers combined (The total volume is \(24 h^{-3} \mathrm{Gpc^{3}}\)). Tests using a reduced covariance matrix and the mean measurement of the 25 \textsc{AbacusSummit} realizations (i.e., corresponding to a volume of \(200 h^{-3} \mathrm{Gpc^{3}}\)) are reserved for the model comparison paper in this series \citep{KP5s1-Maus}. When we combine different tracers, we multiply the Hartlap and Percival factors into the individual covariance matrix first and then generate the block-diagonal combined covariance matrix.

\section{Theory}
\label{sec:theory}
In this section, we will briefly explain the theory behind \textsc{PyBird} and \textit{ShapeFit}. For a more detailed explanation, we will refer the readers to Ref.~\citep{Perko_2016, d_Amico_2020} for \textsc{PyBird} and Ref.~\citep{Brieden_2021} for \textit{ShapeFit}. 
\subsection{\textsc{PyBird}}
\label{sec:Pybird}
Historically, the Standard Perturbation Theory (SPT) for large-scale structures is widely used. However, Ref.~\citep{Carrasco_2012} found the SPT model loses reliability around or even before \(k = 0.1 h \mathrm{Mpc}^{-1}\) when compared to nonlinear power spectrum from \textsc{CAMB} with \textsc{HaloFit} \citep{Smith_2003}. In comparison, the EFTofLSS model is consistent with the \textsc{HaloFit} power spectrum up to \(k = 0.25 h \mathrm{Mpc}^{-1}\). Ref.~\citep{Perko_2016} have derived the equations for the galaxy power spectrum within the EFTofLSS up to one-loop order
\begin{align}
    P_g(k, \mu) &= Z_1(\mu)^2 P_{\mathrm{lin}}(k) + 2 \int \frac{d^3 q}{(2\pi)^3} Z_2 (\boldsymbol{q}, \boldsymbol{k-q}, \mu)^2 P_{\mathrm{lin}}(|\boldsymbol{k-q}|)P_{\mathrm{lin}}(q) \nonumber \\
    & + 6 Z_1(\mu)P_{\mathrm{lin}}(k) \int \frac{d^3 q}{(2\pi)^3} Z_3(\boldsymbol{q}, \boldsymbol{-q}, \boldsymbol{k}, \mu) P_{\mathrm{lin}}(q) + 2 Z_1(\mu)P_{\mathrm{lin}}(k)\nonumber \\
    &\left(c_{ct}\frac{k^2}{k_M^2} + c_{r,1} \mu^2 \frac{k^2}{k_r^2} + c_{r,2} \mu^4 \frac{k^2}{k_r^2}\right) + \frac{1}{\overline{n}_g}\left(c_{\epsilon, 1} + c_{\epsilon, 2}\frac{k^2}{k_M^2} + c_{\epsilon, 3} f\mu^2 \frac{k^2}{k_M^2}\right)
    \label{eq: GPS}
\end{align}
where \(P_{\mathrm{lin}}\) is the linear power spectrum of Cold Dark Matter (CDM) and baryons, usually obtained from Boltzmann codes such as \textsc{CLASS} \citep{Diego_Blas_2011} or \textsc{CAMB} \citep{Lewis_2011, Howlett_2012}. The SPT kernels \(Z_n\) in equation (\ref{eq: GPS}) account for the effect of RSD, galaxy bias and nonlinearities. In EFTofLSS, we use the counter terms \(c_{ct}\), \(c_{r, 1}\), \(c_{r, 2}\) to account for the effect of short wavelength modes on the long wavelength modes, which is absent in the SPT model. Additionally, we have the stochastic terms \(c_{\epsilon, 1}\), \(c_{\epsilon, 2}\), \(c_{\epsilon, 3}\) to account for the stochastic way in which galaxies trace the underlying matter field and the impact of residual shot-noise on the power spectrum. Lastly, \(f\) is the linear growth rate of structure, \(\mu\) is the cosine of the angle between the vector $\boldsymbol{k}$ and the line-of-sight direction, and \(\overline{n}_g\) is the mean number density of galaxies in the survey.\footnote{We follow Ref.~\citep{d_Amico_2020} and set \(k_M\) (Inverse tracer spatial extension scale) \(= k_r\) (Inverse velocity product renormalization scale) \(= 0.7h \mathrm{Mpc}^{-1}\). However, the latest version of \textsc{PyBird} has the default setting of \(k_M = 0.7h \mathrm{Mpc}^{-1}\) and \(k_r = 0.25h \mathrm{Mpc}^{-1}\). We found the change in \(k_r\) has little impact on the constraints on \textit{ShapeFit} and cosmological parameters (see Appendix \ref{sec:version} for more detail).} In addition to the counter terms and stochastic terms, \textsc{PyBird} also has 4 bias parameters: \(b_1\), \(b_2\), \(b_3\), \(b_4\) following the West Coast convention of the bias parameters \citep{Nishimichi_2020}. They enter in the redshift kernel \(Z_n\) calculations. The expressions of the redshift kernels are given in equation (2.2) of Ref.~\citep{d_Amico_2020}. EFTofLSS requires us to resum the effect of long wavelength displacements that cannot be treated perturbatively \citep{d_Amico_2020} to estimate the galaxy power spectrum better. This resummation is required because we cannot model the long wavelength displacement modes perturbatively. The IR-resummation terms of EFTofLSS \citep{Senatore_2015,Lewandowski_2018} in \textsc{PyBird} is given by  \citep{d_Amico_2020, d_Amico_2021}
\begin{equation}
    I_{\mathrm{resum}}^l =  4 \pi \sum_{j=0}^N \sum_{l^*}\sum_{n=1}^{n_{\mathrm{max}}} \sum_{\alpha} (-i)^{l^*} k^{2n} \mathcal{Q}_{|| N-j}^{ll^*}(n, \alpha, f) \int dq \; q^2 \left[\Xi_i(q)\right]^n \xi_j^{l^*} j_{\alpha}(kq),
    \label{eq:resum}
\end{equation}
where \(\mathcal{Q}_{|| N-j}^{ll^*}(n, \alpha, f)\) are the coefficients (see equation (4.7) of Ref.~\citep{d_Amico_2021}), the expressions for \(\left[\Xi_i(q)\right]^n\) are given in equation (4.4) and (4.5) of Ref.~\citep{d_Amico_2021}, \(\xi_j^{l^*}\) is the $j^{\rm th}$-loop-order piece of the Eulerian correlation function (see equation (4.2) of Ref.~\citep{d_Amico_2021}), and \(j_{\alpha}\) is the \(\alpha^{\mathrm{th}}\) order spherical Bessel function. \(l\) and \(l^*\) here denote the multipole of the power spectrum or correlation function, \(n\) is the integer controlling the expansion, and \(N\) denotes the highest loop order for the predictions, which is \(N=1\) (one loop) in this work. The default setting of \textsc{PyBird} is \(n_{\mathrm{max}}=8\) when analysing only monopole and quadrupole and \(n_{\mathrm{max}}=16\) when analysing all three multipoles for the power spectrum if \(k_{\mathrm{max}} < 0.45 h\mathrm{Mpc}^{-1}\), \(n_{\mathrm{max}} = 20\) if \(k_{\mathrm{max}} \geq 0.45 h \mathrm{Mpc}^{-1}\). For correlation function analysis, we set \(n_{\mathrm{max}}=20\) regardless of the multipoles. The galaxy power spectrum (\(P_{g, IRRS}^l\)) after the IR-resummation is given by 
\begin{equation}
    P_{g, IRRS}^l (k) = P_g^l (k) + I_{\mathrm{resum}}^l (k), 
    \label{eq:resum_Pg}
\end{equation}
where the galaxy power spectrum multipoles are 
\begin{equation}
    P_g^l (k) = \frac{2l + 1}{2} \int_{-1}^1 P_g (k, \mu) \mathcal{L}_l(\mu) d\mu. 
\end{equation}
Here, \(\mathcal{L}_l\) is the \(l^{th}\) order Legendre Polynomial. To correct for the Alcock-Paczynski (AP) effect, we first find the galaxy power spectrum \(P_{g, IRRS}\) with 
\begin{equation}
    P_{g, IRRS} (k, \mu) = \sum_l P_{g, IRRS}^l(k) \mathcal{L}_l(\mu). 
    \label{eq:Pgg_IRRS}
\end{equation}
We then applied the AP effect with 
\begin{equation}
    P_{g, IRRS, AP}^l (k, \mu) = \frac{2l + 1}{2f_{\perp}^2(\alpha, q)f_{\parallel}(\alpha, q)} \int_{-1}^{1} P_{g, IRRS} (k^*(k, \mu), \mu^*(\mu)) \mathcal{L}_l(\mu) d\mu,
    \label{eq: Pgg_IRRS_AP}
\end{equation}
where \(f_{\parallel}(\alpha, q)\) is given by 
\begin{equation}
f_{\parallel}(\alpha, q) = \left\{ \begin{array}{rcl}
\alpha_{\parallel} & \mbox{for}
& \textit{Shapefit} \\ q_{\parallel} & \mbox{for} & \textit{Full-Modelling} 
\end{array}\right.
\end{equation}
and \(f_{\perp}(\alpha, q)\) is given by 
\begin{equation}
f_{\perp}(\alpha, q) = \left\{ \begin{array}{rcl}
\alpha_{\perp} & \mbox{for}
& \textit{Shapefit}. \\ q_{\perp} & \mbox{for} & \textit{Full-Modelling}.  
\end{array}\right.
\end{equation}
The AP correction for \textit{Full-Modelling} uses the AP parameters \(q_{\perp}\) and \(q_{\parallel}\) because there is no template in the \textit{Full-Modelling} approach. The AP parameters are given by   
\begin{equation}
    q_{\perp} = \frac{D_M(z)}{D_M^{\mathrm{fid}}(z)}
    \label{eq: q_perp}
\end{equation}
and 
\begin{equation}
    q_{\parallel} = \frac{H^{\mathrm{fid}}(z)}{H(z)},
\end{equation}
where \(D_M\) is the angular diameter distance, \(H\) is the Hubble parameter, and ``\(\mathrm{fid}\)" here denotes the fiducial cosmology used to convert the redshift to distances, and. In contrast, \textit{Shapefit} compares the model power spectrum to the template power spectrum, so we use the \(\alpha\) parameters defined in equation~(\ref{eq: alpha_perp}) and (\ref{eq: alpha_par}).  Additionally, \(k^*\) and \(\mu^*\) in equation~(\ref{eq: Pgg_IRRS_AP}) are given by
\begin{equation}
    k^*(k, \mu) = \frac{k}{f_{\perp}(\alpha, q)}\left[1+\mu^2\left(\frac{f_{\perp}^2(\alpha, q)}{f^2_{\parallel}(\alpha, q)}-1\right)\right]^{\frac{1}{2}},
    \label{eq:k_star}
\end{equation}
and 
\begin{equation}
    \mu^*(\mu) = \mu\frac{f_{\perp}(\alpha, q)}{f_{\parallel}(\alpha, q) }\left[1+\mu^2\left(\frac{f_{\perp}^2(\alpha, q)}{f_{\parallel}(\alpha, q)^2}-1\right)\right]^{-\frac{1}{2}}.
    \label{eq:mu_star}
\end{equation}

\subsubsection{Analytical marginalization}
The previous section shows \textsc{PyBird} can have up to ten free parameters for each redshift bin. We combine data from three different redshift bins in our analysis, so we will have \(3\times10 + 4 = 34\) free parameters assuming a \(\Lambda\)CDM cosmology. More free parameters mean a longer time for the MCMC to converge. To speed up the code, \textsc{PyBird} analytically marginalizes over all nuisance parameters except \(b_1, b_2\) and \(b_4\). We cannot analytically marginalize over them because they are the only parameters that do not enter at linear order in equation (\ref{eq: GPS}). Ref.~\citep{d_Amico_2020} provides the analytical formula for the marginalized likelihood:
\begin{equation}
    \log \mathcal{L} = \frac{1}{2} F_{1, i} F_{2, ij}^{-1} F_{1, j} + F_0 - \frac{1}{2} \ln{|\det(F_2)|} 
    \label{eq: Marg_L}
\end{equation}
where 
\begin{align}
    &F_{2, ij} = (P^W_{,b_{G_{i}}})^T \hat{\icov} P^W_{,b_{G_{j}}} \nonumber \\
    &F_{1, i} =  (P^W_{\mathrm{const}})^T \hat{\icov} P^W_{,b_{G_{i}}} + P_d^T \hat{\icov} P^W_{,b_{G_{i}}} \nonumber \\
    &F_0 = -\frac{1}{2}(P^W_{\mathrm{const}})^T \hat{\icov} P^W_{\mathrm{const}} + (P^W_{\mathrm{const}})^T \hat{\icov} P_d - \frac{1}{2}P_d^T \hat{\icov} P_d, \label{eq:F_fun}
\end{align}
and 
\begin{align}
    P^W_{,b_{G_{i}}} &= \frac{\partial P^W}{\partial b_{G_{i}}}\biggl|_{\Vec{b_G} \rightarrow 0}, \nonumber \\
    P^W_{\mathrm{const}} &= P^W\bigl|_{\Vec{b_G} \rightarrow 0},
 \label{eq: window_power}
\end{align}
with $\Vec{b_G} = \{b_3, c_{ct}, c_{r, 1}, c_{r, 2}, c_{\epsilon, 1}, c_{\epsilon, 2}, c_{\epsilon, 3}\}$ and \(P_d\) denotes the data power spectrum. Generally, \(P^{W}\) denotes the IR resumed model power spectrum multipoles (\(P_{g, IRRS}^l\)) after convolving with the survey window function. However, this work does not have the survey geometry window function because we use cubic box mocks. In this case, the window function will be the binning matrices which account for the binning effect. Lastly, \(\hat{\icov}\) is the inverse covariance matrix after being multiplied by the Hartlap factor in equation (\ref{eq:Hartlap}) and the Percival factor in equation (\ref{eq:m1}). Fig.~\ref{fig:Marg_vs_no_marg} in Appendix \ref{sec:marg} demonstrates numerically that the unmarginalized likelihood gives the same constraints on cosmological parameters as the marginalized likelihood with the mean of the LRG mocks, so for the remainder of this work, we consider only the analytically marginalized constraints. 

\subsection{\textit{ShapeFit}}
Unlike the \textit{Full-Modelling} method, \textit{ShapeFit} first compares the data power spectrum (measured assuming a fiducial redshift to distance relationship) to a template power spectrum generated from a chosen set of template cosmological parameters (noting that the template and fiducial cosmological parameters do not have to be the same, but are usually set to be for ease of analysis). In this study, we set the template cosmology to the fiducial cosmology.\footnote{Ref.~\citep{KP5s3-Noriega} investigates the effect of setting the template cosmology differently from the fiducial cosmology. They find it mainly affects the constraints on \(m\). A 10\% shift in template cosmology from the fiducial cosmology causes a \(1\sigma\) shift for constraints on \(m\) and \(\Omega_{\mathrm{cdm}}h^2\).} By shifting the template power spectrum in specific ways, we can then constrain model-independent parameters \(\alpha_{\perp}\), \(\alpha_{\parallel}\) (which primarily act to shift and stretch the BAO wiggles of the template power spectrum), \(f\sigma_{s8}\) \footnote{In the standard approach, we fix \(\sigma_{s8}\) to its fiducial value and vary \(f\). Ref.~\citep{KP5s2-Maus} suggests varying \(f\) and \(\sigma_{s8}\) separately gives consistent constraints on \(f\sigma_{s8}\) with the standard approach.} (which changes the ratio of the amplitudes of multipoles in the power spectrum), \(m\) (which changes the slope of the power spectrum) \citep{Brieden_2021}.\footnote{There is another Shapefit parameter \(n\) which also changes the slope of the power spectrum. However, \(n\) only quantifies the changes in slope due to different scalar indices \(n_s\). This work studies cosmological models with fixed \(n_s\). Ref.~\citep{KP5s2-Maus, KP5s3-Noriega} investigate models with varying \(n_s\). They find that without the prior from Planck, the degeneracy between \(m\) and \(n\) causes \textit{ShapeFit} to return wider constraints on \(n_s\) than \textit{Full-Modelling}.} These four parameters will be referred to as the \textit{ShapeFit} parameters from hereon. 

As a second step, we can use our knowledge of the fiducial and template cosmological parameters to convert the \textit{ShapeFit} parameters to cosmological parameters encoded in the data. To do this, we start with the definitions of the \textit{ShapeFit} parameters \citep{Brieden_2023}:\footnote{The ``$f\sigma_{s8}$" parameter is actually \(f\sigma_{s8} \approx r_A(f\sigma_{s8})^{\mathrm{tem}} \times \exp{\left(\frac{m}{2a}\tanh{\left(a\ln{(\frac{r_d^{\mathrm{tem}}}{8})}\right)}\right)}\) because \(m\) not only changes the tilt of the power spectrum, but also the amplitude. Therefore, if we integrate over the power spectrum, we get this extra exponential factor from integration by parts. However, we only use the amplitude at the pivot scale during the conversion. Equation (\ref{eq: m}) demonstrates \(m\) will not change the amplitude at the pivot scale, so we do not need to multiply this extra factor during the conversion \citep{Brieden_2021, Brieden_2023}.} 
\begin{equation}
    \alpha_{\perp} = \frac{q_{\perp}}{s}, 
    \label{eq: alpha_perp}
\end{equation}
\begin{equation}
    \alpha_{\parallel} = \frac{q_{\parallel}}{s},
    \label{eq: alpha_par}
\end{equation}
\begin{align}
    f\sigma_{s8} &= \frac{(f\sigma_{s8})^{\mathrm{tem}}}{\left(f \left( P_{\mathrm{\mathrm{lin}}}\left(k_p\right)\right)^{1/2}\right)^{\mathrm{tem}}}f\left(\frac{1}{s^3} P_{\mathrm{\mathrm{lin}}}\left(k_p/s\right)\right)^{1/2} \nonumber \\  
    &= r_A(f\sigma_{s8})^{\mathrm{tem}},
    \label{eq: fsigma8}
\end{align}
and 
\begin{equation}
    \ln{\left(\frac{P_{\mathrm{lin}}(k)}{P^{\mathrm{tem}}_{\mathrm{lin}}(k)}\right)} = \frac{m}{a} \tanh{\left[a\ln{\left(\frac{k}{k_p}\right)}\right]} + n \ln{\left(\frac{k}{k_p}\right)}, 
    \label{eq: m}
\end{equation}

\begin{equation}
    s = \frac{r_d}{r_d^{\mathrm{tem}}}
    \label{eq: s_ratio}. 
\end{equation}
Here, \(P_{\mathrm{lin}}^{\mathrm{tem}}\) denotes the template linear power spectrum, \(r_d\) is the sound horizon at the drag epoch, and the pivot scale \(k_p \approx \frac{\pi}{r_d} \approx 0.03 h \mathrm{Mpc}^{-1}\). We follow Ref.~\citep{Brieden_2021} to set \(a = 0.6\), and \(n = 0\) because we fix the tilt of the primordial power spectrum \(n_s\). The unit for \(D_M, H(z)\), and \(r_d\) are all in \(h^{-1} \mathrm{Mpc}\). Lastly, ``\(\mathrm{tem}\)" denotes the template cosmology used to generate the template power spectrum.  

Several differences exist between implementing \textit{ShapeFit} and \textit{Full-Modelling} even in the same model. Firstly, \textit{ShapeFit} does not require recalculating the linear matter power spectrum for each iteration of MCMC. Secondly, although the scaling term in the right-hand-side of equation~(\ref{eq: m}) depends on \(k\), Ref.~\citep{Brieden_2021} suggests we can take the scaling terms out of the loop integrals. Therefore, we can multiply the scaling terms by the template loop terms to get the new ones similar to scaling the power spectrum with \(\sigma_8\) (see equation (3.9) of Ref.~\citep{Brieden_2021} for more detail). Equation (\ref{eq:resum}) shows only the matrix \(\mathcal{Q}_{|| N-j}^{ll^*}(n, \alpha, f)\) depends on the growth rate \(f\), so we only have to recompute the matrix \(\mathcal{Q}_{|| N-j}^{ll^*}(n, \alpha, f)\) for each iteration. Ref.~\citep{d_Amico_2021} derived the analytical expression for \(\mathcal{Q}_{|| N-j}^{ll^*}(n, \alpha, f)\), so the computation is relatively fast. Then, we can rescale the rest of equation (\ref{eq:resum}) with \(\left(\frac{P_{\mathrm{lin}}(k)}{P_{\mathrm{lin}}(k)^{\mathrm{tem}}}\right)^{n + 1}\) for the IR resummation of the linear and counter terms and \(\left(\frac{P_{\mathrm{lin}}(k)}{P_{\mathrm{lin}}(k)^{\mathrm{tem}}}\right)^{n + 2}\) for the IR resummation of the loop terms. The definition of the expansion order \(n\) is in equation (\ref{eq:resum}). Whether or not the constraints from \textit{ShapeFit} can reproduce those from the more complete \textit{Full-Modelling} approach hence depends on the validity of this approximation as well as the degree to which the four compressed \textit{ShapeFit} parameters capture all the cosmological model dependence in the power spectrum.

Furthermore, the first step of \textit{ShapeFit} is model-independent, so we only have to assume a cosmological model when we convert the \textit{ShapeFit} parameters to the cosmological parameters. Therefore, if we want to fit different models, we only need to redo the conversion and do not need to repeat the first step. On the other hand, we can fit the cosmological parameters directly for \textit{Full-Modelling}. The balance of speed between the two methods hence depends on how long it takes to fit the power spectrum model and how long a given conversion to cosmological parameters takes. In section \ref{sec:comparison}, we compare the speed of \textit{Full-Modelling} and \textit{ShapeFit} with or without the Taylor expansion emulators for \(\Lambda\)CDM and \(w\)CDM cosmologies with our setting of \textsc{PyBird}. 

In \textsc{PyBird}, \textit{ShapeFit} is implemented as follows. Firstly, we generate the linear power spectrum from \textsc{CLASS} \citep{Diego_Blas_2011} or \textsc{CAMB} \citep{Lewis_2011} using the template cosmological parameters. We then compute and store the power spectrum's loop, counter, and IR-resummation terms with one-loop EFTofLSS. MCMC will propose different \textit{ShapeFit} parameters for each iteration, and we first use equation (\ref{eq: m}) to calculate the new linear power spectrum with the new slope \(m\). We then use the new slope to rescale the template loop and IR resummation terms described in the previous paragraph. Next, we add the IR resummation terms to the linear and loop power spectra. Lastly, we substitute the input \(\alpha\) parameters in equation~(\ref{eq: Pgg_IRRS_AP}) and convolve the window function with the linear and loop power spectra.\footnote{We are using the cubic box mocks so that we will multiply the pre-computed binning matrix to power spectra.} We repeat the last three steps for each MCMC iteration until the chain converges. 

\subsection{Converting \textit{ShapeFit} parameters to cosmological parameters}
Firstly, we need to determine the desired cosmological model before converting \textit{ShapeFit} parameters into cosmological parameters. In this work, we will use \textsc{PyBird} to constrain parameters in the \(\Lambda\)CDM, \(w\)CDM, and \(o\)CDM models. For all models, we put a flat prior on all cosmological parameters except a Big Bang Nucleosynthesis (BBN) prior on \(\omega_b\).\footnote{Ref.~\citep{KP5s2-Maus, KP5s3-Noriega} investigate the effect of relaxing the BBN prior on \textit{ShapeFit} and \textit{Full-Modelling}, finding that this significantly weakens the constraints on \(\Omega_bh^2\) and \(h\). The wider constraints for \textit{ShapeFit} are because the distances extracted from the BAO feature only constrain \(r_d \times h\). Without a prior on BBN to constrain \(r_d\), we obtain a weaker constraint on \(h\). \textit{Full-Modelling}, on the other hand, can constrain \(\Omega_bh^2\) (and \(r_d\) by extension) directly from the amplitude of the BAO wiggles. However, this constraint is much weaker than that from the BBN prior.} The detail of the priors is in section \ref{sec: config}. Next, we approximate the posteriors for the Shapefit parameters to be Gaussian distributed with covariance matrix \(C\) and mean \(q_{\mathrm{mean}}\). To calculate the covariance matrix and the mean, we first remove the burn-in with \textsc{ChainConsumer} and then calculate the covariance matrix and the mean for the \textit{ShapeFit} parameters after marginalizing over the nuisance parameters. With this approach, the log-likelihood function for the conversion is given by 
\begin{equation}
    \log{\mathcal{L}} = -\frac{1}{2}(q - q_{\mathrm{mean}})C^{-1}(q-q_{\mathrm{mean}})^T. 
    \label{eq: lnL}
\end{equation}
For each iteration, we use \textsc{CLASS} to find \(r_d, D_M(z), H(z), P_{\mathrm{lin}}\) based on the cosmological parameters proposed by the MCMC. Then, we use the Eisenstein-Hu fitting formula \citep{Eisenstein_1999} to find \(T_{\mathrm{lin}}^{NW}\) (the no-wiggle transfer function). The \(\alpha\) parameters are found with equation (\ref{eq: alpha_perp}) and (\ref{eq: alpha_par}). \(f\sigma_{s8}\) can be found by computing equation ~(\ref{eq: fsigma8}). Lastly, the slope of the power spectrum is given by 
\begin{equation}
      m = \frac{d}{d \ln{k}}\left.\left(\frac{(T_{\mathrm{lin}}^{NW})^2}{s^3 (T_{\mathrm{lin}}^{NW, \mathrm{fid}})^2}\right)\right\vert_{k_p}.
    \label{eq: m_convert}
\end{equation}
There are three different ways of calculating the no-wiggle transfer function. First is through the Eisenstein-Hu fitting formula \citep{Eisenstein_1999} implemented in the \textsc{Nbodykit} package \citep{Hand_2018}. The second is through the polynomial fitting formula in Ref.~\citep{Hinton_2017}. The third is through the spectral decomposition from Ref.~\citep{Hamann_2010, Wallisch_2018}. We use the implementation of the polynomial fitting formula and the spectral decomposition from \textsc{BARRY} \citep{Hinton_2020}. Appendix \ref{sec:conversion} shows that these three de-wiggle algorithms give consistent results for \(\Lambda\)CDM, \(w\)CDM, and \(o\)CDM. Therefore, we decided to use the Eisenstein-Hu fitting formula for the rest of this work. 

This conversion assumes the posteriors of the \textit{ShapeFit} parameters are Gaussian distributed. To test this, we also directly interpolate the posterior of the Shapefit chain as the new likelihood during the MCMC rather than assuming Gaussianity, which enables us to capture any non-Gaussian information. However, we generally find consistent constraints with the Gaussian approximation and the linear interpolation methods. We decided to use the Gaussian approximation for the rest of this work because creating the linear interpolator for \textit{ShapeFit} parameters takes longer than computing the covariance matrix and the mean of the \textit{ShapeFit} parameters.     

\subsection{Computational efficiency}
\label{sec:speed}

Generally, the bottleneck of the \textit{Full-Modelling} methodology is calculating the model power spectrum. To overcome this problem, Ref.~\citep{Colas_2020} uses Taylor expansion to approximate the model power spectrum from \textsc{PyBird} during MCMC. They find the maximum deviation between the Taylor expansion and exact calculation is under 1\% for the monopole and 10\% for the quadrupole of the power spectrum for the \(\Lambda\)CDM cosmological model with massive neutrinos. They concluded that the errors have a negligible effect on the constraints of cosmological parameters. In this work, we follow Ref.~\citep{Colas_2020} to approximate the model power spectrum during the MCMC with a third-order Taylor expansion about models evaluated at equally spaced points within the cosmological parameter space. Since Ref.~\citep{Colas_2020} find the Taylor expansion has a negligible effect on the constraints on the cosmological parameters in the \(\Lambda\)CDM and \(\nu \Lambda\)CDM models, we only verify the accuracy of the Taylor expansion for the $w$CDM and \(o\)CDM models in Appendix \ref{sec:Taylor_grid}. Our finding shows that the Taylor expansion's grid spacing does not impact our final constraints of cosmological parameters in \(w\)CDM and \(o\)CDM.

Similarly, the conversion from \textit{ShapeFit} parameters to cosmological parameters can be computationally expensive. To speed this process up, we similarly compute the \textit{ShapeFit} parameters from a grid of cosmological parameters, again using the third-order Taylor expansion to interpolate between grid points during the MCMC. We find this interpolation gives consistent constraints \citep{KP5s1-Maus}. We provide the configuration of the grids for the various models in Section \ref{sec: config}. 

\section{Summary of main results}
\label{sec:result}
This section introduces the configurations utilized to fit the DESI mocks. We test four different prior configurations in \textsc{PyBird}, with a primary focus on the ``BOSS MaxF" configuration as detailed in Ref.~\citep{d_Amico_2020}. Subsequently, we provide an overview of the constraints on cosmological parameters in the \(\Lambda\)CDM model for both \textit{ShapeFit} and \textit{Full-Modelling} using a combination of three tracers. This analysis confirms that both methods return consistent and unbiased results at a level suitable for DESI analyses.
\subsection{Configuration}
\label{sec: config}
For \textit{ShapeFit}, the template cosmological parameters in this work are \(\ln{(10^{10} A_s)} = 3.0364\), \(h = 0.6736\), \(\omega_{cdm} = 0.12\), \(\omega_b = 0.02237\), \(w = -1.0\), and \(\Omega_k = 0.0\). These parameters are the same as the fiducial parameters used to make the clustering measurements and the true cosmological parameters of the \textsc{AbacusSummit} mocks introduced in Section~\ref{sec:mock}. For the grid computation, we use nine grid points (one at the centre, four at each positive/negative direction) for each cosmological or \textit{ShapeFit} parameter. 

The priors of the cosmological and \textit{ShapeFit} parameters in this work are summarized in Table \ref{tab: pybird_priors}. The Gaussian prior on \(\omega_b\) represents current results from BBN \citep{Cooke_2018}. The four different prior configurations for our bias and nuisance parameters are summarized in Table \ref{tab:bias_prior}. This work mainly uses the ``BOSS MaxF" and ``BOSS MinF" priors taken from Ref.~\citep{d_Amico_2020} with small changes to fit the DESI mocks. The results using the ``MaxF" and ``MinF" priors, where the priors for the counter terms and stochastic terms are completely free, are mainly adopted in the comparison paper \citep{KP5s1-Maus} to ensure consistent degrees of freedom between different models of the power spectrum. In both ``BOSS MinF" and ``MinF" configurations, \(b_2\) and \(b_3\) are calculated using the local Lagrangian approximation \citep{Desjacques_2018}. The derivations of these local Lagrangian relations are in Appendix \ref{sec:bias_transform}. 

\begin{table}[t!]
\centering                          
\renewcommand{\arraystretch}{1.4} 
\begin{tabular}{c|c}        
Full-Modeling & ShapeFit\\ \hline \hline
H$_0$ &  $f\sigma_8$\\
$\mathcal{U}[55.36,79.36]$ & $\mathcal{U}[0,1]$ \\
\hline
$\omega_{\mathrm{b}}$ &  $\alpha_{\parallel}$ \\
$\mathcal{N}[0.02237,0.00037]$ & $\mathcal{U}[0.9,1.1]$ \\
\hline
$\omega_{\mathrm{cdm}}$ &  $\alpha_{\perp}$ \\
$\mathcal{U}[0.08,0.16]$ & $\mathcal{U}[0.9,1.1]$ \\
\hline
$\log(10^{10} A_\mathrm{s})$& $m$\\
$\mathcal{U}[2.0364,4.0364]$ & $\mathcal{U}[-0.4,0.4]$ \\
\hline
$w$& \\
$\mathcal{U}[-1.5, -0.5]$ & \\
\hline
$\Omega_k$&\\
$\mathcal{U}[-0.25, 0.25]$& 
\end{tabular}
\caption{Priors on the \textit{ShapeFit} and cosmological parameters in this work. We use \(\mathcal{U}\) (\(\mathcal{N}\)) to denote the uniform (Gaussian) prior. The first value is the lower bound (mean), and the second is the upper bound (standard deviation). The boundaries for the flat priors on cosmological parameters are also the boundaries of the grid used to interpolate the power spectrum.}    
\label{tab: pybird_priors} 
\end{table}

\begin{table}[t!]
    \centering
    {\renewcommand{\arraystretch}{1.4}
    \begin{tabular}{c|c|c|c|c}
         & BOSS MaxF  & BOSS MinF & MaxF & MinF\\ \hline \hline
        $b_1$ & $\mathcal{U}[0.0,4.0]$ & $\mathcal{U}[0.0,4.0]$ & $\mathcal{U}[0.0,4.0]$ & $\mathcal{U}[0.0,4.0]$\\
        \hline
        $b_2$ & $\mathcal{U}[-15.0,15.0]$ & $1.0$ & $\mathcal{U}[-15.0,15.0]$ & $1.0$\\
        \hline
        $b_3$ & $\mathcal{N}[0,2]$ & $\frac{882-3045(b_1 - 1)}{882}$ & $\mathcal{N}[0,10]$ & $\frac{882-3045(b_1 - 1)}{882}$\\
        \hline
        $b_4$ & $b_2$ & $\mathcal{U}[-15.0,15.0]$ & $\mathcal{U}[-15.0,15.0]$ & $\mathcal{U}[-15.0,15.0]$ \\
        \hline
        $c_{ct}$ & $\mathcal{N}[0,2]$ & $\mathcal{N}[0,2]$ & $\mathcal{U}[-\infty,\infty]$ & $\mathcal{U}[-\infty,\infty]$\\
        \hline
        $c_{r,1}$ & $\mathcal{N}[0,8]$ & $\mathcal{N}[0,8]$ & $\mathcal{U}[-\infty,\infty]$ & $\mathcal{U}[-\infty,\infty]$\\
        \hline
        $c_{r,2}$ & 0 & 0 & 0 & 0\\
        \hline
        $c_{\epsilon, 1}/\overline{n_g}$ & $\mathcal{N}[0,800]$ & $\mathcal{N}[0,800]$ & $\mathcal{U}[-\infty,\infty]$ & $\mathcal{U}[-\infty,\infty]$\\
        \hline
        $c_{\epsilon, 2}/\overline{n_g}$ & $\mathcal{N}[0,\frac{20000}{3}]$ & $0$ & 0 & 0 \\
        \hline
        $c_{\epsilon, 3}/\overline{n_g}$ & $\mathcal{N}[0,\frac{20000}{3}]$ & $\mathcal{N}[0,\frac{20000}{3}]$ & $\mathcal{U}[-\infty,\infty]$ & $\mathcal{U}[-\infty,\infty]$\\
    \end{tabular}}
    \caption{The priors on the bias, counter, and stochastic terms for different prior configurations when fitting only the \textbf{monopole and quadrupole}. This paper mainly uses the ``BOSS MaxF" and ``BOSS MinF" configurations. The "MaxF" and "MinF" configurations are used in the comparison paper \citep{KP5s1-Maus} to compare with other pipelines in DESI. For both ``BOSS MinF" and ``MinF" configurations, the values for \(b_2\) and \(b_3\) are determined by the local Lagrangian approximation. The derivation is in the Appendix \ref{sec:bias_transform}. When we also fit the \textbf{hexadecapole}, for ``BOSS MaxF" and ``BOSS MinF" configurations, we follow Ref.~\citep{d_Amico_2020} to change \(c_{r,1}\) prior to $\mathcal{N}[0,4]$ and put a $\mathcal{N}[0,4]$ prior on \(c_{r,2}\). For the ``MaxF" and ``MinF" configurations, we impose an infinite flat prior on \(c_{r,2}\). }
    \label{tab:bias_prior}
\end{table}

There are 25 different mocks for each redshift bin from the \textsc{AbacusSummit} simulation suite. The volume of the \textsc{AbacusSummit} simulation box is 8 \(\mathrm{Gpc}^3/h^3\), which is similar to the volume expected for the complete DESI survey. As such, we primarily use the covariance matrix corresponding to a single realization (hereon referred to as the single-box covariance matrix) to calculate the possible systematics in the pipeline. Tests using a $25\times$ reduced covariance matrix (representing the error on the mean) are mainly used to compare the differences between different pipelines in the comparison paper \citep{KP5s1-Maus}. Doing so enables us to detect if there are any systematics that will affect our measurements at the level of the Year 1 analysis (containing approximately 1/5 of the complete DESI data) while avoiding misleading results that may arise from the fact that the simulations are not necessarily converged to a sufficient degree on non-linear scales (i.e., Ref.~\citep{Grove_2022}, which shows that different simulation codes can differ in their predictions for the power spectrum on non-linear scales at a level comparable to the precision using our $25\times$ reduced covariance matrix).

\subsection{Main results}
\begin{figure}
    \centering
	\includegraphics[width=1.0\textwidth, trim={20pt 15pt 20pt 20pt}, clip]{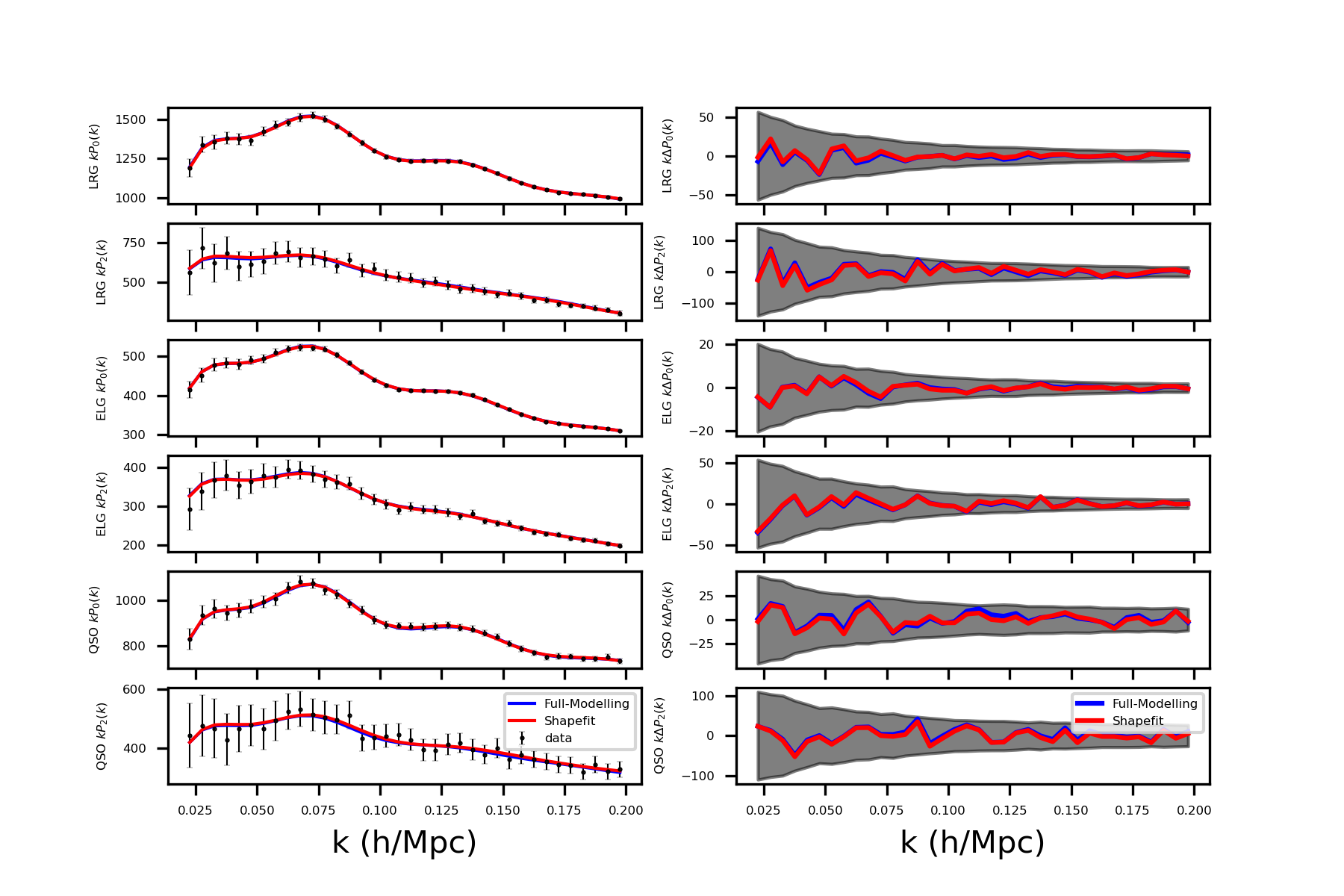}
    \caption{This plot shows the best-fit model of the power spectrum from the LRG, ELG, and QSO mocks for both \textit{Full-Modelling} and \textit{ShapeFit} methods with the second fitting configuration. We use \(k_{\mathrm{max}} = 0.20 h \mathrm{Mpc}^{-1}\), ``BOSS MaxF" prior, and without the hexadecapole to generate this plot. On the left, the error bar is given by the single box covariance matrix, which indicates the statistical error from the DESI Y5 survey. On the right, the shaded area is the data power spectrum uncertainty, and the lines are the difference between the best-fit and data power spectrum. The best-fit power spectrum from both \textit{ShapeFit} and \textit{Full-Modelling} are in good agreement. The maximum deviation between the two is from the QSO because the power spectrum from QSO has a larger statistical error, which gives these two models more freedom. Lastly, the plot also shows that the best-fit models from both methods are well within the uncertainty of the measurement, indicating that our model is a good fit for the data. We present the reduced \(\chi^2\) of the best-fit power spectra in Table~\ref{tab:chi2}.}
    \label{fig:bestfit}
\end{figure}

\begin{figure}
	\includegraphics[width=0.75\textwidth]{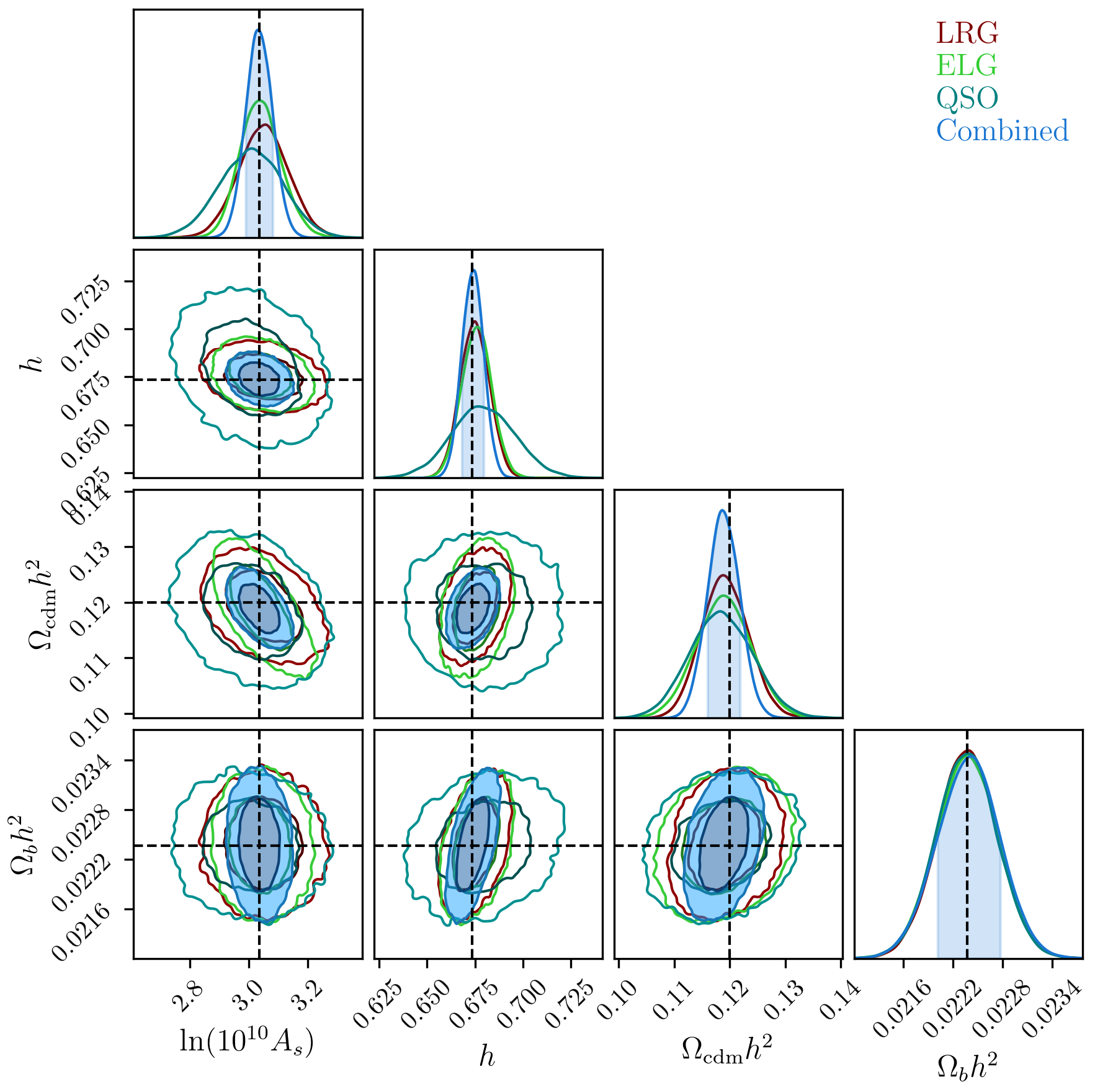}
    \caption{This plot illustrates the constraints on \(\Lambda\)CDM cosmological parameters using \textit{Full-Modelling} with \(k_{\mathrm{max}} = 0.20 h \mathrm{Mpc}^{-1}\), the ``BOSS MaxF" prior, and without the hexadecapole using the LRG, ELG, and QSO mocks, and their combination. The dashed line indicates the truth values of the cosmological parameters. The constraints on \(\omega_b\) are dominated by the BBN prior. The QSO mock has weaker constraints than the other two tracers because it has a much lower number density. Combining different tracers can significantly tighten the constraints on the cosmological parameters. Furthermore, in all cases, \textsc{PyBird} can produce unbiased constraints on all cosmological parameters.}
    \label{fig:FS_bestfit}
\end{figure}

\begin{figure}
    \centering
	\includegraphics[width=0.495\textwidth]{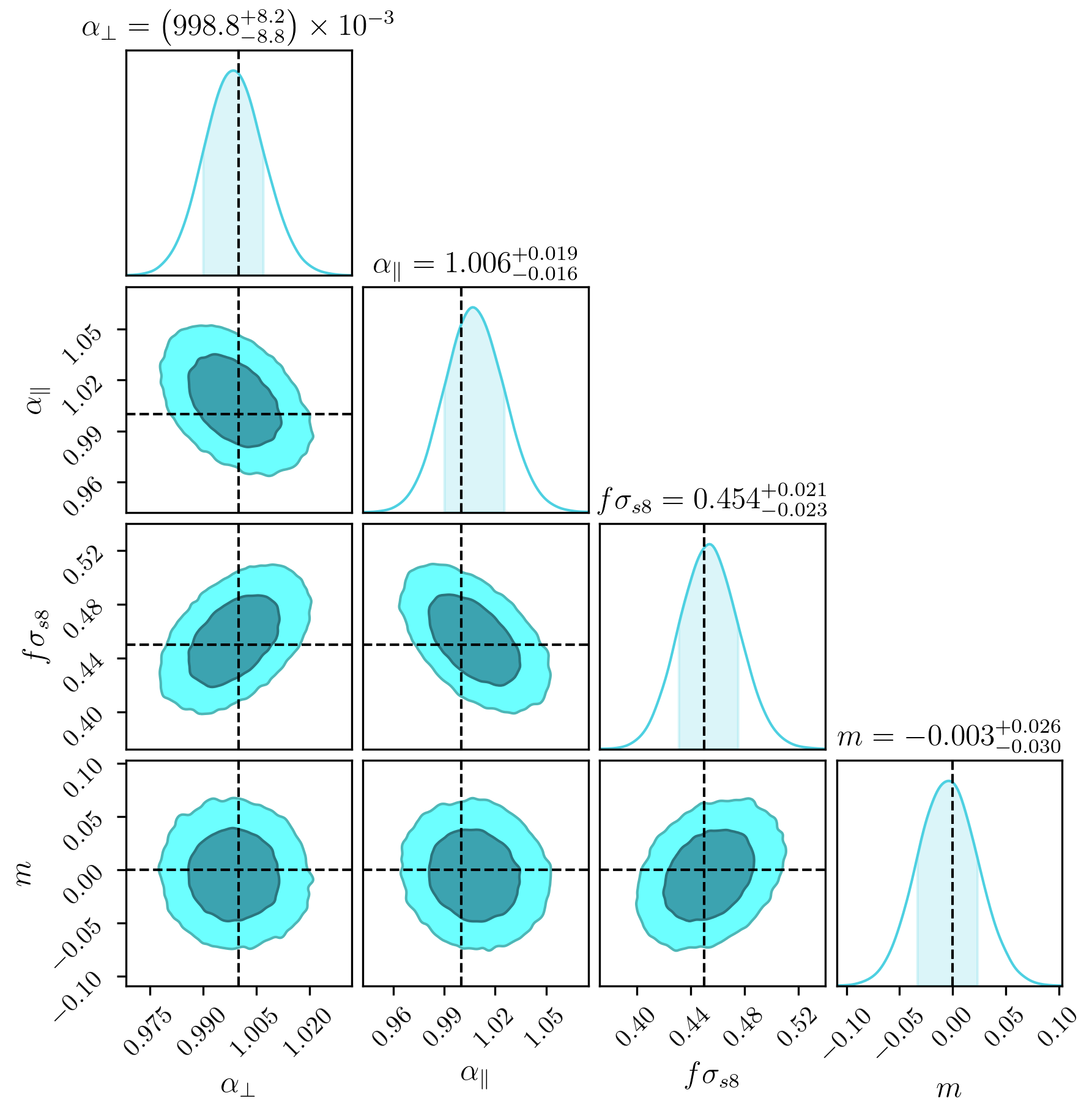}
    \includegraphics[width=0.495\textwidth]{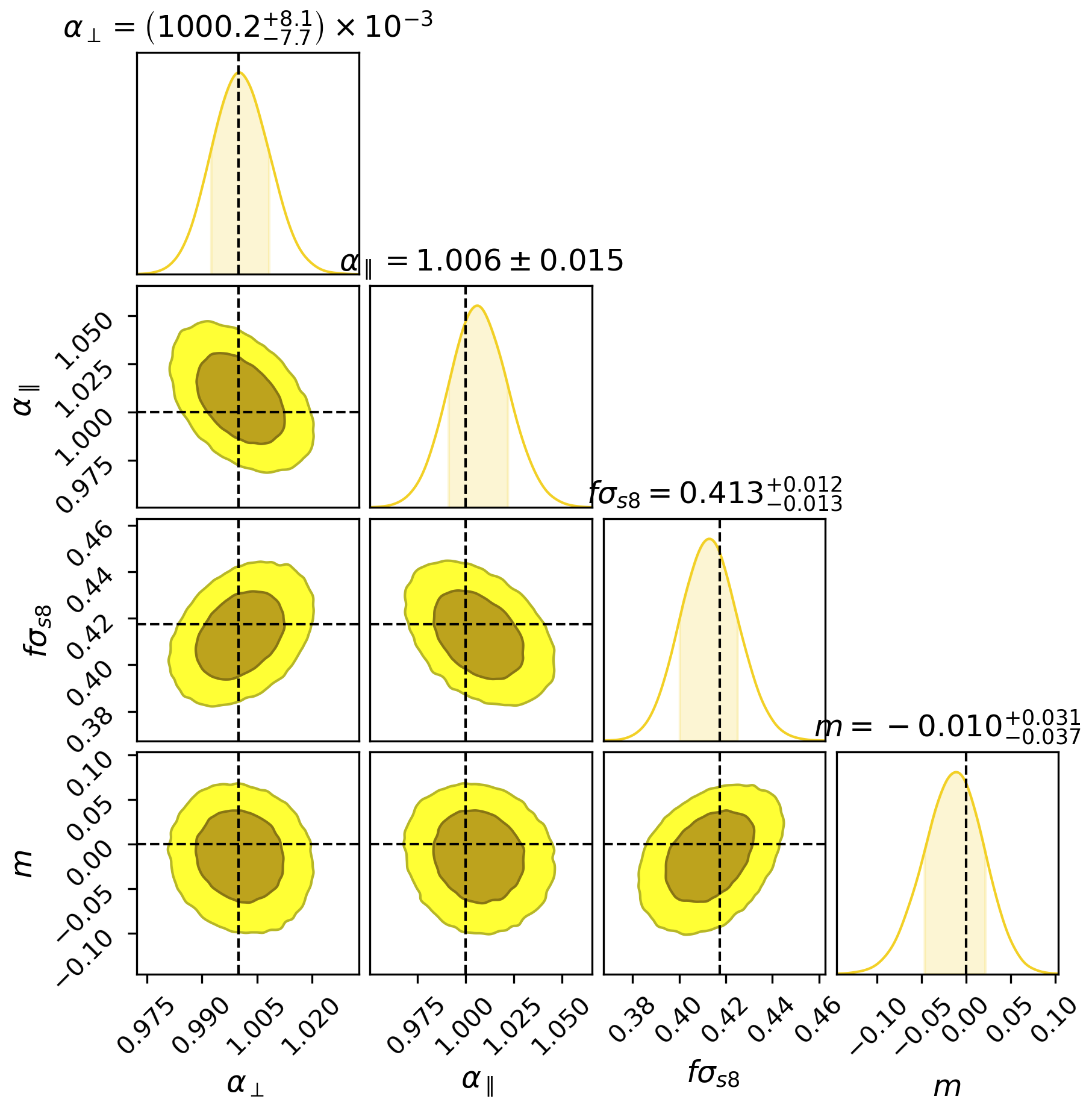}
    \includegraphics[width=0.495\textwidth]{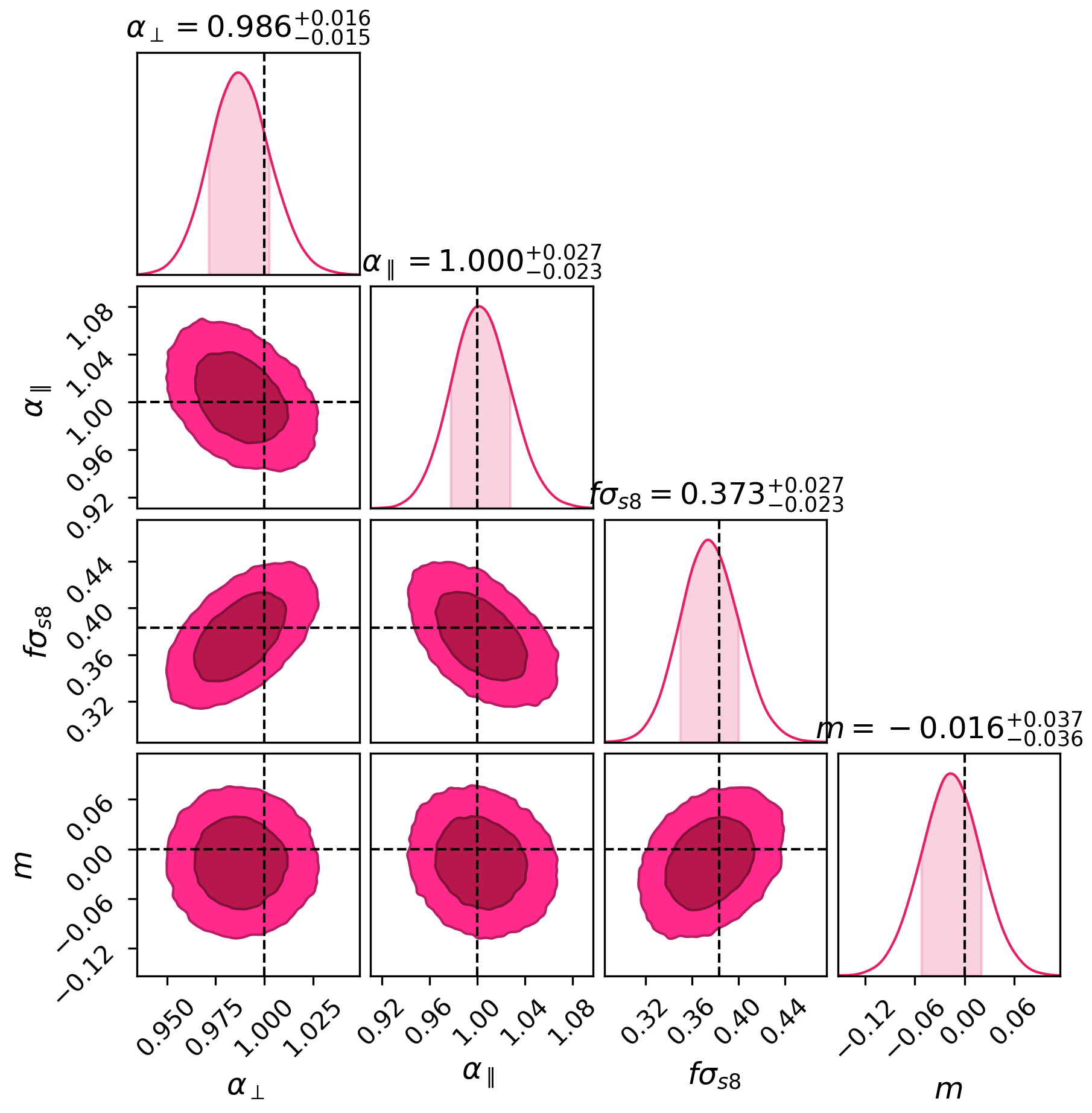}
    
    \caption{Constraints on the \textit{ShapeFit} parameters from three different tracers with \(k_{\mathrm{max}} = 0.20 h \mathrm{Mpc}^{-1}\), the ``BOSS MaxF" prior, and without the hexadecapole. The top left is the constraints from LRG, the top right is from ELG, and the bottom is from QSO. Dashed lines correspond to the simulation expectations. Similar to the \textit{Full-Modelling} constraints, \textsc{PyBird} can produce unbiased constraints on the \textit{ShapeFit} parameters for all three tracers.}
    \label{fig:SF_bestfit}
\end{figure}

\begin{figure}
    \centering
	\includegraphics[width=0.7\textwidth]{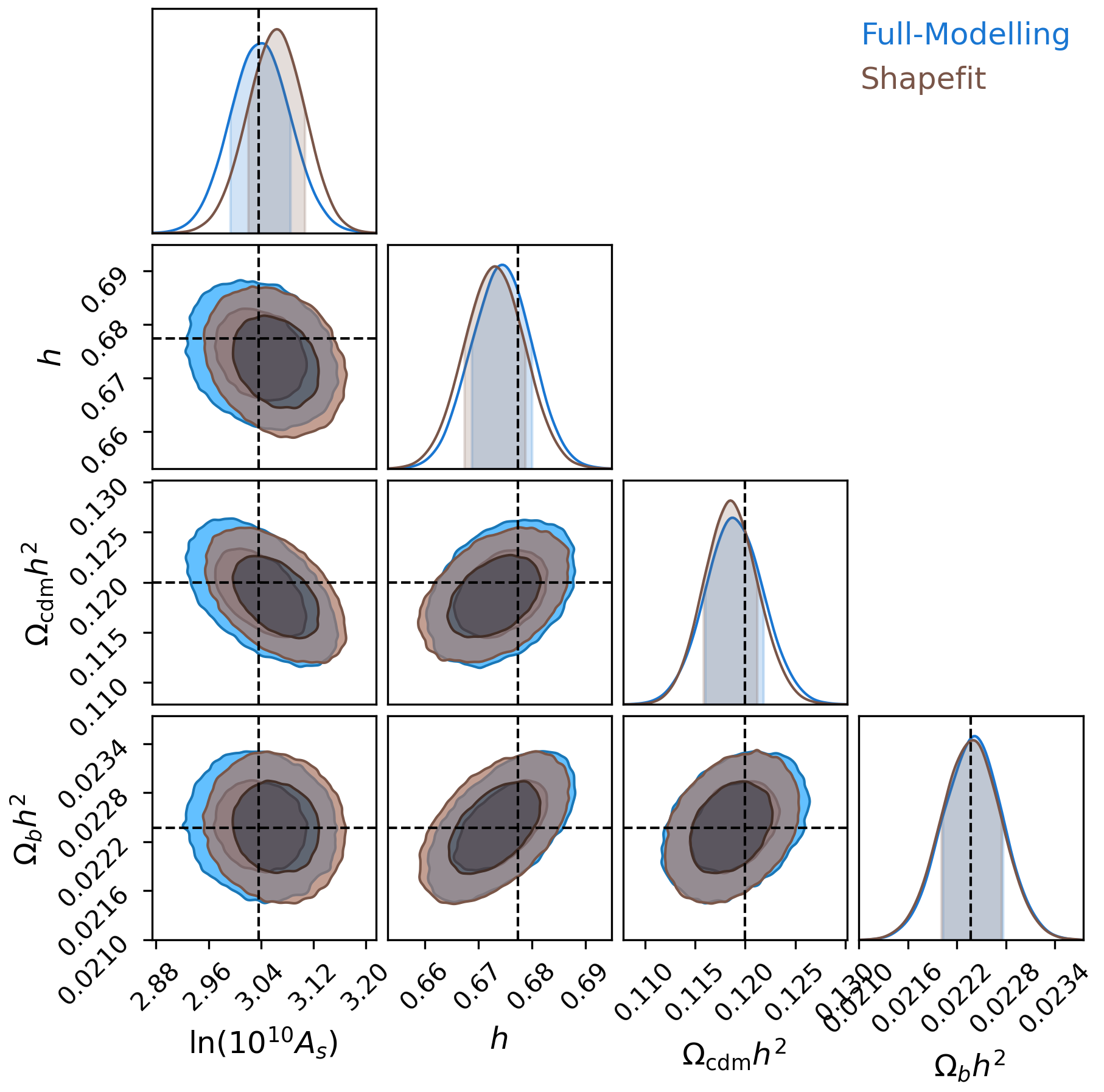}
    \caption{This plot compares the constraints on cosmological parameters using all tracers from \textit{ShapeFit} and \textit{Full-Modelling} with \(k_{\mathrm{max}} = 0.20 h \mathrm{Mpc}^{-1}\), the ``BOSS MaxF" prior, and without the hexadecapole. The constraints are in good agreement with each other except a \(\sim 0.4\sigma\) shift in \(\ln{(10^{10}A_s)}\). This systematic shift is much less than the statistical uncertainty, so it will not dominate the error bar. Table.~\ref{tab:LCDM_shift} summarizes the systematic shifts in this plot.}
    \label{fig:FS_vs_SF_bestfit}
\end{figure}

Given the above configuration, we start by presenting the main findings of this work. We examine the more nuanced effects (and potential systematic error) of fitting to different \(k_{\mathrm{max}}\) scales, with/without including the power spectrum hexadecapole, and of different prior configurations in sections \ref{sec:Shapefit} and \ref{sec:Full-Modelling}. For Fig~\ref{fig:bestfit}, \ref{fig:FS_bestfit}, \ref{fig:SF_bestfit}, and \ref{fig:FS_vs_SF_bestfit} in this section. We use \(k_{\mathrm{max}} = 0.20 h \mathrm{Mpc}^{-1}\), ``BOSS MaxF" prior, and without the hexadecapole. We use the second fitting configuration in this section. 

\begin{table}[]
\begin{tabular}{c|c|c|c|c|c|c}
Tracer   & $\chi^2$ (FM) & $D_f^{\mathrm{FM}}$ & $R_{\chi^2}$ (FM) & $\chi^2$ (SF) & $D_f^{\mathrm{SF}}$ & $R_{\chi^2}$ (SF) \\ \hline \hline
LRG      & 8.82 (70.56)     & 60       & 0.15 (1.18)              & 8.86 (70.88)     & 60       & 0.15 (1.18)              \\ \hline
ELG      & 8.78 (70.24)     & 60       & 0.15 (1.17)              & 8.71 (69.68)     & 60       & 0.15 (1.16)              \\ \hline
QSO      & 9.17 (73.36)    & 60       & 0.16 (1.22)              & 9.19 (73.52)     & 60       & 0.15 (1.23)              \\ \hline 
Combined & 28.84 (230.72)    & 188      & 0.15 (1.23)              & 26.76 (214.08)    & 180      & 0.15 (1.19)             
\end{tabular}
\caption{This table shows the goodness of fit with the \textit{Full-Modelling} (FM) and the \textit{ShapeFit} (SF) with monopole and quadrupole, the ``BOSS MaxF" prior, and the second fitting configuration. Furthermore, since we use the single box covariance matrix to fit the mean of the data average over eight mocks, we expect the reduced \(\chi^2\) (\(R_{\chi^2}\)) to be much less than \(1\) because the averaging removes most of the noise. Therefore, we also rescaled the \(\chi^2\) by eight in the bracket to show the proper reduced \(\chi^2\). The data length minus the number of free parameters in our model gives the number of degrees of freedom (\(D_f\)). We have four cosmological/\textit{ShapeFit} parameters and eight bias parameters for the \(\Lambda\)CDM cosmology with the "BOSS MaxF" prior. Furthermore, we only fit the monopole and quadrupole using \(0.02 h\mathrm{Mpc}^{-1} \leq k \leq 0.20 h\mathrm{Mpc}^{-1}\) with the bin width of \(0.005 h\mathrm{Mpc}^{-1}\). This bin width means we have 72 data points in total for each tracer. One advantage of \textit{Full-Modelling} over \textit{ShapeFit} is that it can fit multiple tracers simultaneously. For the combined fit, there are \(72\times3 = 216\) data points, \textit{Full-Modelling} has \(4 + 8\times 3 = 28\) free parameters (4 cosmological parameters plus eight bias parameters in each redshift bin). On the other hand, the \textit{ShapeFit} parameters are redshift dependent, so its total number of free parameters for the combined fit is equal to the sum of the free parameters for a single tracer fit. Nevertheless, the reduced \(\chi^2\) from both \textit{Full-Modelling} and \textit{ShapeFit} are almost the same, showing they are equally good fit to the data.}
\label{tab:chi2}
\end{table}

\begin{table}[]
\setlength{\tabcolsep}{6pt}
\begin{tabular}{c|c|c|c|c}
Method   & $\ln{(10^{10}A_s)}$ & $100h$ & $100\Omega_{\mathrm{cdm}}h^2$ & $100\Omega_b h^2$  \\ \hline \hline
FM     & $3.043_{-0.051}^{+0.040}$(3.051)     & $67.45_{-0.60}^{+0.54}$(67.31)     & $11.87_{-0.27}^{+0.30}$(11.84)      & $2.241_{-0.039}^{+0.036}$(2.243)  \\ \hline
SF      & $3.064_{-0.044}^{+0.042}$(3.065)     & $67.32_{-0.59}^{+0.57}$(67.25)       & $11.85_{-0.26}^{+0.27}$(11.85)     &  $2.240_{-0.039}^{+0.035}$(2.235)  \\ \hline
$\Delta_{\Lambda\mathrm{CDM}}$      & $0.45\sigma (0.31\sigma)$     & $0.23\sigma (0.10\sigma)$       & $0.09\sigma (0.02\sigma)$              & $0.04\sigma (0.21\sigma)$ 
\end{tabular}
\caption{This table demonstrates the constraints on cosmological parameters with \textit{Full-Modelling} (FM) and \textit{ShapeFit} (SF). We also provide the value of the best-fit parameters inside the bracket. Both \textit{Full-Modelling} and \textit{ShapeFit} provide equally precise and accurate constraints on the cosmological parameters. The differences \(\Delta_{\Lambda\mathrm{CDM}}\) between the means of the posteriors of the cosmological parameter for the two methods are less than \(0.5\sigma\), and the differences between the best-fits are around \(0.3\sigma\).}
\label{tab:LCDM_shift}
\end{table}

Fig.~\ref{fig:bestfit} illustrates the best-fit power spectra from both \textit{ShapeFit} and \textit{Full-Modelling} methods. Both are in good agreement with each other and with the data. We show the reduced \(\chi^2\) of the best-fits from \textit{Full-Modelling} and \textit{ShapeFit} in Table.~\ref{tab:chi2}. The reduced \(\chi^2\) are low because we use the mean of eight different mocks, which cancels out most of the noise in the data power spectrum. In the bracket, we re-scale the \(\chi^2\) by eight to estimate the \(\chi^2\) without removing the noise. We find both \textit{ShapeFit} and \textit{Full-Modelling} are excellent fits to the data power spectrum with similar reduced \(\chi^2\). The constraints on the cosmological parameters from \textit{Full-Modelling} are in Fig.~\ref{fig:FS_bestfit}. It shows the constraints from \textit{Full-Modelling} with \textsc{PyBird} are unbiased relative to the truth values indicated by the dashed lines. As expected, combining all tracers gives the most robust constraints on the cosmological parameters. \textsc{PyBird} can provide unbiased constraints for both the single tracer case and the combination of all tracers. Fig.~\ref{fig:SF_bestfit} illustrates the constraints of \textit{ShapeFit} parameters from three different tracers. On the top left is LRG, the top right is ELG, and the bottom is QSO. Similar to Fig.~\ref{fig:FS_bestfit}, \textsc{PyBird} can produce unbiased constraints on the \textit{ShapeFit} parameters. 

Fig.~\ref{fig:FS_bestfit} and Fig.~\ref{fig:SF_bestfit} illustrate that \textit{Full-Modelling} and \textit{ShapeFit} with \textsc{PyBird} can return unbiased constraints. Fig.~\ref{fig:FS_vs_SF_bestfit} and Table~\ref{tab:LCDM_shift} compare the constraints of cosmological parameters from \textit{ShapeFit} and \textit{Full-Modelling} in \(\Lambda\)CDM. The constraints of all cosmological parameters are within \(1\sigma\) of their respective truth values, albeit \(\Omega_{\mathrm{cdm}}h^2\) is slightly more biased than the others. Generally, they are in good agreement with each other. There is around \(0.3\sigma\) shift between the best-fits \(\ln{(10^{10}A_s)}\) of \textit{ShapeFit} and \textit{Full-Modelling}, which leads to the small difference between \textit{ShapeFit} and \textit{Full-Modelling} best-fit in Fig.~\ref{fig:bestfit}. The difference in best-fit power spectra in Fig.~\ref{fig:bestfit} is less than \(0.3\sigma\), implying the difference in \(\ln{(10^{10}A_s)}\) is being absorbed by the nuisance parameters. Additionally, \textit{ShapeFit} and \textit{Full-Modelling} may `weigh' the data in different ways, for instance \textit{ShapeFit} does not care about the BAO relative amplitude, which will be varied during a \textit{Full-Modelling} fit. This difference could also contribute to the difference in the \(\ln{(10^{10}A_s)}\) constraints. Furthermore, the degeneracies between the nuisance and cosmological/\textit{ShapeFit} parameters are also different, which could also shift the constraints on the cosmological parameters. Nonetheless, because the shift here is small compared to the recovered precision on the cosmological parameters, we expect it to have only a small impact on the constraints from DESI. 

In summary, we find that \textsc{PyBird} can give unbiased constraints on cosmological parameters for the \(\Lambda\)CDM model with \(k_{\mathrm{max}} = 0.20 h \mathrm{Mpc}^{-1}\), without the hexadecapole, and with the ``BOSS MaxF" configurations for both \textit{ShapeFit} and \textit{Full-Modelling} approaches. In section \ref{sec:Shapefit} and \ref{sec:Full-Modelling}, we extend these analyses for the \(w\)CDM and the \(o\)CDM model. For these extended cosmological models, including the hexadecapole or changing to the ``BOSS MinF" prior will significantly improve the constraints. Furthermore, the constraints from \textit{ShapeFit} and \textit{Full-Modelling} are in good agreement, and the systematic shift in the best-fit parameters are less than \(0.7\sigma\) for \(w\)CDM and \(0.5\sigma\) for \(o\)CDM. 

\section{\textit{ShapeFit} tests}
In this section, we test the impact of fitting up to different \(k_{\mathrm{max}}\) scales, including the hexadecapole, and different configurations for priors when applying the \textit{ShapeFit} method. 
\label{sec:Shapefit}
\subsection{Effect of $k_{\mathrm{max}}$}
\begin{figure}
	\includegraphics[width=1.0\textwidth]{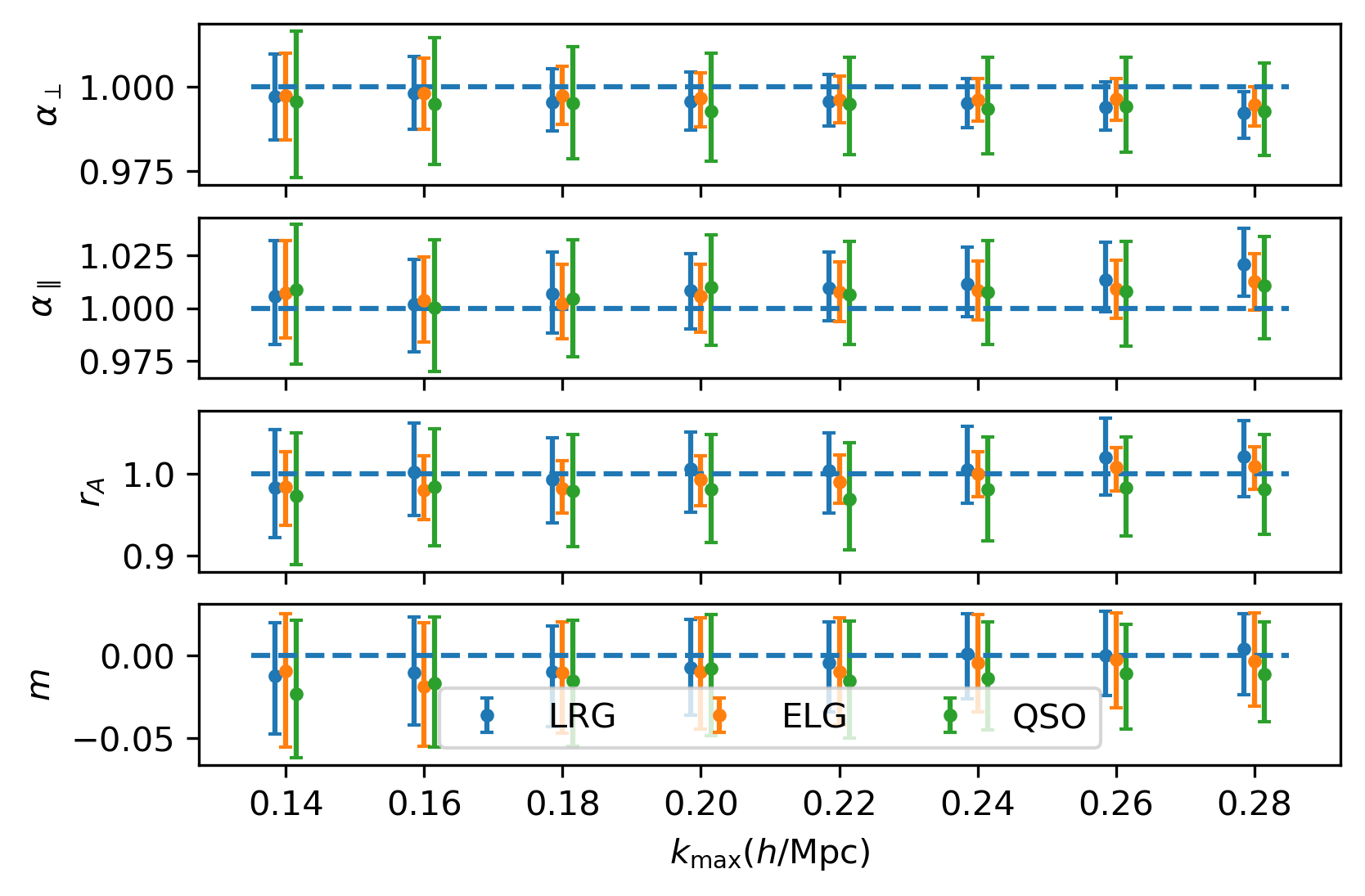}
    \caption{Constraints on the \textit{ShapeFit} parameters with different \(k_{\mathrm{max}}\) with the mean of the LRG mocks (blue), ELG mocks (orange), and the QSO mocks (green) with the ``BOSS MaxF" prior and without the hexadecapole. The template parameters are fixed to the true cosmology. \textsc{PyBird} can produced constraints on the \textit{ShapeFit} parameters within $1\sigma$ of the truth for all three tracers up to \(k_{\mathrm{max}} = 0.26 h \mathrm{Mpc}^{-1}\). However, the constraints on the \textit{ShapeFit} parameters do not improve much beyond \(k_{\mathrm{max}} = 0.20h \mathrm{Mpc}^{-1}\) while the systematic deviations from the truth values increase to $>0.5\sigma$. Therefore, we decide to use \(k_{\mathrm{max}} = 0.20h \mathrm{Mpc}^{-1}\) for \textit{ShapeFit} analysis in this work.}
    \label{fig:kmax_nohex}
\end{figure}

Fig.~\ref{fig:kmax_nohex} illustrates how the constraints on the \textit{ShapeFit} parameters change with respect to \(k_{\mathrm{max}}\). We use the single-box covariance matrix for this plot. The constraints at \(k_{\mathrm{max}} = 0.20 h \mathrm{Mpc}^{-1}\) are slightly different from Fig~\ref{fig:SF_bestfit}, because we use our second fitting configuration for Fig~\ref{fig:SF_bestfit}, but the first fitting configuration for Fig.~\ref{fig:kmax_nohex}. Fig.~\ref{fig:kmax_nohex} shows increasing \(k_{\mathrm{max}}\) makes the constraints on the \textit{ShapeFit} parameters tighter, but it also increases the modelling errors, particularly for the \(\alpha\) parameters. We expect this result because, on the one hand, increasing \(k_{\mathrm{max}}\) provides more information. On the other hand, our model may break down on small scales. Table~\ref{tab:LRG_SF_kmax_nohex_tab} compiles the constraints and the modelling error for the mean of the LRG, ELG, and QSO mocks with \textit{ShapeFit}. Both Table~\ref{tab:LRG_SF_kmax_nohex_tab} and Fig.~\ref{fig:kmax_nohex} demonstrate the constraints on the \textit{ShapeFit} parameters (particularly $\alpha_{\perp}$ and $\alpha_{\parallel}$) deviate from the truth by more than $0.5\sigma$ for \(k_{\mathrm{max}} \geq 0.22 h \mathrm{Mpc}^{-1}\), which could indicate our modelling fails beyond \(k_{\mathrm{max}} = 0.20 h \mathrm{Mpc}^{-1}\). Furthermore, Fig.~\ref{fig:FS_kmax_nohex} illustrates the same trend for the cosmological parameters with \textit{Full-Modelling}. This result indicates that this systematic behaviour is not introduced by \textit{ShapeFit} but by the theoretical model itself. To eliminate this potential error, we decided to use \(k_{\mathrm{max}} = 0.20 h \mathrm{Mpc}^{-1}\). Furthermore, Table~\ref{tab:LRG_SF_kmax_nohex_tab} demonstrates the constraints of \textit{ShapeFit} parameters improve little beyond \(k_{\mathrm{max}} = 0.20 h \mathrm{Mpc}^{-1}\). Therefore, using \(k_{\mathrm{max}} = 0.20 h \mathrm{Mpc}^{-1}\) enables us to obtain tight constraints on the \textit{ShapeFit} parameters while avoiding potential modelling biases.

\begin{table}
    \centering
    \renewcommand{\arraystretch}{1.4} 
    \caption{Summary of the relative biases in the \textit{ShapeFit} parameters with the single-box covariance matrix using the mean of LRG, ELG, and QSO mocks. The QSO mocks give much larger uncertainty because their number density is much lower than LRG and ELG, so their effective volume is smaller than LRG or ELG. Since the template cosmology is equal to the true cosmology, the truth values are \(\alpha_{\perp} = \alpha_{\parallel} = r_A = 1.0\) and \(m = 0.0\). The constraints on \textit{ShapeFit} parameters become tighter when we use higher \(k_{\mathrm{max}}\), but does start to saturate for the largest \(k_{\mathrm{max}}\) we test, where the counter and stochastic contributions to the best-fit power spectrum model become equal to or larger than the (cosmologically informative) linear or loop terms. Furthermore, the best fits also deviate further from the truth value, possibly due to the model failing to predict the small-scale clustering accurately. We found for \(k_{\mathrm{max}} = 0.20 h \mathrm{Mpc}^{-1}\), the best-fit parameters are less than \(0.5 \sigma\) from the truth, while the constraints do not improve much beyond \(k_{\mathrm{max}} = 0.20 h \mathrm{Mpc}^{-1}\). Therefore, we choose to use \(k_{\mathrm{max}} = 0.20 h \mathrm{Mpc}^{-1}\) for future fittings.}
    \label{tab:LRG_SF_kmax_nohex_tab}
    \begin{tabular}{cccccc}
        \hline\hline
		Tracer & $k_{\mathrm{max}} (h \mathrm{Mpc}^{-1})$ & $\Delta \alpha_{\perp} \%$ & $\Delta \alpha_{\parallel} \%$ & $\Delta r_A \%$ & $m$ \\ 
		\hline
		LRG & 0.14 & $-0.3^{+1.2}_{-1.3}$ & $0.6^{+2.7}_{-2.3}$ & $-0.4^{+6.5}_{-5.9}$ & $-0.009^{+0.031}_{-0.038}$ \\ 
		LRG & 0.16 & $-0.28^{+1.16}_{-0.98}$ & $0.0^{+2.3}_{-2.1}$ & $1.0^{+6.1}_{-5.1}$ & $-0.011^{+0.034}_{-0.031}$ \\ 
		LRG & 0.18 & $-0.32^{+0.93}_{-0.94}$ & $0.6^{+2.0}_{-1.8}$ & $-0.2^{+5.6}_{-4.5}$ & $-0.014^{+0.030}_{-0.031}$ \\ 
		LRG & 0.20 & $-0.38^{+0.85}_{-0.86}$ & $0.8^{+1.7}_{-1.8}$ & $0.1^{+4.9}_{-4.7}$ & $-0.012^{+0.032}_{-0.026}$ \\ 
		LRG & 0.22 & $-0.37^{+0.71}_{-0.79}$ & $1.0^{+1.5}_{-1.7}$ & $0.8^{+4.9}_{-4.4}$ & $-0.005^{+0.024}_{-0.031}$ \\ 
		LRG & 0.24 & $-0.47^{+0.69}_{-0.75}$ & $1.3^{+1.5}_{-1.9}$ & $1.2\pm 4.6$ & $-0.002^{+0.026}_{-0.025}$ \\ 
		LRG & 0.26 & $-0.47^{+0.63}_{-0.78}$ & $1.4^{+1.7}_{-1.5}$ & $1.9^{+4.7}_{-4.3}$ & $0.001^{+0.026}_{-0.024}$ \\ 
		LRG & 0.28 & $-0.80^{+0.66}_{-0.70}$ & $2.2\pm 1.6$ & $1.8^{+4.4}_{-4.5}$ & $0.003^{+0.023}_{-0.026}$ \\ \hline
		ELG & 0.14 & $-0.5^{+1.5}_{-1.1}$ & $0.7^{+2.6}_{-2.1}$ & $-0.8^{+4.2}_{-4.3}$ & $-0.014^{+0.041}_{-0.039}$ \\ 
		ELG & 0.16 & $-0.24^{+1.13}_{-0.96}$ & $0.5^{+1.9}_{-2.1}$ & $0.0^{+4.0}_{-4.3}$ & $-0.013^{+0.033}_{-0.042}$ \\ 
		ELG & 0.18 & $-0.26^{+0.91}_{-0.86}$ & $0.2^{+1.8}_{-1.7}$ & $-1.1^{+3.9}_{-2.7}$ & $-0.010^{+0.029}_{-0.037}$ \\ 
		ELG & 0.20 & $-0.41^{+0.83}_{-0.75}$ & $0.3^{+1.9}_{-1.3}$ & $-0.2^{+3.4}_{-3.0}$ & $-0.018^{+0.037}_{-0.029}$ \\ 
		ELG & 0.22 & $-0.31^{+0.63}_{-0.73}$ & $0.9^{+1.3}_{-1.5}$ & $-0.1^{+3.3}_{-3.0}$ & $-0.015^{+0.037}_{-0.027}$ \\ 
		ELG & 0.24 & $-0.39^{+0.65}_{-0.63}$ & $0.7^{+1.4}_{-1.3}$ & $0.3^{+3.1}_{-2.9}$ & $-0.001^{+0.027}_{-0.033}$ \\ 
		ELG & 0.26 & $-0.41^{+0.62}_{-0.64}$ & $1.0\pm 1.4$ & $0.7\pm 2.9$ & $-0.002^{+0.027}_{-0.030}$ \\ 
		ELG & 0.28 & $-0.64^{+0.64}_{-0.51}$ & $1.0^{+1.5}_{-1.2}$ & $1.1^{+2.5}_{-3.0}$ & $0.003^{+0.023}_{-0.033}$ \\ \hline
		QSO & 0.14 & $-0.5^{+2.0}_{-2.3}$ & $0.4^{+3.6}_{-3.1}$ & $-2.3^{+8.5}_{-7.6}$ & $-0.013^{+0.037}_{-0.047}$ \\ 
		QSO & 0.16 & $-0.7^{+2.2}_{-1.7}$ & $0.0^{+3.3}_{-2.8}$ & $-1.8^{+8.2}_{-6.0}$ & $-0.018^{+0.041}_{-0.037}$ \\ 
		QSO & 0.18 & $-0.6^{+1.7}_{-1.6}$ & $0.7^{+2.6}_{-2.9}$ & $-0.7^{+6.4}_{-6.8}$ & $-0.015^{+0.035}_{-0.041}$ \\ 
		QSO & 0.20 & $-0.2^{+1.3}_{-1.9}$ & $0.2^{+3.1}_{-2.1}$ & $-1.6^{+6.8}_{-6.0}$ & $-0.013^{+0.036}_{-0.037}$ \\ 
		QSO & 0.22 & $-0.6^{+1.5}_{-1.4}$ & $0.4^{+2.8}_{-2.1}$ & $-1.4^{+6.4}_{-6.2}$ & $-0.010^{+0.031}_{-0.039}$ \\ 
		QSO & 0.24 & $-0.5^{+1.3}_{-1.5}$ & $0.5^{+2.7}_{-2.2}$ & $-0.6^{+6.1}_{-6.3}$ & $-0.018^{+0.036}_{-0.030}$ \\ 
		QSO & 0.26 & $-0.5^{+1.3}_{-1.6}$ & $0.6^{+2.5}_{-2.3}$ & $-0.1^{+5.6}_{-6.5}$ & $-0.008^{+0.027}_{-0.035}$ \\ 
		QSO & 0.28 & $-0.6^{+1.3}_{-1.4}$ & $1.1^{+2.3}_{-2.6}$ & $-0.6\pm 5.9$ & $-0.009^{+0.029}_{-0.031}$ \\ 
		\hline\hline
    \end{tabular}
\end{table}

\subsection{Including the hexadecapole}
\begin{figure}
    \centering
    \includegraphics[width=0.495\textwidth]{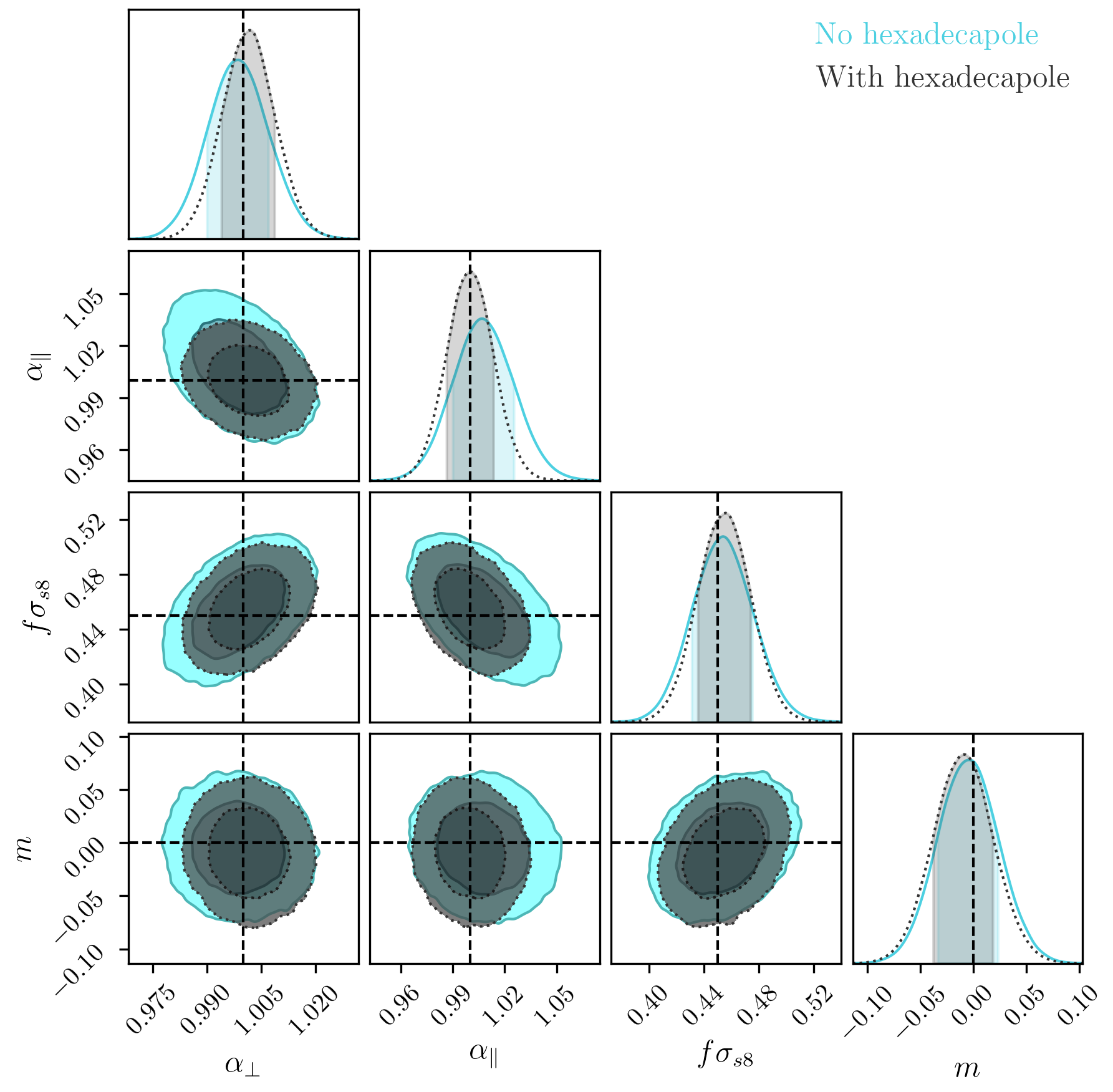}
    \includegraphics[width=0.495\textwidth]{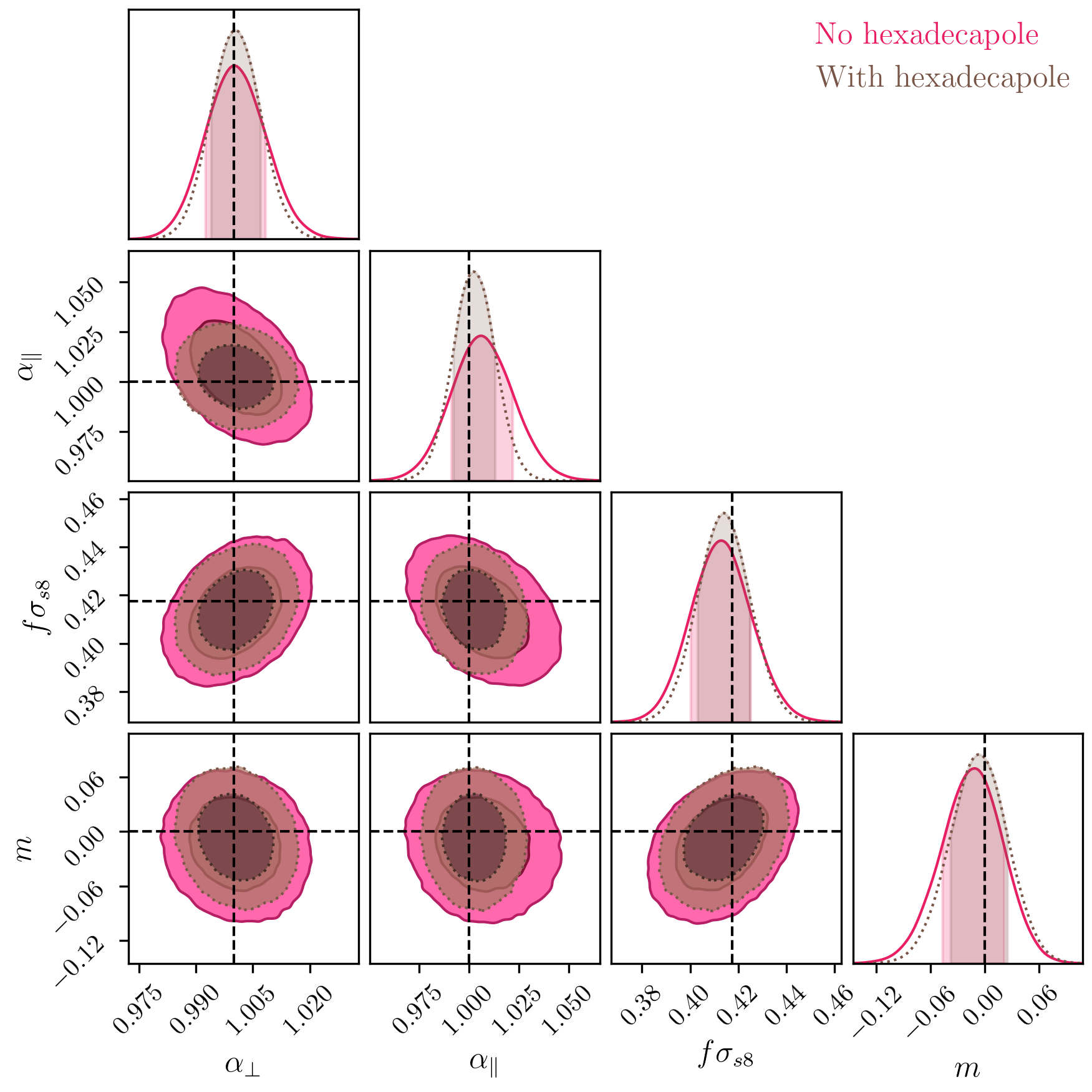}
    \includegraphics[width=0.495\textwidth]{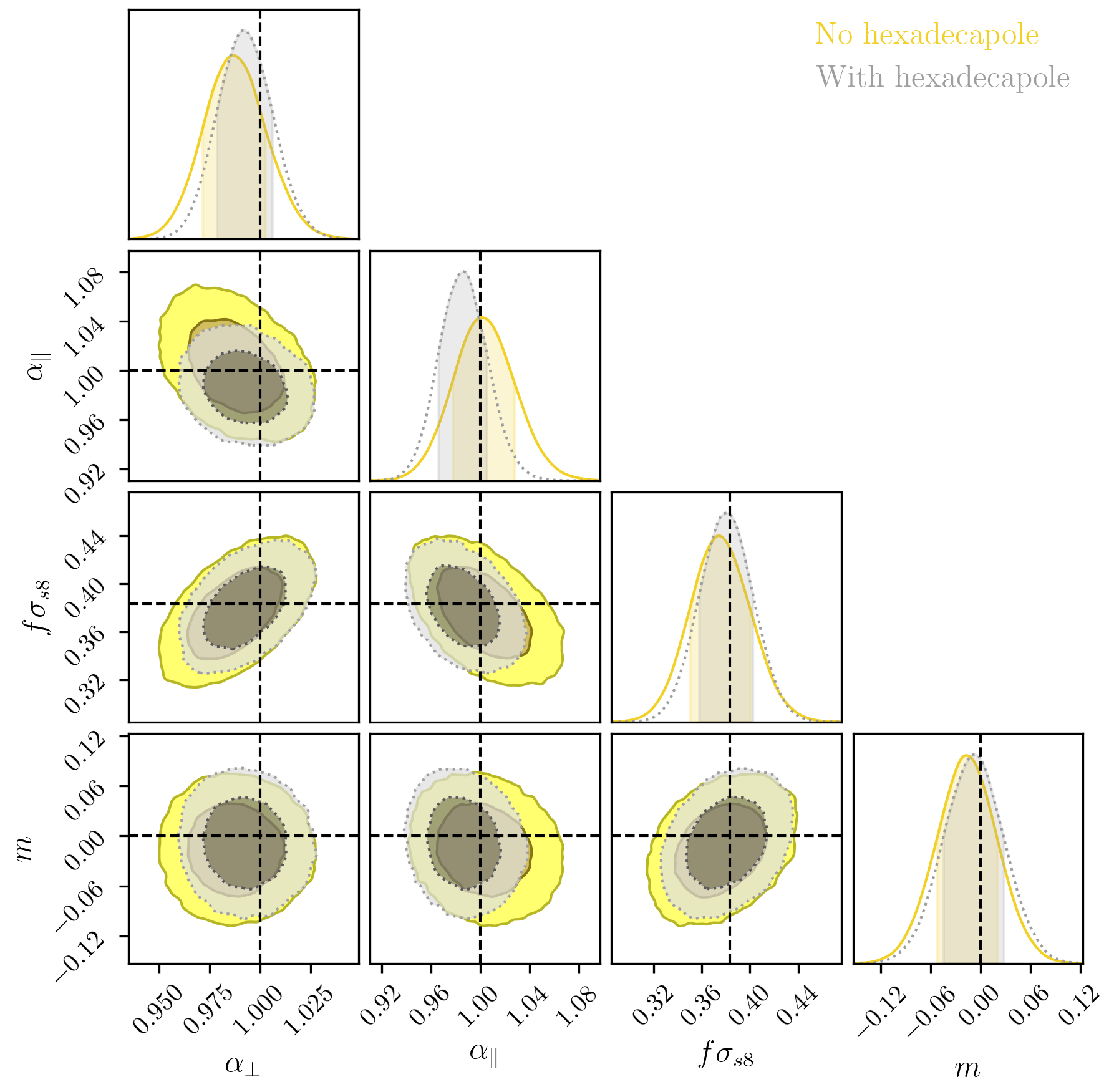}
    \caption{This plot compares constraints on the \textit{ShapeFit} parameters with \(k_{\mathrm{max}} = 0.20 h \mathrm{Mpc}^{-1}\) with or without the hexadecapole using the LRG mocks (top left), ELG mocks (top right), and QSO mocks (bottom) with ``BOSS MaxF" prior. These plots demonstrate that including the hexadecapole can significantly tighten the constraints for the \(\alpha\) parameters and drive them closer to the truth value. Including the hexadecapole can also slightly tighten the constraints for \(f\sigma_{s8}\). Overall, the constraints on \textit{ShapeFit} parameters after including the hexadecapole show no substantial biases for this choice of \(k_{\mathrm{max}}\).}
    \label{fig:hex_LRG}
\end{figure}

Fig.~\ref{fig:hex_LRG} illustrates the effect of adding hexadecapole on the constraints on cosmological parameters with the means of the LRG (top left), ELG (top right), and QSO (bottom) mocks. These constraints are calculated with the single-box covariance matrix. For all three tracers, the constraints for \(\alpha\) parameters are tightened significantly after including the hexadecapole. There are also minor improvements for \(f\sigma_{s8}\) but not for \(m\) after adding the hexadecapole. This lack of improvement in \(m\) is likely because the hexadecapole has larger error bars, so the tilt of the hexadecapole is not well constrained. Hence, it makes sense that the hexadecapole (and even the quadrupole) contain very little information on \(m\). The constraints of \textit{ShapeFit} parameters remain consistent with the truth value after adding the hexadecapoles.  

\subsection{Maximum and minimum freedom}
\begin{figure}
    \centering
    \includegraphics[width = 0.495\textwidth]{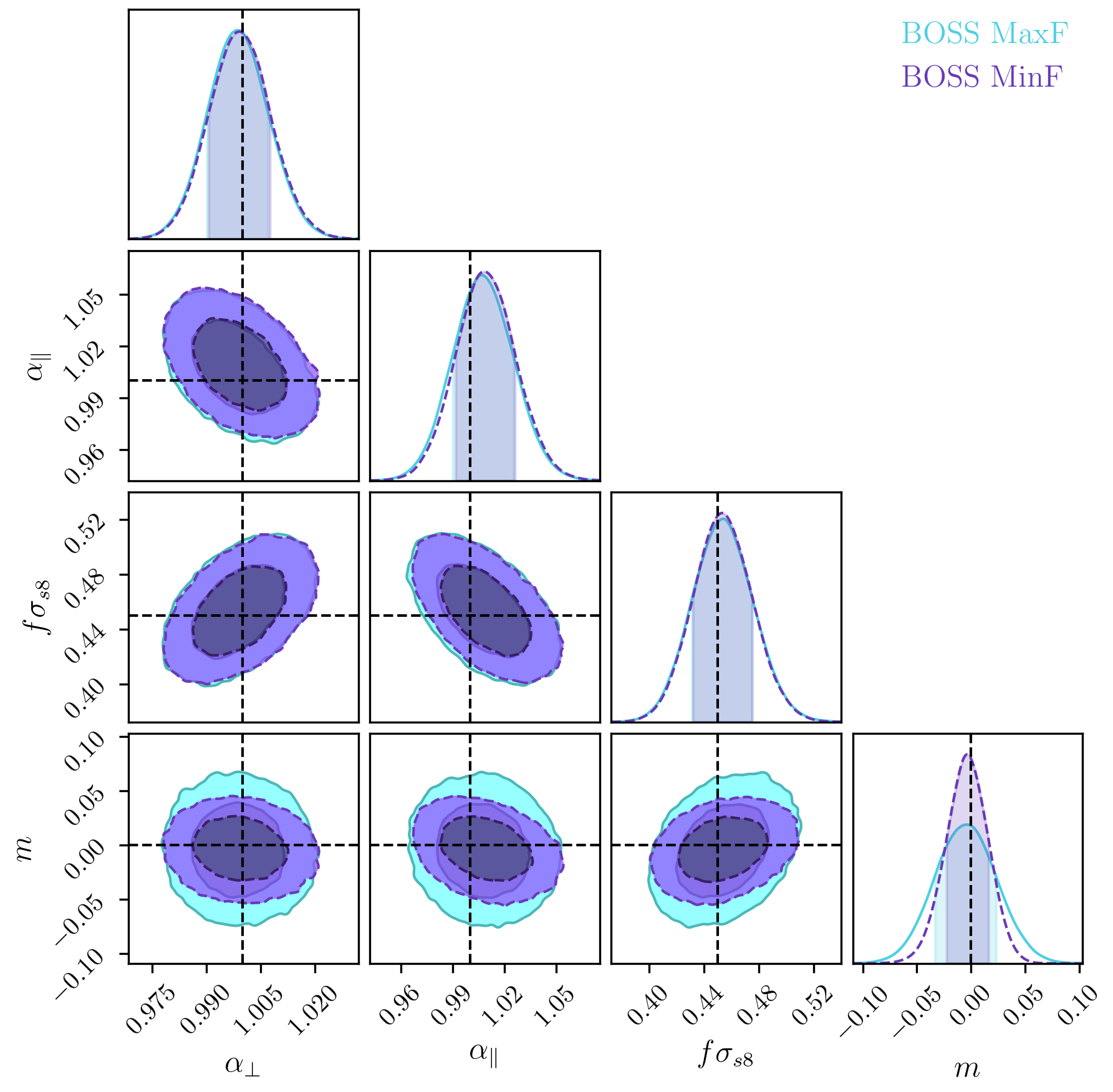}
    \includegraphics[width = 0.495\textwidth]{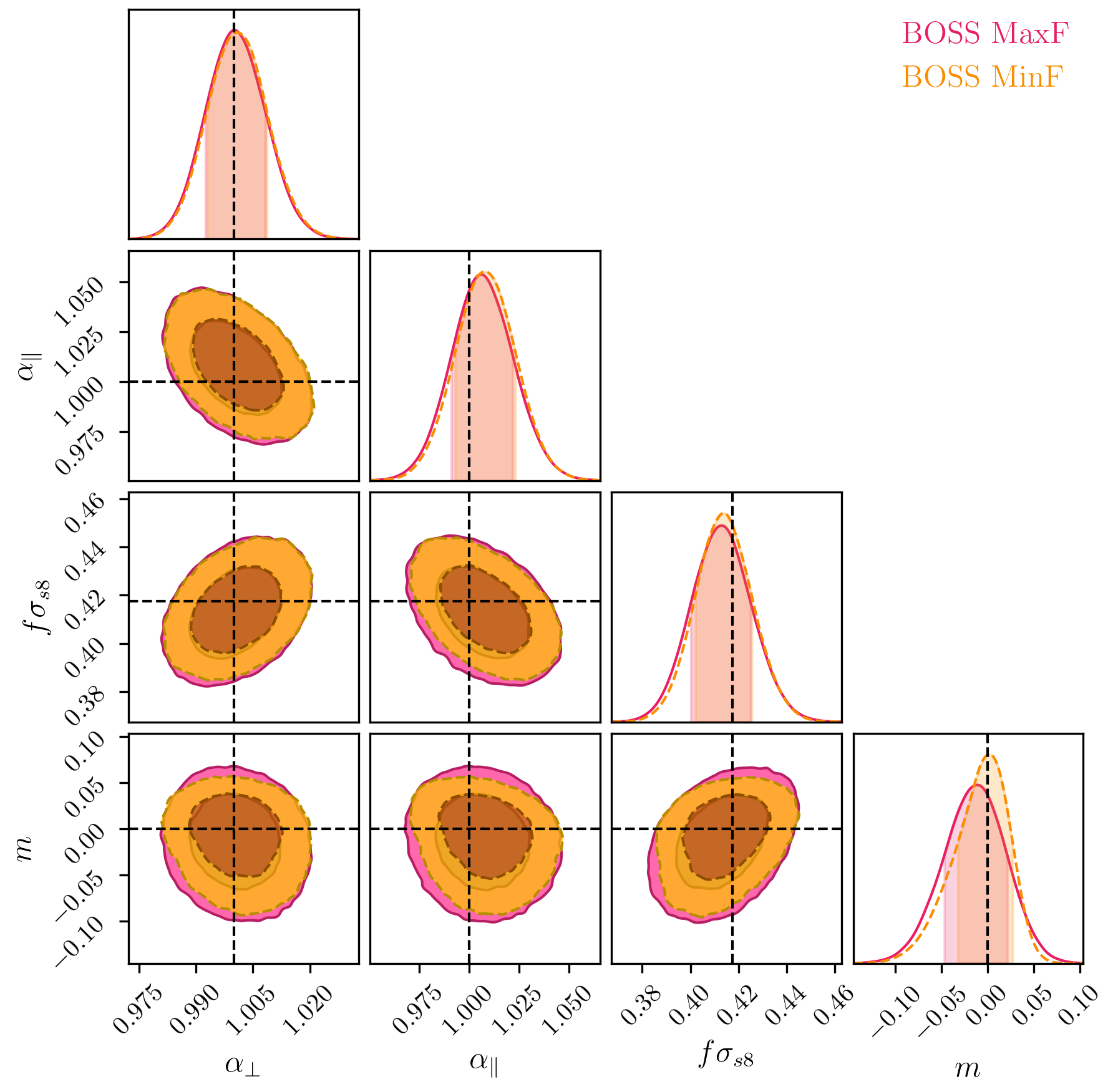}
    \includegraphics[width = 0.495\textwidth]{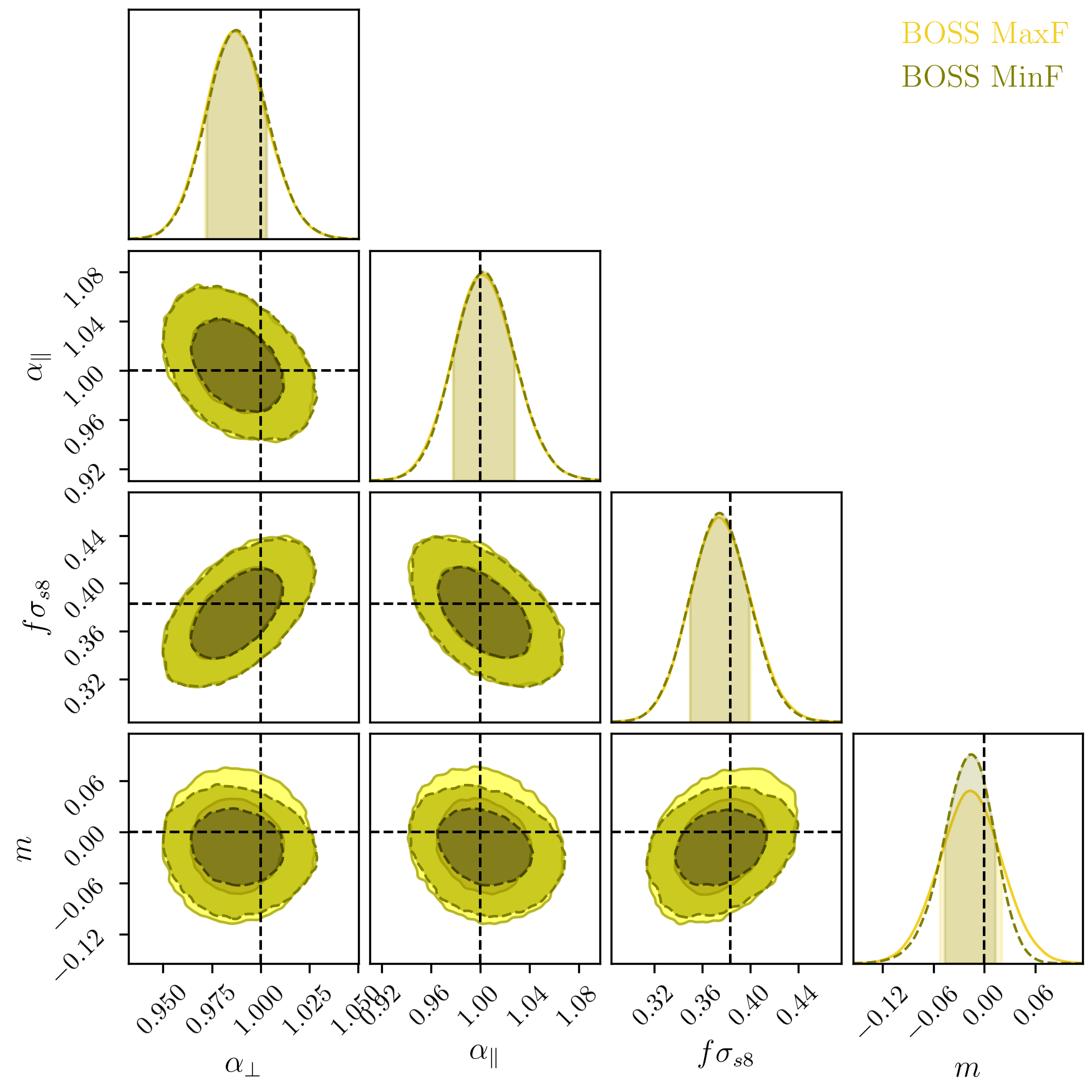}
    \caption{The effect of changing the prior configurations from ``BOSS MaxF" to ``BOSS MinF" for LRG (top left), ELG (top right), and QSO (bottom) with \(k_{\mathrm{max}} = 0.20 h^{-1} \mathrm{Mpc}\). Changing the prior configuration provides a much tighter constraint on \(m\). This improvement is because \(b_3\) is degenerate with \(m\), which is fixed in the ``BOSS MinF" configuration.} 
    \label{fig:Fixed_bias}
\end{figure}

Fig.~\ref{fig:Fixed_bias} shows the constraints on the \textit{ShapeFit} parameters after fixing \(b_2\) and \(b_3\) based on the local Lagrangian relation and \(c_{\epsilon, 2}\) to zero (``BOSS MinF" configuration). We choose to fix \(c_{\epsilon, 2}\) to zero to compare with other pipelines in DESI since the other models do not include a stochastic term with similar scale-dependence to \(c_{\epsilon, 2}\). We find that fixing these parameters tightens the constraint on \(m\), and the final constraints remain consistent with the truth values. The improvement in \(m\) mainly arises because of the degeneracy between \(m\) and \(b_3\) when the latter is allowed to be free \citep{KP5s1-Maus}.

\section{\textit{Full-Modelling} tests}
Similar to Section \ref{sec:Shapefit}, we now test the same configurations for the \textit{Full-Modelling} fit. Furthermore, we also compare the constraints from \textit{ShapeFit} and \textit{Full-Modelling} fit with different cosmological models.

\label{sec:Full-Modelling}
\subsection{Effect of $k_{\mathrm{max}}$}

\begin{figure}
	\includegraphics[width=1.0\textwidth]{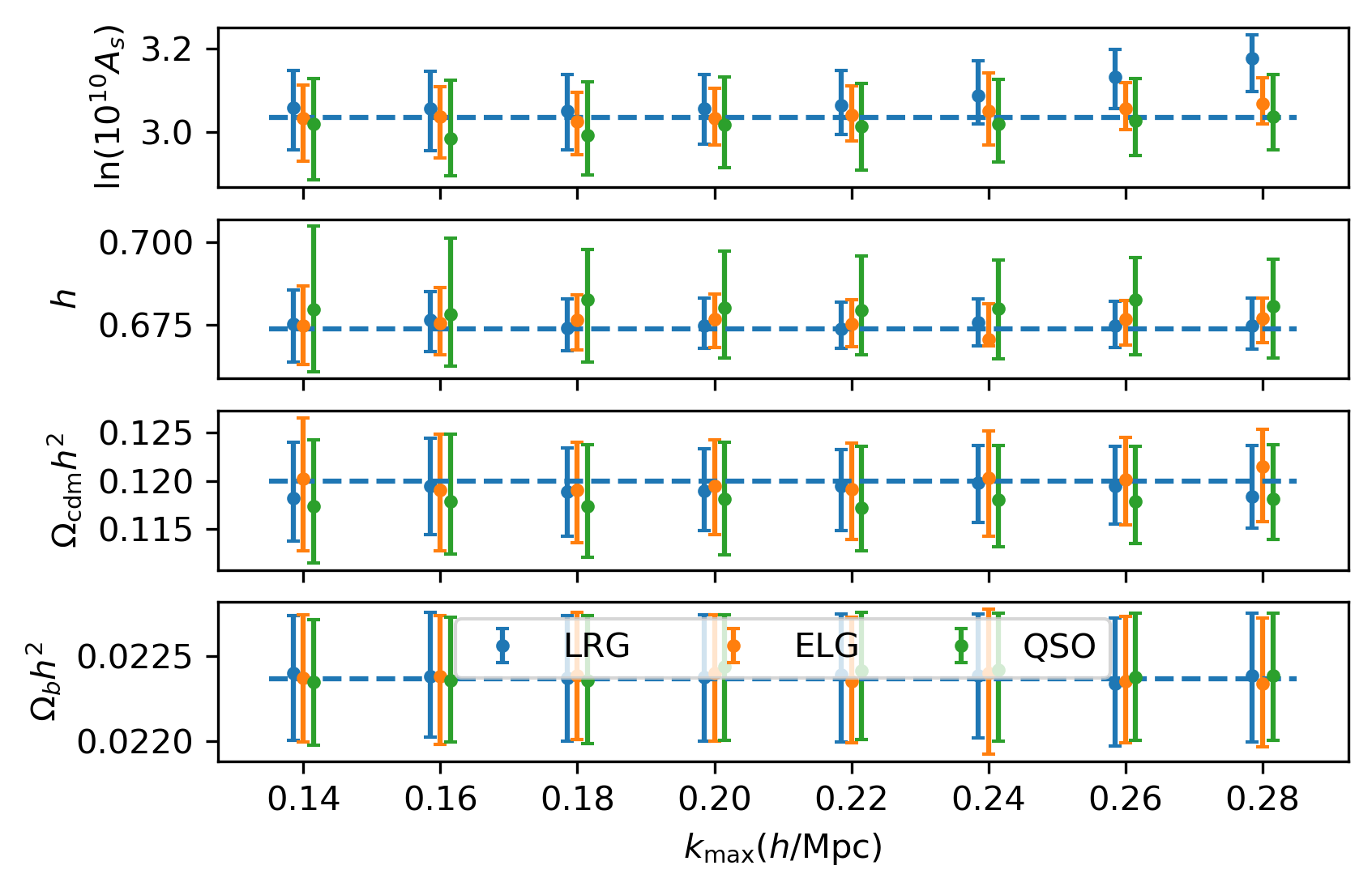}
    \caption{The constraints of cosmological parameters with the mean of the LRG mocks (blue), ELG mocks (orange), and QSO mocks (green) using the \textit{Full-Modelling} fit with the ``BOSS MaxF" prior and without the hexadecapole. The dashed lines are the truth values. For \(\Omega_b h^2\), the constraints are dominated by its Gaussian prior. The best-fits of \(\ln{(10^{10}A_s)}\) increase as the \(k_{\mathrm{max}}\) increases for large \(k_{\mathrm{max}}\). This bias is probably because our model fails to model the small scale accurately. }  
    \label{fig:FS_kmax_nohex}
\end{figure}

Fig.~\ref{fig:FS_kmax_nohex} shows the constraints from \textit{Full-Modelling} with the first fitting configuration using different choices of \(k_{\mathrm{max}}\). Similar to the \textit{ShapeFit} results, increasing \(k_{\mathrm{max}}\) will increase the constraining power but also increase the systematic bias. The constraints on cosmological parameters from different tracers are consistent with the truth. QSO mocks give larger errors similar to \textit{ShapeFit} because of its lower number density. There is a trend that the bias of best-fit \(\ln{(10^{10}A_s)}\) increases as \(k_{\mathrm{max}}\) increases. This only occurs beyond \(k_{\mathrm{max}} = 0.22 h \mathrm{Mpc}^{-1}\), where the small-scale systematics begins to creep in. For consistency, we also choose to fix \(k_{\mathrm{max}} = 0.20 h \mathrm{Mpc}^{-1}\) for future analysis, the same as \textit{ShapeFit}. Table~\ref{tab:LRG_FS_kmax_nohex_tab} summarises the deviations of the constraints from the truth.

\begin{table}
    \centering
    \renewcommand{\arraystretch}{1.4} 
    \caption{This table shows the constraints on the cosmological parameters in the \(\Lambda\)CDM model with \textit{Full-Modelling} using the single-box covariance matrix. Similar to \textit{ShapeFit}, the constraints on cosmological parameters are tighter when we use higher $k_{\mathrm{max}}$ because it contains more information. However, it will increase the systematic error because our model fails at the small scale. We choose to fit with $k_{\mathrm{max}} = 0.20 h \mathrm{Mpc}^{-1}$ for later tests because it provides one of the tightest constraints and with systematic error less than \(0.5\sigma\).}
    \label{tab:LRG_FS_kmax_nohex_tab}
    \begin{tabular}{cccccc}
        \hline\hline
		Tracer & $k_{\mathrm{max}} (\mathrm{Mpc}^{-1})$ & $\Delta \ln(10^{10} A_s) \%$ & $\Delta h \%$ & $\Delta \Omega_{\mathrm{cdm}} h^2 \%$ & $\Delta \Omega_bh^2 \%$ \\ 
		\hline
		LRG & 0.14 & $0.7^{+3.0}_{-3.3}$ & $0.2^{+1.5}_{-1.7}$ & $-1.5^{+4.8}_{-3.7}$ & $0.1^{+1.5}_{-1.8}$ \\ 
		LRG & 0.16 & $0.7^{+2.9}_{-3.3}$ & $0.4^{+1.3}_{-1.4}$ & $-0.5^{+4.1}_{-4.2}$ & $0.1^{+1.7}_{-1.6}$ \\ 
		LRG & 0.18 & $0.5^{+2.9}_{-3.1}$ & $0.1^{+1.3}_{-1.0}$ & $-0.9\pm 3.8$ & $0.0^{+1.6}_{-1.7}$ \\ 
		LRG & 0.20 & $0.7^{+2.6}_{-3.0}$ & $0.2^{+1.1}_{-1.2}$ & $-1.0^{+3.7}_{-3.3}$ & $0.1^{+1.6}_{-1.7}$ \\ 
		LRG & 0.22 & $0.9^{+2.7}_{-2.3}$ & $0.02^{+1.22}_{-0.88}$ & $-0.4^{+3.1}_{-3.8}$ & $0.1^{+1.6}_{-1.8}$ \\ 
		LRG & 0.24 & $1.7^{+2.7}_{-2.2}$ & $0.3^{+1.1}_{-1.0}$ & $-0.2^{+3.2}_{-3.4}$ & $0.1^{+1.6}_{-1.7}$ \\ 
		LRG & 0.26 & $3.2^{+2.1}_{-2.5}$ & $0.2^{+1.1}_{-1.0}$ & $-0.5^{+3.5}_{-3.3}$ & $-0.1^{+1.7}_{-1.6}$ \\ 
		LRG & 0.28 & $4.6^{+1.8}_{-2.6}$ & $0.1^{+1.3}_{-1.0}$ & $-1.4^{+4.4}_{-2.7}$ & $0.1^{+1.6}_{-1.8}$ \\ \hline
		ELG & 0.14 & $-0.1^{+2.6}_{-3.3}$ & $0.2^{+1.8}_{-1.7}$ & $0.2^{+5.3}_{-6.2}$ & $0.0\pm 1.7$ \\ 
		ELG & 0.16 & $0.0^{+2.3}_{-3.2}$ & $0.3^{+1.6}_{-1.4}$ & $-0.8^{+4.8}_{-5.3}$ & $0.1^{+1.6}_{-1.8}$ \\ 
		ELG & 0.18 & $-0.3^{+2.3}_{-2.6}$ & $0.3^{+1.3}_{-1.2}$ & $-0.9^{+4.5}_{-4.3}$ & $-0.2^{+1.8}_{-1.5}$ \\ 
		ELG & 0.20 & $-0.1\pm 2.3$ & $0.3^{+1.2}_{-1.1}$ & $-1.2^{+4.4}_{-3.9}$ & $0.1\pm 1.7$ \\ 
		ELG & 0.22 & $0.1^{+2.3}_{-2.0}$ & $0.3^{+1.1}_{-1.0}$ & $-0.7^{+4.0}_{-4.3}$ & $-0.1^{+1.7}_{-1.6}$ \\ 
		ELG & 0.24 & $0.3^{+2.2}_{-1.8}$ & $0.23^{+1.02}_{-0.96}$ & $0.3^{+3.4}_{-4.4}$ & $0.2^{+1.4}_{-1.9}$ \\ 
		ELG & 0.26 & $0.7^{+2.1}_{-1.6}$ & $0.44^{+0.85}_{-1.15}$ & $0.1^{+3.6}_{-4.0}$ & $-0.1^{+1.7}_{-1.6}$ \\ 
		ELG & 0.28 & $1.0^{+2.0}_{-1.6}$ & $0.49^{+0.91}_{-1.09}$ & $1.2^{+3.2}_{-4.7}$ & $-0.1\pm 1.7$ \\ \hline
		QSO & 0.14 & $-0.5^{+3.5}_{-4.4}$ & $0.9^{+3.7}_{-2.8}$ & $-2.2^{+5.7}_{-4.9}$ & $-0.1^{+1.6}_{-1.7}$ \\ 
		QSO & 0.16 & $-1.7^{+4.6}_{-2.9}$ & $0.7^{+3.4}_{-2.3}$ & $-1.8^{+5.8}_{-4.5}$ & $0.0^{+1.7}_{-1.6}$ \\ 
		QSO & 0.18 & $-1.4^{+4.2}_{-3.2}$ & $1.3^{+2.2}_{-2.8}$ & $-2.2^{+5.3}_{-4.4}$ & $-0.1\pm 1.7$ \\ 
		QSO & 0.20 & $-0.9^{+3.6}_{-3.5}$ & $0.3^{+2.9}_{-2.0}$ & $-1.3^{+4.7}_{-4.8}$ & $0.0^{+1.5}_{-1.7}$ \\ 
		QSO & 0.22 & $-0.7\pm 3.4$ & $0.8^{+2.5}_{-2.0}$ & $-2.3^{+5.4}_{-3.7}$ & $0.2^{+1.5}_{-1.8}$ \\ 
		QSO & 0.24 & $-0.6^{+3.5}_{-3.0}$ & $0.9^{+2.2}_{-2.3}$ & $-1.6^{+4.6}_{-4.1}$ & $0.2^{+1.5}_{-1.9}$ \\ 
		QSO & 0.26 & $-0.3^{+3.3}_{-2.8}$ & $1.3^{+1.9}_{-2.5}$ & $-1.7^{+4.7}_{-3.6}$ & $0.0^{+1.7}_{-1.6}$ \\ 
		QSO & 0.28 & $0.0^{+3.3}_{-2.6}$ & $1.0^{+2.1}_{-2.3}$ & $-1.5^{+4.7}_{-3.5}$ & $0.1^{+1.6}_{-1.7}$ \\ 
		\hline\hline
    \end{tabular}
\end{table}

\begin{figure}
    \centering
    \includegraphics[width = 0.495\textwidth]{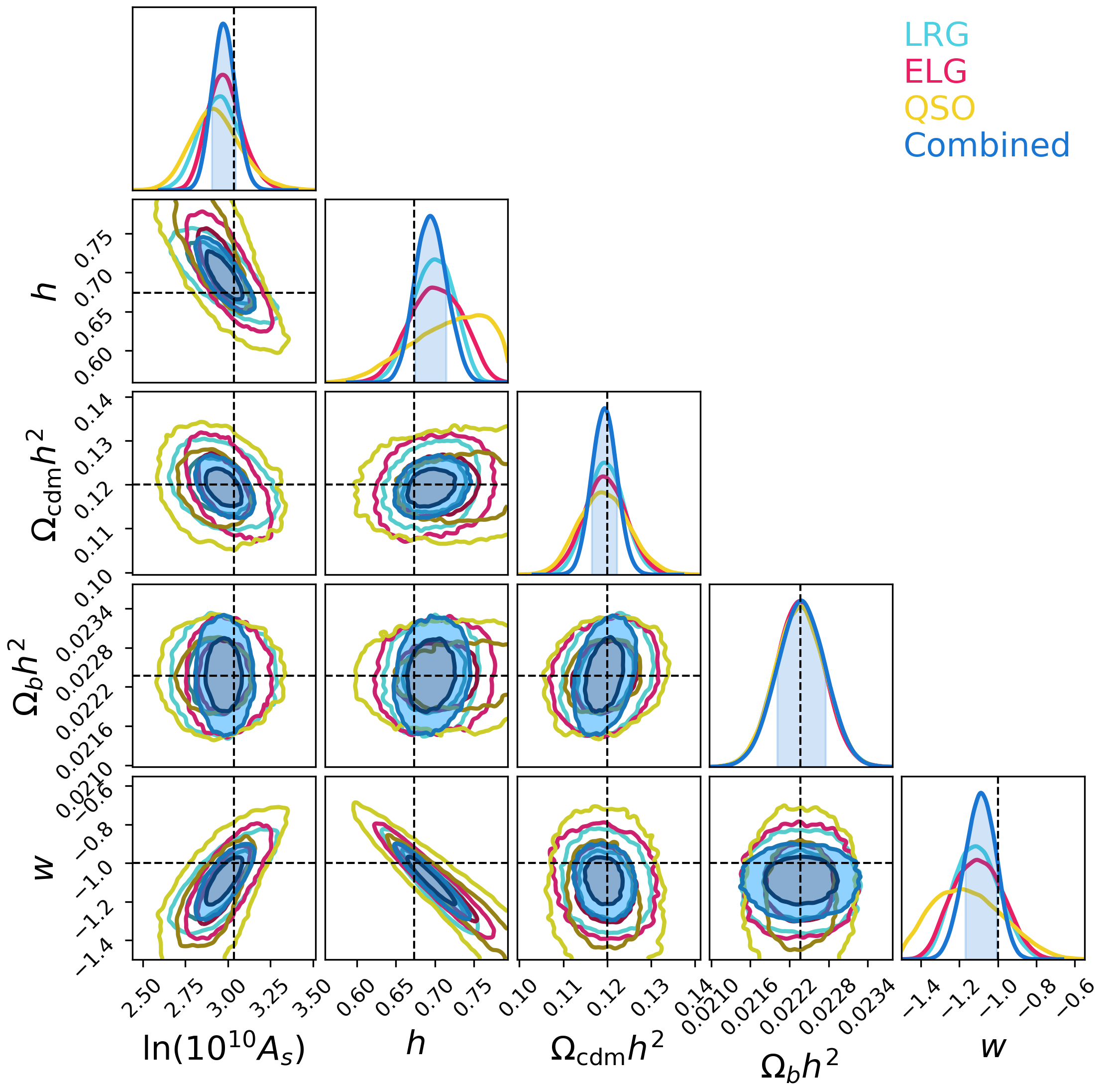}
    \includegraphics[width = 0.495\textwidth]{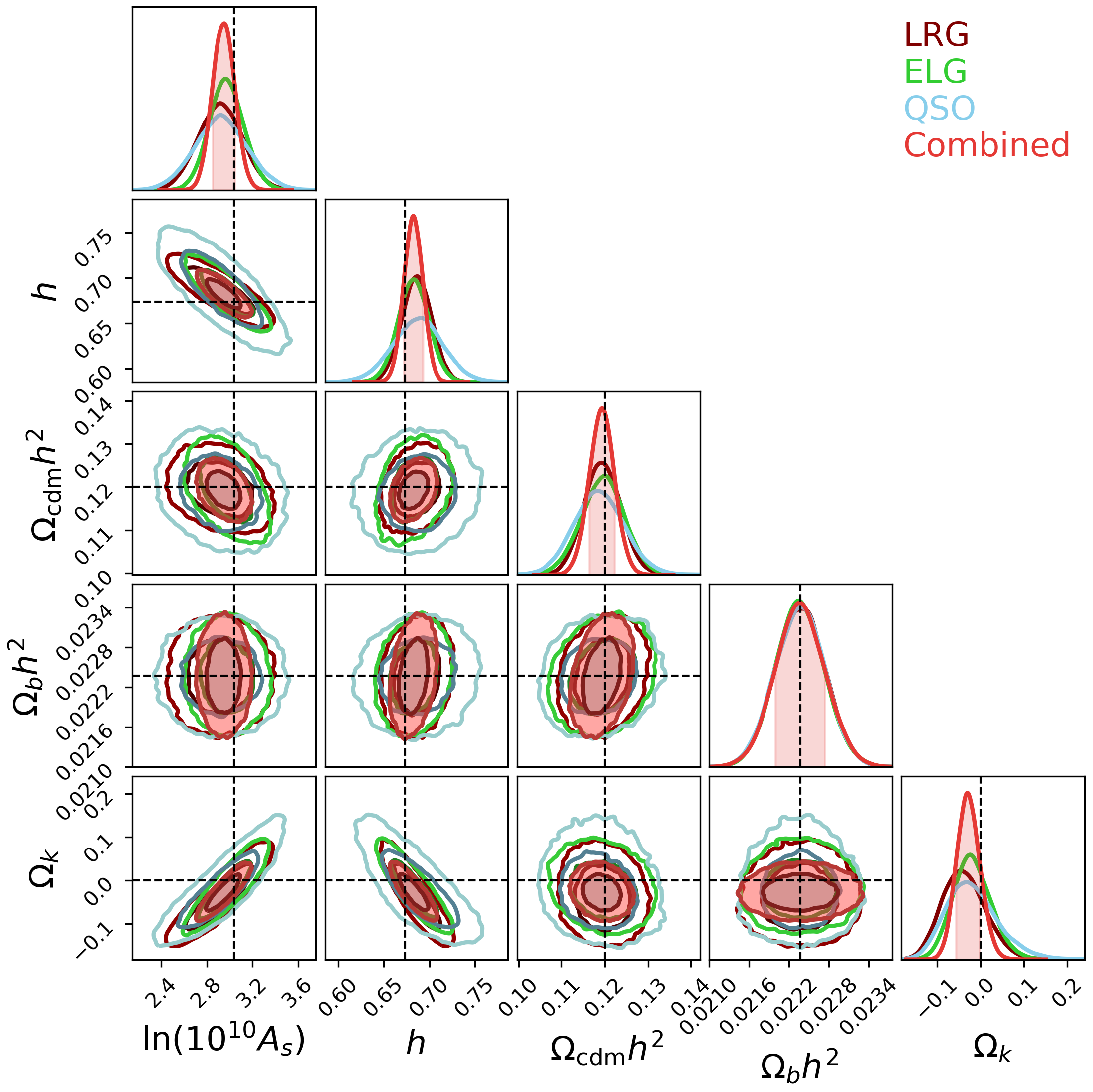}
    \caption{Constraints on the cosmological parameters in the $w$CDM (left) and \(o\)CDM (right) models with different tracers using the \textit{Full-Modelling} fitting methodology with the ``BOSS MaxF" prior, \(k_{\mathrm{max}} = 0.20 h \mathrm{Mpc}^{-1}\), and without the hexadecapole. Similar to Fig.~\ref{fig:FS_bestfit}, adding different tracers together significantly improves the constraints on the cosmological parameters. All the constraints are still consistent with the truth value at around \(1 \sigma\), but we see more significant deviations from the truth for \(\ln{(10^{10}A_s)}\) and \(h\). These two parameters are highly degenerate with \(w\) and $\Omega_{k}$, which moves them further away from the truth in the marginalized one-dimensional posterior. However, since these parameters are highly degenerate with each other, the two-dimensional contour demonstrates that the truth values are still within \(1\sigma\).} 
    \label{fig:wCDM_FS}
\end{figure}

Using \(k_{\mathrm{max}}=0.20 h \mathrm{Mpc}^{-1}\), Fig.~\ref{fig:wCDM_FS} illustrates the constraints on cosmological parameters in the $w$CDM (left) and \(o\)CDM models (right) with the combination of LRG, ELG, and QSO mocks. The corresponding plot for $\Lambda$CDM was shown earlier in Fig.~\ref{fig:FS_bestfit}. Switching from \(\Lambda\)CDM to $w$CDM or oCDM significantly weakens the constraints on the base $\Lambda$CDM parameters. In the case of $w$CDM and \(o\)CDM, \(\ln{(10^{10}A_s)}\) and \(h\) are weakened compared to Fig.~\ref{fig:FS_bestfit} because they are both degenerate with the dark energy equation of state \(w\) and curvature \(\Omega_k\). For $w$CDM, this also shifts \(h\) to a higher value and \(\ln{(10^{10}A_s)}\) to a lower value. However, the error bars on these parameters are large enough such that in both extended models, the constraints remain consistent with the truth values at around the \(1\sigma\) level.

\subsection{Including the hexadecapole}
\begin{figure}
    \centering
    \includegraphics[width=0.495\textwidth]{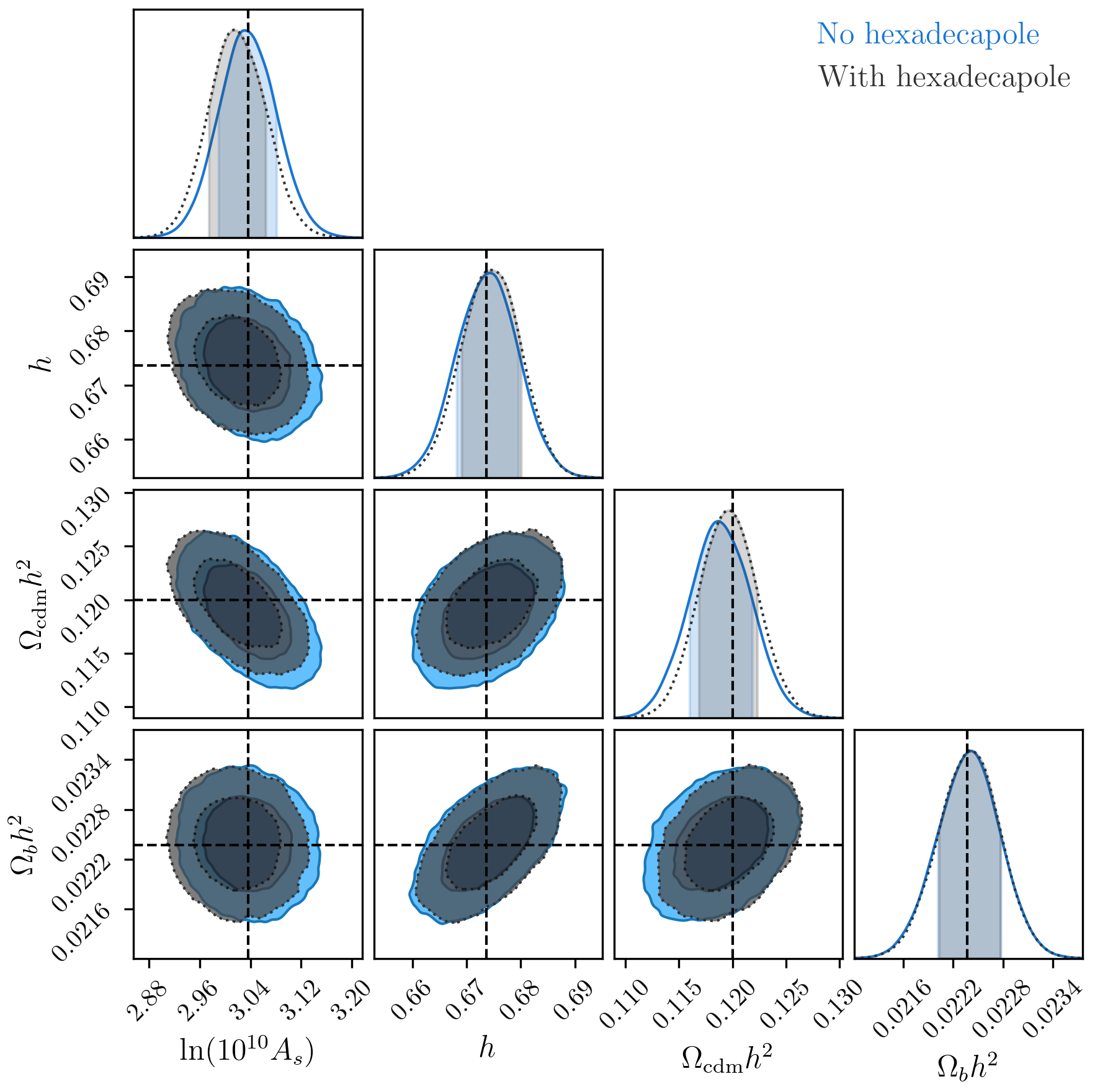}\\
    \includegraphics[width=0.495\textwidth]{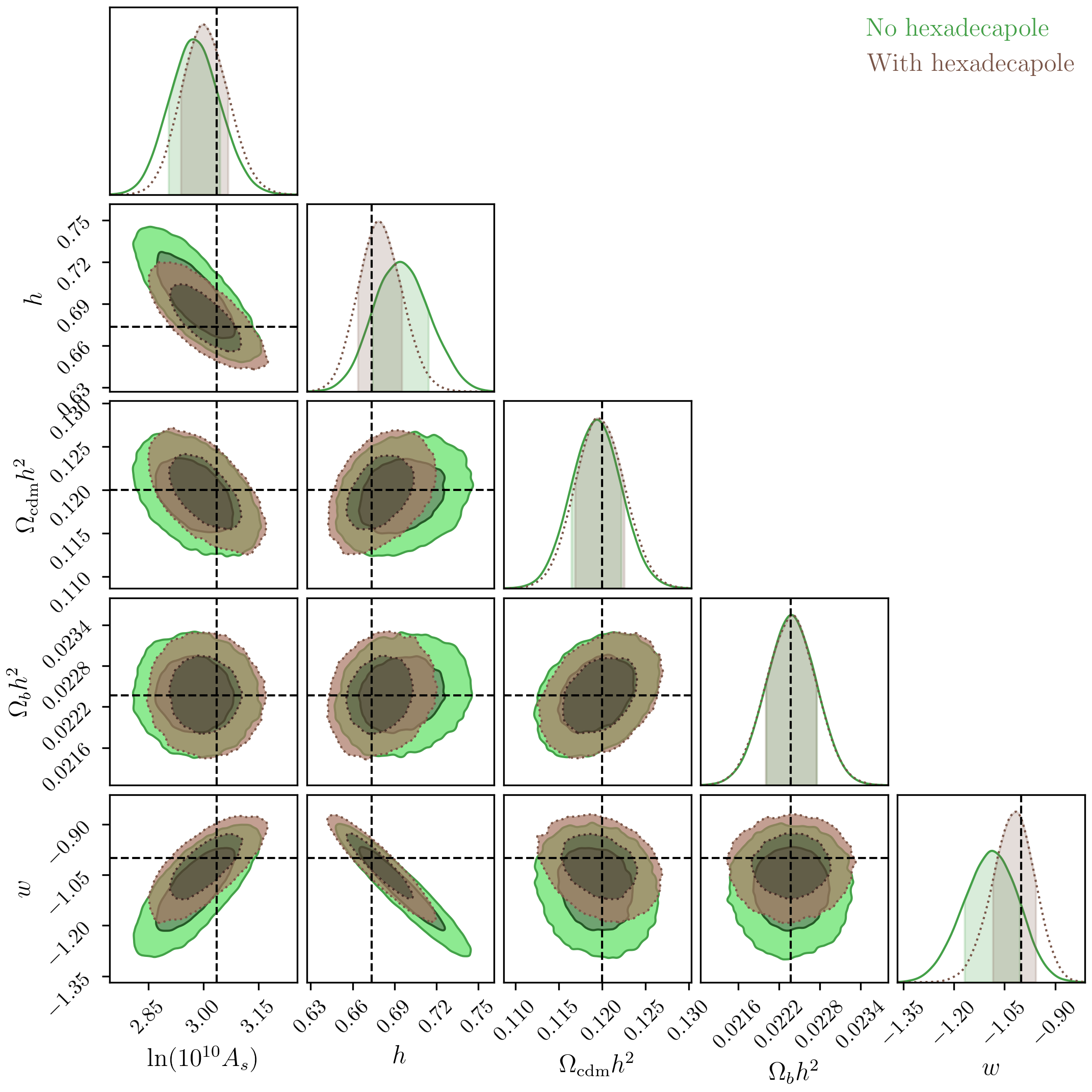}
    \includegraphics[width=0.495\textwidth]{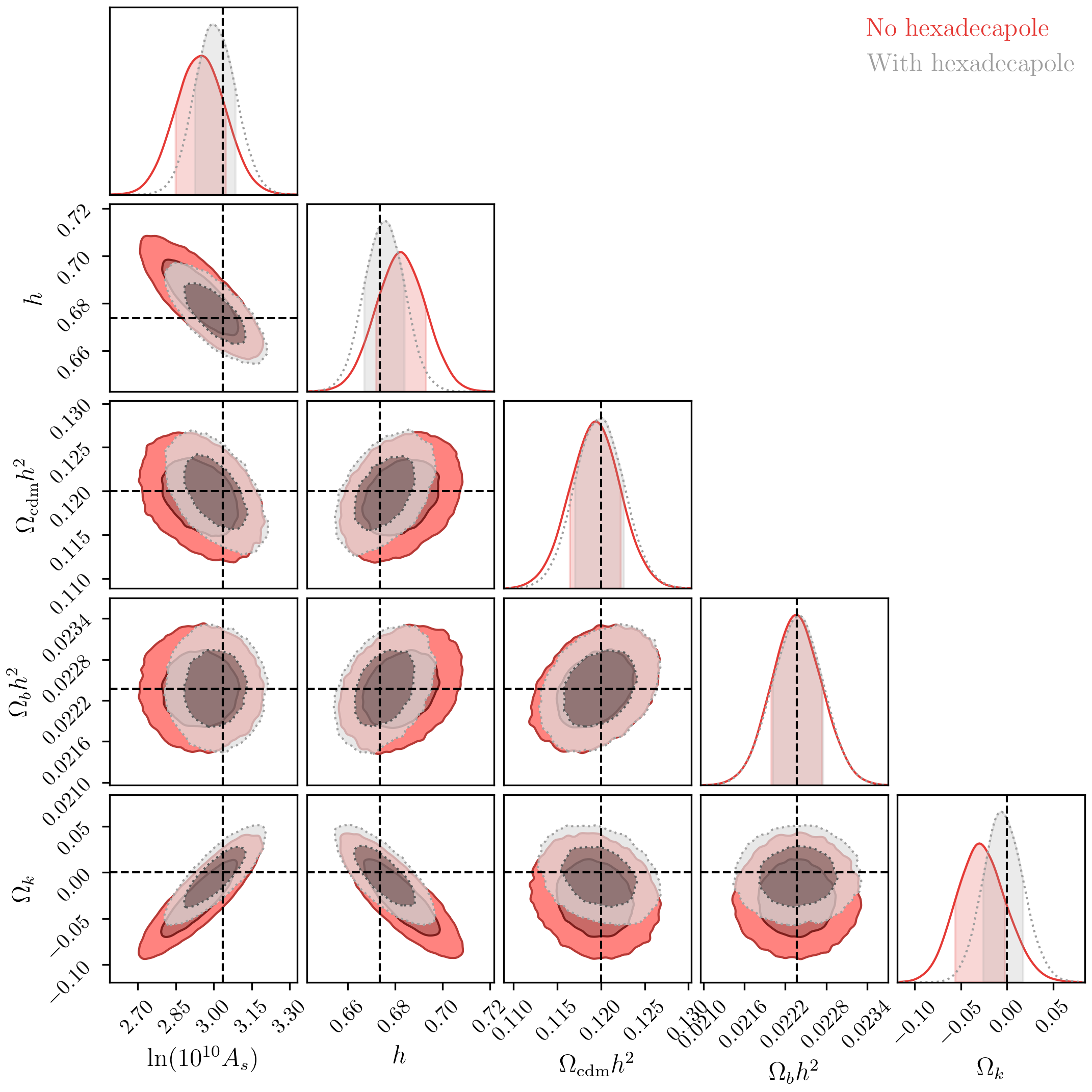}
    \caption{This plot shows the effect of adding hexadecapole on the constraints on cosmological parameters. The top one is for \(\Lambda\)CDM, the bottom left is for \(w\)CDM, and the bottom right is for \(o\)CDM. Here, we combine LRG, ELG, and QSO mocks. Contrary to \textit{ShapeFit}, including the hexadecapole does not improve the constraints of cosmological parameters for the $\Lambda$CDM model probably because the internal prior of \(\Lambda\)CDM on the \textit{ShapeFit} parameters allows little information can be gained from the hexadecapole. However, for $w$CDM and \(o\)CDM, the improvement from adding the hexadecapole is significant and reduces the systematic bias.} 
    \label{fig:hex_comparison_LRG_FS}
\end{figure}

Fig.~\ref{fig:hex_comparison_LRG_FS} illustrates the constraints on cosmological parameters after adding the hexadecapole for the combination of the LRG, ELG, and QSO mocks. In general, adding the hexadecapole has little impact on the constraints of cosmological parameters in the \(\Lambda\)CDM model. This finding is contrary to the results from \textit{ShapeFit} (Fig.~\ref{fig:hex_LRG}) where we see improved constraints on all \textit{ShapeFit} parameters except \(m\). This difference between \textit{ShapeFit} and \textit{Full-Modelling} is due to the internal prior of \(\Lambda\)CDM on the \textit{ShapeFit} parameters, which allows little additional information to be gleaned from this low amplitude multipole.

However, for other cosmological models, this is different. The constraints for \(h\) and \(w\) are tighter and closer to the truth value after including the hexadecapole for \(w\)CDM. It also shifts \(\ln{(10^{10}A_s)}\) slightly because it is highly degenerate with \(h\) and \(w\). Similarly, adding the hexadecapole when fitting the \(o\)CDM model also significantly improves the constraints for \(\ln{(10^{10}A_s)}\), \(h\), and \(\Omega_k\). In both cases, adding the hexadecapole reduces the systematic bias for $k_{\mathrm{max}} = 0.20 h \mathrm{Mpc}^{-1}$. Therefore, we recommend including it in cosmological analyses for extended cosmological models.

\subsection{Maximum and minimum freedom}

\begin{figure}
    \centering
    \includegraphics[width=0.5\textwidth]{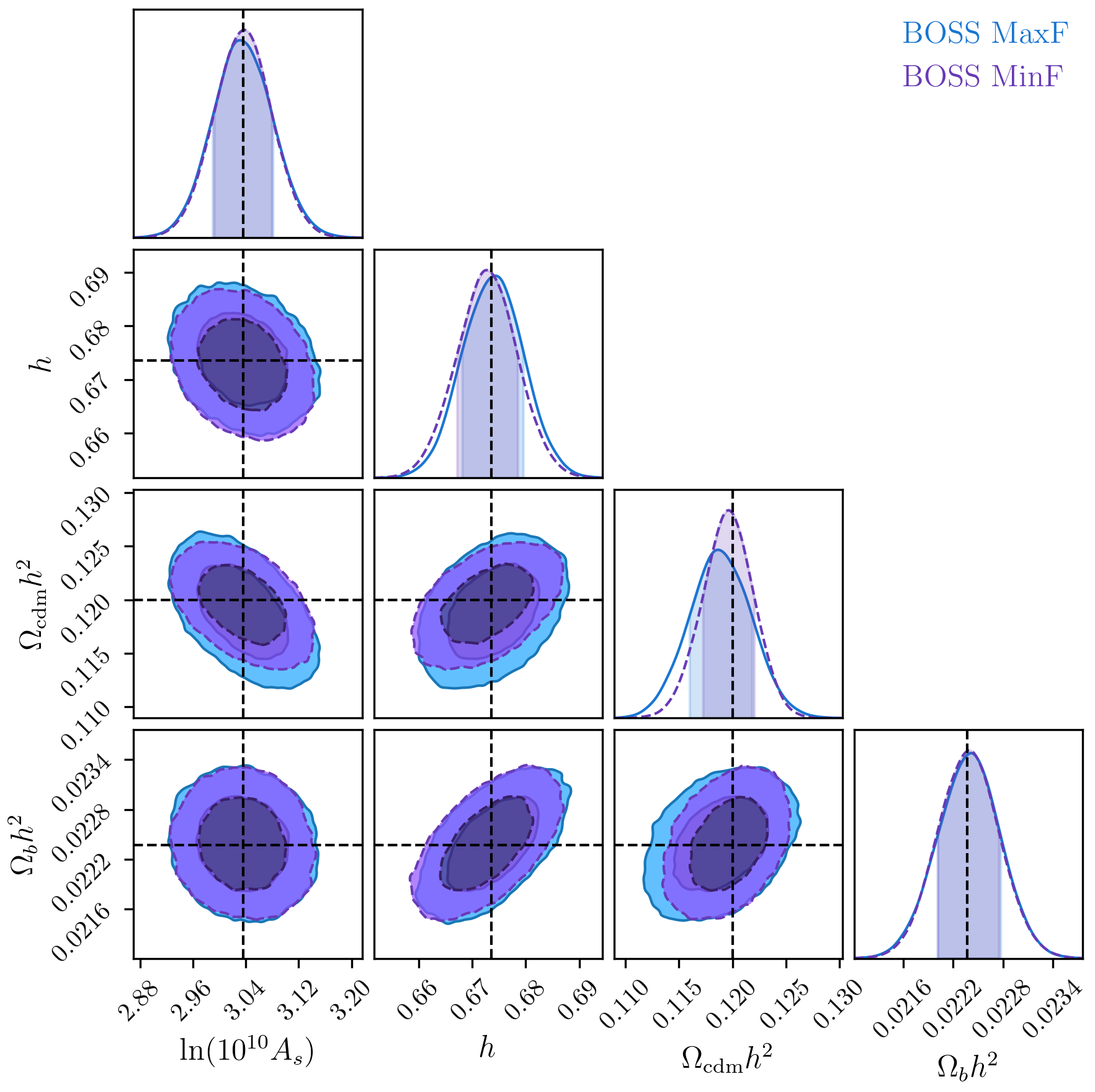}\\
    \includegraphics[width=0.49\textwidth]{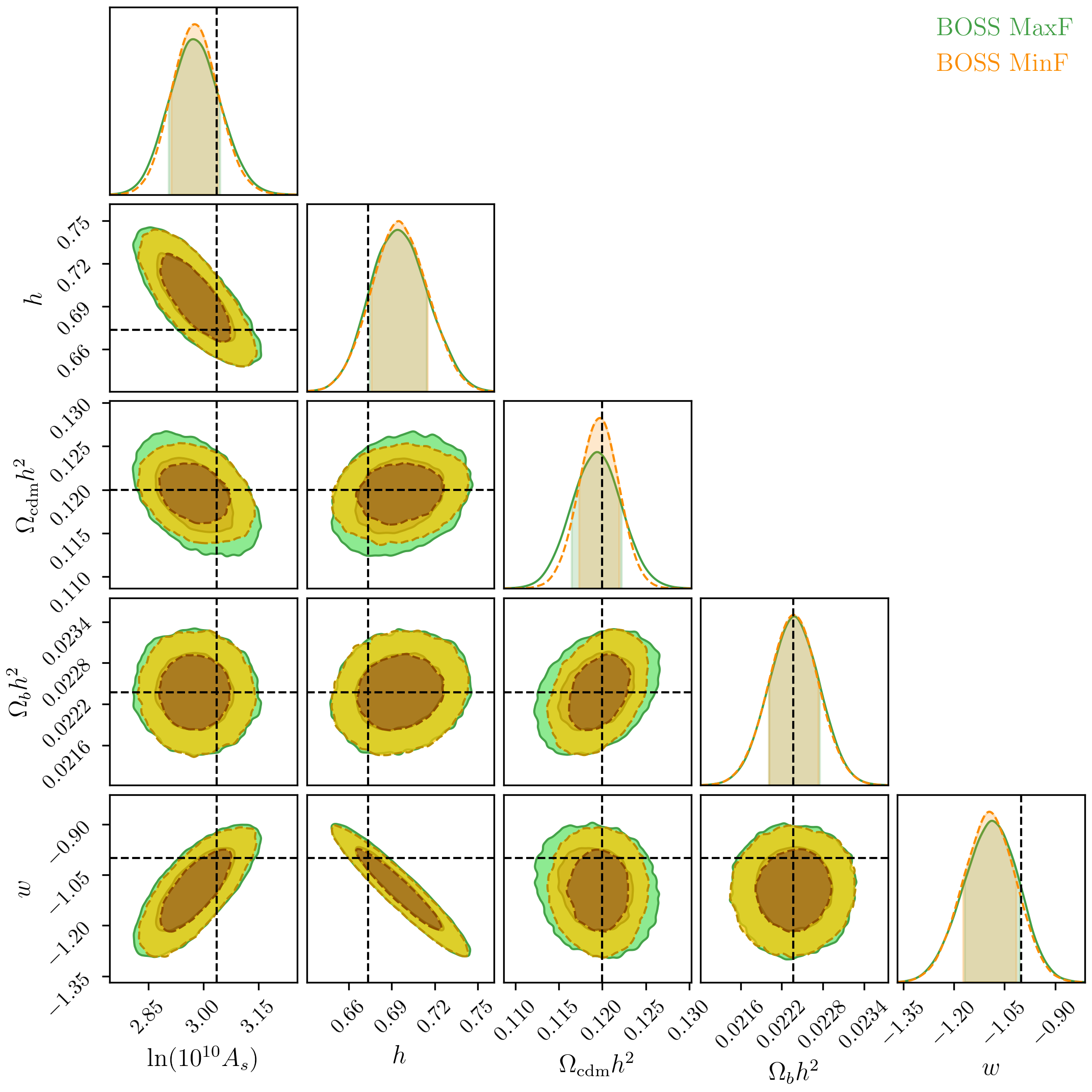}
    \includegraphics[width=0.49\textwidth]{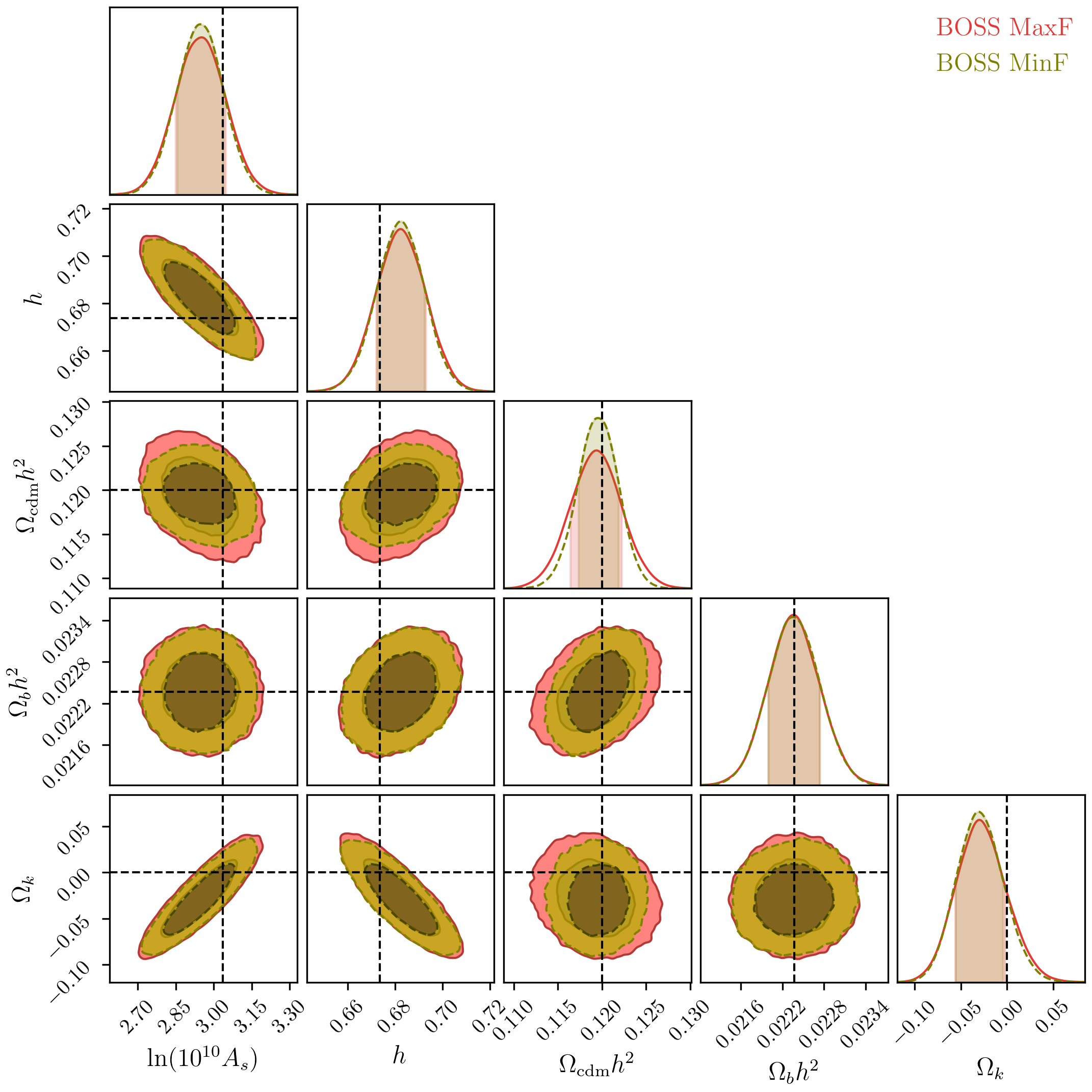}
    \caption{Cosmological constraints after changing the prior configuration from ``BOSS MaxF" to ``BOSS MinF" in \textit{Full-Modelling} fitting. We produce both contours using the combination of the LRG, ELG, and QSO mocks without the hexadecapole. Within $\Lambda$CDM (top panel), $w$CDM (lower left) and oCDM (lower right), the only impact is that the constraint on \(\Omega_{\mathrm{cdm}}h^2\) is tighter than for the ``BOSS MaxF" configuration. This is consistent with Fig.~\ref{fig:hex_LRG} where the constraint on the \textit{ShapeFit} parameter \(m\) is tighter with the ``BOSS MinF" configuration.} 
    \label{fig:fixed_bias_FS}
\end{figure}

Fig.~\ref{fig:fixed_bias_FS} demonstrates the effect of fixing \(b_2\) and \(b_3\) to the local Lagrangian relation, and \(c_{\epsilon, 2}\) to zero for the \textit{Full-Modelling} fit with the combination of the LRG, ELG, and QSO mocks. Fixing these parameters provides a tighter constraint for \(\Omega_{\mathrm{cdm}}h^2\). This improvement is consistent with the results from \textit{ShapeFit} where the constraint on \(m\) is tighter with the ``BOSS MinF" configuration. This effect also applies to the cases of $w$CDM and oCDM. However, the constraints on the extended cosmological parameters $w$ and $\Omega_{k}$ themselves are largely unaffected, as these primarily change the overall amplitude of the power spectrum and the redshift-distance relationship rather than the power spectrum tilt. All cosmological parameters remain consistent with their respective truth values with the ``BOSS MinF" configuration.



\begin{figure}
	\includegraphics[width=0.49\textwidth]{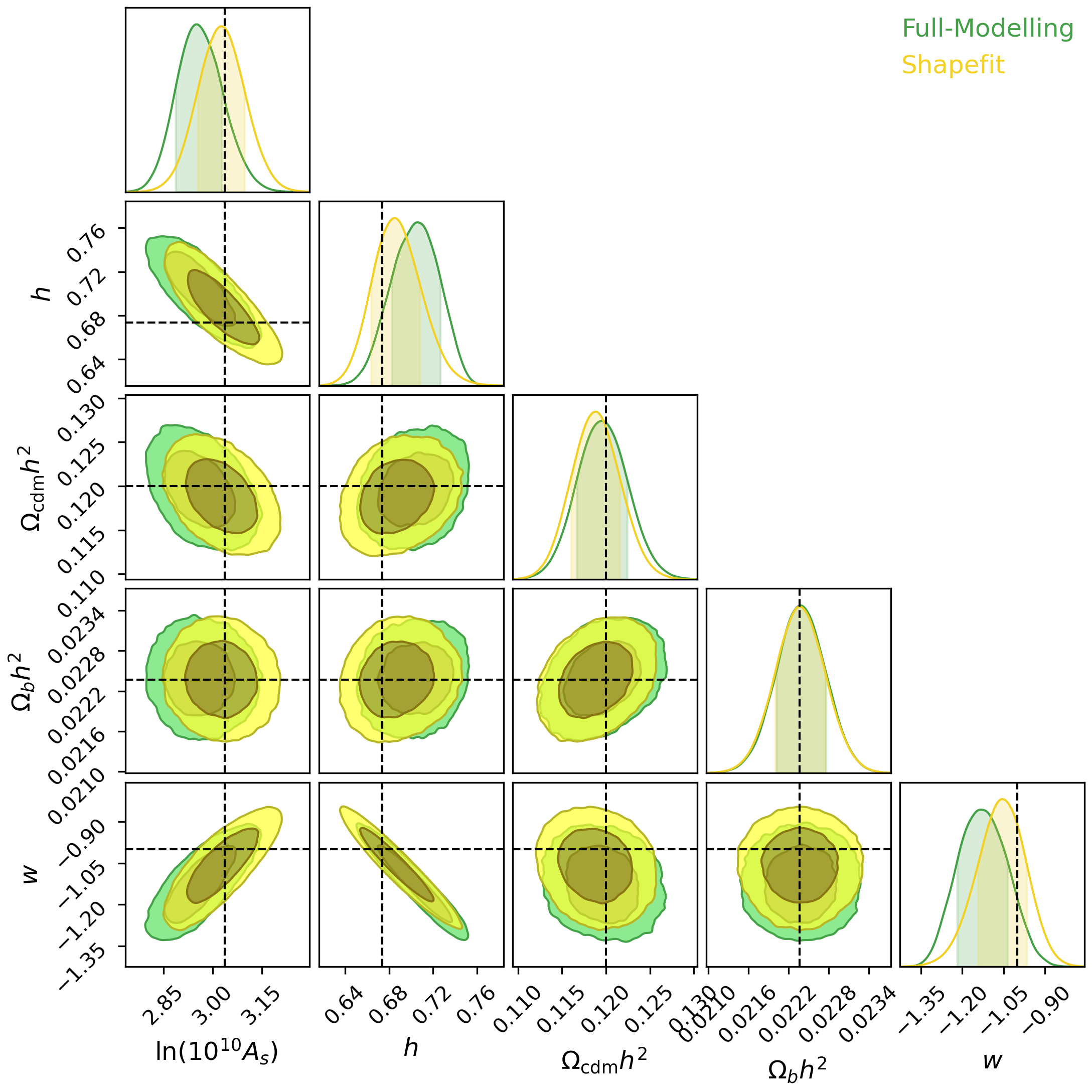}
    \includegraphics[width=0.49\linewidth]{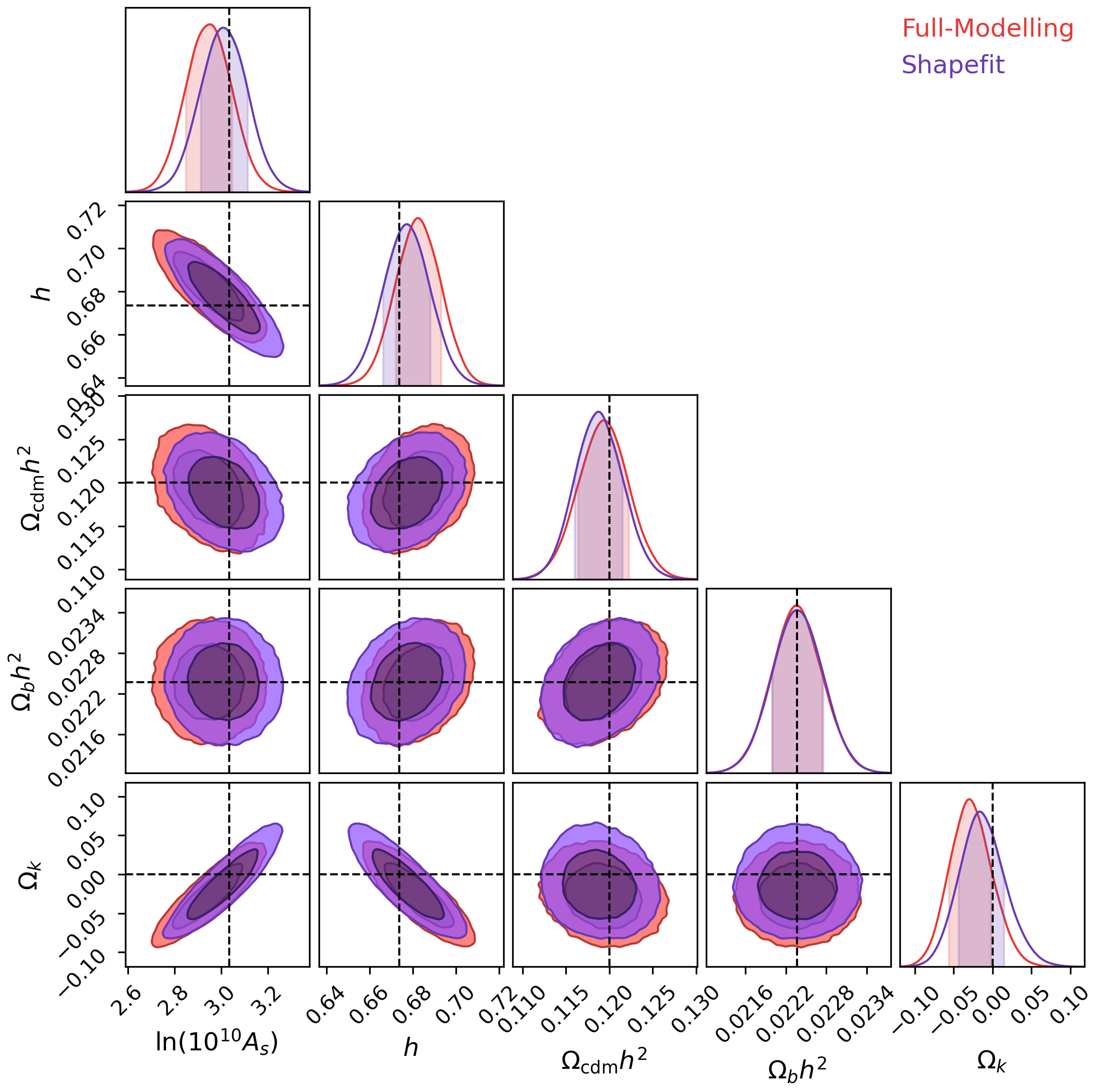}
    \caption{This figure compares the constraints from \textit{ShapeFit} and \textit{Full-Modelling} within the $w$CDM (left) and \(o\)CDM (right) cosmological models. The constraints are consistent with the truth from both methods except for shifts in the \(\ln{(10^{10}A_s)}\) and \(h\) parameters, which in turn propagate through into $w$ and $\Omega_{k}$ because these three parameters are degenerate with one another. Table~\ref{tab:wCDM_shift} summarizes the constraints on cosmological parameters from \textit{ShapeFit} and \textit{Full-Modelling} for \(w\)CDM and \(o\)CDM. Furthermore, this table also presents the shifts between \textit{ShapeFit} and \textit{Full-Modelling} constraints.}
    \label{fig:FS_vs_SF_wCDM_combined}
\end{figure}

\begin{table}[]
\setlength{\tabcolsep}{1pt}
\begin{tabular}{c|c|c|c|c|c}
\hline
Method   & $\ln{(10^{10}A_s)}$ & $100h$ & $100\Omega_{\mathrm{cdm}}h^2$ & $100\Omega_b h^2$ & $w$  \\ \hline \hline
FM     & $2.951_{-0.065}^{+0.076}$(2.982)     & $70.6_{-2.3}^{+2.1}$(69.0)     & $11.95_{-0.29}^{+0.29}$(12.01)      & $2.238_{-0.036}^{+0.038}$(2.234) & $-1.140_{-0.079}^{+0.104}$(-1.068) \\ \hline
SF      & $3.026_{-0.072}^{+0.071}$(3.030)     & $68.5_{-2.2}^{+2,3}$(68.5)       & $11.88_{-0.28}^{+0.28}$(11.89)     &  $2.236_{-0.036}^{+0.039}$(2.238) & $-1.055_{-0.088}^{+0.089}$(-1.050) \\ \hline
$\Delta_{w\mathrm{CDM}}$      & $1.05\sigma (0.68\sigma)$     & $0.91\sigma (0.25\sigma)$       & $0.25\sigma (0.40\sigma)$              & $0.05\sigma (0.11\sigma)$ & $0.93\sigma (0.19\sigma)$ \\ \hline
\end{tabular}
\newline
\vspace*{1 cm}
\newline
\begin{tabular}{c|c|c|c|c|c}
\hline
Method   & $\ln{(10^{10}A_s)}$ & $100h$ & $100\Omega_{\mathrm{cdm}}h^2$ & $100\Omega_b h^2$ & $\Omega_k$  \\ \hline \hline
FM     & $2.956_{-0.108}^{+0.090}$(2.963)     & $68.3_{-1.1}^{+1.0}$(68.2)     & $11.93_{-0.29}^{+0.29}$(11.92)      & $2.236_{-0.036}^{+0.038}$(2.250) & $-0.030_{-0.026}^{+0.027}$(-0.025) \\ \hline
SF      & $3.007_{-0.094}^{+0.107}$(3.009)     & $67.7_{-1.1}^{+1.1}$(67.7)       & $11.88_{-0.28}^{+0.27}$(11.88)     &  $2.235_{-0.036}^{+0.040}$(2.237) & $-0.016_{-0.038}^{+0.029}$(-0.014) \\ \hline
$\Delta_{o\mathrm{CDM}}$      & $0.51\sigma (0.43\sigma)$     & $0.53\sigma (0.43\sigma)$       & $0.17\sigma (0.13\sigma)$              & $0.00\sigma (0.35\sigma)$ & $0.51\sigma (0.37\sigma)$ \\ \hline
\end{tabular}

\caption{These tables demonstrate the constraints on cosmological parameters with \textit{Full-Modelling} (FM) and \textit{ShapeFit} (SF) for the $w$CDM model (top) and the \(o\)CDM model (bottom). The best-fit parameters are inside the bracket. \textit{ShapeFit} provides unbiased constraints on all cosmological parameters while the constraints on \(\ln{(10^{10}A_s)},\) \(h, \) and \(w\) for \(w\)CDM or \(\Omega_k\) for \(o\)CDM with the \textit{Full-Modelling} are slightly more than \(1\sigma\) away from the true. This shift is likely due to the effect of the prior volume for the nuisance parameters \citep{Simon_2023,KP5s2-Maus}. Since the best-fit parameters are not as affected by the prior volume effect \citep{KP5s2-Maus}, we use it to quote the discrepancy between the constraints from \textit{ShapeFit} and \textit{Full-Modelling}. Compared to the \(\Lambda\)CDM model, we see more significant discrepancies between \textit{ShapeFit} and \textit{Full-Modelling} in the \(w\)CDM model. Nonetheless, they all agree within \(0.7\sigma\). For the \(o\)CDM, the discrepancies between \textit{ShapeFit} and \textit{Full-Modelling} are within \(0.5\sigma\).}  
\label{tab:wCDM_shift}
\end{table}

\subsection{Comparing \textit{ShapeFit} vs \textit{Full-Modelling}}
\label{sec:comparison}
Having investigated how different analysis choices affect both the \textit{ShapeFit} and \textit{Full-Modelling} fitting methodologies, we now try to validate whether the two methods give consistent results for the extended cosmology. The case for $\Lambda$CDM was presented in Fig.~\ref{fig:FS_vs_SF_bestfit}. Fig.~\ref{fig:FS_vs_SF_wCDM_combined} illustrates that the constraints on cosmological parameters in both $w$CDM and \(o\)CDM cosmologies from \textit{Full-Modelling} and \textit{ShapeFit} are also consistent with each other except for shifts in \(w\) (for $w$CDM, $\Omega_{k}$ for \(o\)CDM), \(\ln{(10^{10}A_s)}\) and \(h\) parameters.

Table \ref{tab:wCDM_shift} summarizes the constraints on the cosmological parameters and the shifts between \textit{ShapeFit} and \textit{Full-Modelling} for \(w\)CDM (top) and \(o\)CDM (bottom). Inside the brackets, we also quote the best-fit parameters. Fig.~\ref{fig:FS_vs_SF_wCDM_combined} illustrates the constraints on the cosmological parameters from \textit{Full-Modelling} deviate further from the truth than \textit{ShapeFit}. Ref. \citep{KP5s2-Maus} demonstrate similar shifts in the \(w\)CDM cosmology with DESI mocks with \textsc{Velocileptors}. Furthermore, Ref.~\citep{Simon_2023} also observes a `prior volume' effect when fitting BOSS data using \textsc{PyBird} due to the non-Gaussian posteriors of the nuisance parameters that are marginalized over. Therefore, the shift observed in \(w\)CDM and \(o\)CDM constraints relative to the truth here for \textit{Full-Modelling} is also likely due to a prior volume effect. To the best of the authors' knowledge, this effect has not been examined in \textit{ShapeFit} before. However, Table~\ref{tab:wCDM_shift} demonstrates the difference between the mean of posterior and the best-fit from \textit{ShapeFit} is small, which may suggest \textit{ShapeFit} is less affected by the presence of uncontrolled nuisance parameters. 

The best-fit parameters are less affected by the prior volume effect than the mean of the posterior \citep{KP5s2-Maus}. Therefore, we quote the shift between \textit{Full-Modelling} and \textit{ShapeFit} with the best-fit parameters. In summary, the discrepancy of the best-fit parameters between the two is less than \(0.7\sigma\) for the \(w\)CDM model and less than \(0.5\sigma\) for the \(o\)CDM model. These shifts for \(h\) and the extended cosmological parameter are possibly due to the shift of \(\ln{(10^{10}A_s)}\) observed in \(\Lambda\)CDM propagating along the degeneracy between \(\ln{(10^{10}A_s)}\), $h$ and the extended cosmological parameters. We conclude that constraints from \textit{ShapeFit} and \textit{Full-Modelling} are largely consistent, so the choice of which to use/quote depends on the exact question one wishes to answer. In section \ref{sec:Taylor_no} and \ref{sec:Taylor_yes}, we investigate the speed of \textit{ShapeFit} and \textit{Full-Modelling} using our setting of \textsc{PyBird} without and with the Taylor expansion emulator, respectively. The tests were done on a laptop with an Intel i7 2.5GHz quad-core processor with 16 threads and 64 GB of RAM. We use the \textsc{emcee} sampler \citep{Foreman_Mackey_2013} with the "stretch move" \citep{Goodman_2010} to update new proposals. We also use the default convergence criteria in \textsc{emcee} where the MCMC is terminated when the number of iterations is more than 50 times larger than the integrated autocorrelation time. 

\subsubsection{Without the Taylor expansion emulator}
\label{sec:Taylor_no}

Table \ref{tab:time_LCDM_no} demonstrates the total run time \(T\) for \textit{ShapeFit} and \textit{Full-Modelling} analysis without Taylor expansion emulators with the \(\Lambda\)CDM model (top half) and the \(w\)CDM model (bottom half). We break down both analyses into two steps. Firstly, the run time for a single likelihood evaluation during MCMC to find the constraints on cosmological/\textit{ShapeFit} parameters (\(T_M\)). Secondly, the run time for converting the \textit{ShapeFit} parameters to cosmological parameters (\(T_C\)). \textit{Full-Modelling} has the advantage of directly returning the cosmological parameters, so it does not need the conversion step. Furthermore, \(N_M\)/\(N_C\) denotes the total number of iterations for the MCMC/conversion until the chain is converged. Additionally, the time it takes to run a single likelihood evaluation of the MCMC is around 0.45 seconds for \textit{ShapeFit} and 13.4 seconds for \textit{Full-Modelling}. It also takes \textit{ShapeFit} around 5.3 seconds to convert the \textit{ShapeFit} parameters to cosmological parameters. These times are estimated with the \textsc{time} module in \textsc{Python} averaging over ten runs. Here, we assume the convergence of the MCMC chain does not depend on the Taylor expansion, so \(N_M\) and \(N_C\) are taken from Table \ref{tab:time_LCDM_yes} for Table \ref{tab:time_LCDM_no}. It takes around 5.3 seconds per likelihood evaluation to find the \textit{ShapeFit} parameters; around 5 seconds were spent running \textsc{CLASS} with the input cosmological parameters. On the other hand, for \textit{Full-Modelling}, it takes around 13.4 seconds to find the model power spectrum per likelihood evaluation, 5 seconds for running \textsc{CLASS} and around 8 seconds for calculating the IR-resummation terms.\footnote{A careful reader may notice the run time for generating the model power spectrum with \textit{Full-Modelling} when fitting all three tracers does not change compared to the single tracer case. This is because once we know the IR-resummation terms at one redshift bin, we can rescale the IR-resummation terms, the linear terms, and loop terms together to other redshift bins using their respective \(\sigma_8\) and \(f\) similar to the process outline in section \ref{sec:Pybird}. This process is negligible compared to generating new IR-resummation terms, so the speed stays approximately the same. Similarly, by just calling the Boltzmann code once, we can generate the \textit{ShapeFit} parameters and the linear power spectrum for multiple redshift bins.} 

In general, \textit{ShapeFit} is around three times faster than \textit{Full-Modelling} when fitting only one redshift bin within the \(\Lambda\)CDM cosmology. However, when fitting all three tracers, \textit{ShapeFit} becomes around ten times faster than \textit{Full-Modelling}. This speed advantage of \textit{ShapeFit} is driven by 1) the fact it only takes around one-third of the time per iteration to compute \textit{ShapeFit} parameters compared to the \textit{Full-Modelling} model power spectrum and 2) the number of iterations required to obtain the same convergence in cosmological results is less because the overall dimensionality of the problem at the point in the pipeline when we fit for cosmological parameters is less --- \textit{Full-Modelling} requires us to fit for bias parameters at the same time as cosmological parameters, whereas we do not worry about the bias parameters during the conversion from \textit{ShapeFit} parameters to cosmological parameters, which is the most time-consuming step. 

\begin{table}[]
\setlength{\tabcolsep}{12pt}
\begin{tabular}{c|c|c|c|c|c|c|}
\hline
Method  & Model  & $10^{-5}N_M$ &  $T_M$ (h) & $10^{-5}N_C$ & $T_C$ (h) & $T$ (h) \\ \hline \hline
SF (LRG)  &$\Lambda$CDM              & 1.20                    & 15               & 1.92                     & 283          & 298         \\ \hline
SF (ELG)  &$\Lambda$CDM              & 1.92                    & 24                & 2.24                     & 330          & 354         \\ \hline
SF (QSO)  &$\Lambda$CDM              & 1.92                    & 24               & 1.60                     & 236          & 260         \\ \hline
SF (all) &$\Lambda$CDM          & 5.04                    & 63               & 0.96                      & 141          & 204         \\ \hline
FM (LRG)  &$\Lambda$CDM            & 2.30                    & 858              & -                          & -            & 858         \\ \hline
FM (ELG)  &$\Lambda$CDM              & 3.17                    & 1179            & -                          & -            & 1179        \\ \hline
FM (QSO)  &$\Lambda$CDM             & 2.21                    & 822              & -                          & -            & 822         \\  \hline
FM (all)  &$\Lambda$CDM           & 6.40                    & 2382             & -                          & -            & 2382 \\ \hline \hline      

SF (LRG)   &$w$CDM              & 1.20                    & 15               & 2.40                     & 353          & 368         \\ \hline
SF (ELG)   &$w$CDM             & 1.92                    & 24               & 2.40                     & 353          & 377         \\ \hline
SF (QSO)  &$w$CDM             & 1.92                    & 24              & 2.40                     & 353          & 377         \\ \hline
SF (all)  &$w$CDM             & 5.04                    & 63               & 2.40                      & 353          & 416         \\ \hline
FM (LRG) &$w$CDM              & 1.68                    & 625             & -                          & -            & 625        \\ \hline
FM (ELG) &$w$CDM               & 2.52                    & 938            & -                          & -            & 938        \\ \hline
FM (QSO) &$w$CDM               & 2.52                    & 938              & -                          & -            & 938        \\  \hline
FM (all) &$w$CDM           & 5.72                   & 2129           & -                          & -            & 2129  \\ \hline     
\end{tabular}
\caption{The total run time (\(T\)) for \textit{ShapeFit} (SF) and \textit{Full-Modelling} (FM) analysis without Taylor expansion emulators with the \(\Lambda\)CDM model (top half) and the \(w\)CDM model (bottom half). We break down both analyses into two steps. Firstly, we use (\(T_M\)) to denote the time it takes for MCMC to find the constraints on the cosmological/\textit{ShapeFit} parameters. Secondly, we use (\(T_C\)) to denote the time it takes to convert the \textit{ShapeFit} parameters to cosmological parameters. Furthermore, \(N_M\)/\(N_C\) denotes the number of iterations for the MCMC/conversion. Lastly, the time it takes to run each likelihood evaluation of the MCMC is around 0.45 seconds for \textit{ShapeFit} and 13.4 seconds for \textit{Full-Modelling}. It also takes \textit{ShapeFit} around 5.3 seconds per likelihood evaluation to convert the \textit{ShapeFit} parameters to cosmological parameters.} 
\label{tab:time_LCDM_no}
\end{table}

The bottom half of Table~\ref{tab:time_LCDM_no} compares the computational speed between \textit{ShapeFit} and \textit{Full-Modelling} with the \(w\)CDM model. Generally, \textit{ShapeFit} is still faster than \textit{Full-Modelling}. However, the advantage is less in the \(w\)CDM cosmology than in the \(\Lambda\)CDM cosmology. This result is mainly because \textit{ShapeFit} requires larger \(N_C\) to converge while \(N_M\) for \textit{Full-Modelling} is similar or even less than its respective \(\Lambda\)CDM value. Overall, without the Taylor expansion emulator, \textit{ShapeFit} is faster than \textit{Full-Modelling} in the case of \(\Lambda\)CDM and \(w\)CDM. We do not show the result for \(o\)CDM here because it takes a similar amount of iterations \(N_M\) to \(w\)CDM. Therefore, we expect the result for \(o\)CDM will be the same as \(w\)CDM.

\subsubsection{With the Taylor expansion emulator}
\label{sec:Taylor_yes}

In this section, we compare the speed performance of \textit{ShapeFit} and \textit{Full-Modelling} with the Taylor expansion emulator. We use the same notation as Table~\ref{tab:time_LCDM_no}. Furthermore, We use \(T_G\), \(L_M\), and \(L_C\) to the total time to generate the grid, the time it takes to run each iteration of MCMC for \textit{ShapeFit} or \textit{Full-Modelling}, and the time it takes to convert the \textit{ShapeFit} parameters to cosmological parameters per iteration.\footnote{There is also time associated with calculating the derivatives of the power spectrum/\textit{ShapeFit} parameters for the Taylor expansion. For \textit{Full-Modelling}, this is around 5 minutes for \(\Lambda\)CDM and 50 minutes for \(w\)CDM for each redshift bin. For \textit{ShapeFit}, it is under 1 minute for both \(\Lambda\)CDM and \(w\)CDM. This is a significant difference, but since the computation time for the derivatives is much less than the grid computation time, we do not include this in Table~\ref{tab:time_LCDM_yes}.} From section~\ref{sec:Taylor_no}, it takes \textit{ShapeFit} around 5.3 seconds and \textit{Full-Modelling} 13.4 seconds to generate a single grid point. We use nine grid points centred around the truth for each parameter. We have four cosmological parameters in \(\Lambda\)CDM, so its total number of grid points is \(9^4 = 6561\). For \(w\)CDM, it will be \(9^5 = 59049\). The number of grid points is the same for \textit{ShapeFit} and \textit{Full-Modelling}. Different from Table ~\ref{tab:time_LCDM_no}, Table~\ref{tab:time_LCDM_yes} shows \textit{Full-Modelling} with the Taylor expansion emulator in \(\Lambda\)CDM is around 30\% faster than \textit{ShapeFit} in the single tracer case and more than two times faster when combining all three tracers. This is because the only part of both analysis pipelines where we have not employed any emulation (the actual fitting of the \textit{ShapeFit} template to the data) becomes the most significant computational cost, even more extensive than the time taken to produce our grid of power spectra for \textit{Full-Modelling}.

On the other hand, in the \(w\)CDM cosmology, Table \ref{tab:time_LCDM_yes} demonstrates \textit{ShapeFit} is again around two times faster than \textit{Full-Modelling}. This result is because now the grid computation time \(T_G\) does dominate for both \textit{ShapeFit} and \textit{Full-Modelling} once we add one extra dimension. Since each grid point evaluation in \textit{ShapeFit} is faster than \textit{Full-Modelling} (requiring only a call to \textsc{CLASS} rather than \textsc{CLASS} and \textsc{PyBird}), the total time is reduced. We expect the same result for the \(o\)CDM cosmology because it needs the same number of points to construct the grid for Taylor expansion. Therefore, when \(T_G\) dominates, \textit{ShapeFit} will be faster than \textit{Full-Modelling}.  


The above timings aside, there are several major caveats in the analysis here. Firstly, one may want to trade the accuracy of the modelling for speed, so the conclusion here will depend on the setting for the Boltzmann code and \textsc{PyBird}. For example, we set the maximum wavenumber for the linear power spectrum from the Boltzmann code to be \(100 h \mathrm{Mpc}^{-1}\). Suppose we reduce this to \(10 h \mathrm{Mpc}^{-1}\), the time per grid point evaluation for \textit{ShapeFit} reduces from 5.3 seconds to around 1.6 seconds and \textit{Full-Modelling} from 13.4 seconds to 9.7 seconds. Furthermore, we are doing the IR-resummation over the full power spectrum. \textsc{PyBird} can also perform IR-resummation only around the BAO peak \citep{d_Amico_2021}. This feature could reduce the time for IR-resummation from around 8 seconds to around 4 seconds. Furthermore, although the default setting for \(n_{\mathrm{max}}\) is 8 for only fitting monopole and quadrupole. When we generate the grid for \textit{Full-Modelling} and template power spectrum for \textit{Shapefit}, we set \(k_{\mathrm{max}} = 0.5 h \mathrm{Mpc}^{-1}\) and with all three multipoles. Therefore, \(n_{\mathrm{max}} = 20\) for the IR-resummation calculation. We chose this setting because we do not need to generate a new grid when we change \(k_{\mathrm{max}}\) or the number of multipoles. However, the higher \(n_{\mathrm{max}}\) made the IR-resummation calculation about three times slower. Changing these settings may significantly impact the conclusion here. Secondly, machine learning emulators such as \textsc{matryoshka} \citep{Donald_McCann_2022} exist. The time it may take to train such an emulator for \textit{Full-Modelling} model power spectrum could be similar to or even faster than training the same emulator for \textit{ShapeFit} parameters. Then, in this case, \textit{Full-Modelling} will always be faster than \textit{ShapeFit}. Thirdly, one could also emulate the model power spectrum based on the \textit{ShapeFit} parameters. \textsc{Velocileptor} and \textsc{FOLPS\(\nu\)} \citep{KP5s2-Maus, KP5s3-Noriega} both use such an approach, and we expect this would again result in \textit{ShapeFit} running faster than \textit{Full-Modelling} even for the $\Lambda$CDM model. We leave this improvement for future work. Lastly, we only investigated cases where we only fit one cosmological model using both approaches. If we are fitting multiple cosmological models, since \textit{ShapeFit} does not need to refit the power spectrum and the grid computation for \textit{ShapeFit} is faster than \textit{Full-Modelling}, \textit{ShapeFit} will have an overall speed advantage when fitting multiple cosmologies. Of course, for cosmologies beyond those tested here, one would want to investigate the validity of the \textit{ShapeFit} approach --- and we have not taken into account in this analysis the considerable computational cost such testing may require. In summary, the speed of \textit{Full-Modelling} and \textit{ShapeFit} depend heavily on the setting of the Boltzmann code, \textsc{PyBird}, and even emulators. Therefore, we encourage the readers to perform their own test if they are using a different setting.

\begin{table}[]
\setlength{\tabcolsep}{4pt}
\begin{tabular}{c|c|c|c|c|c|c|c|c|c|}
Method &Model  & $T_G$ (h) & $L_M$ (s) & $10^{-5}N_M$ & $T_M$ (h) & $L_C$ (s) & $10^{-5}N_C$ & $T_C$ (h) & $T$ (h) \\ \hline \hline
SF (LRG) &$\Lambda$CDM               & 9.66               & 0.45          & 1.20                     & 15     & 0.008         & 1.92                    & 0.43            & 25      \\ \hline
SF (ELG) &$\Lambda$CDM                & 9.66               & 0.45          & 1.92                     & 24     & 0.008         & 2.24                    & 0.50            & 34      \\ \hline
SF (QSO) &$\Lambda$CDM                 & 9.66               & 0.45          & 1.92                     & 24     & 0.008         & 1.60                    & 0.36            & 34      \\ \hline
SF (all) &$\Lambda$CDM           & 9.66               & 0.45          & 5.04                     & 63     & 0.012         & 0.96                     & 0.32            & 73     \\ \hline
FM (LRG)  &$\Lambda$CDM              & 24               & 0.014         & 2.30                     & 0.9      & -           & -                         & -            & 25      \\ \hline
FM (ELG) &$\Lambda$CDM                & 24               & 0.014         & 3.17                     & 1.2      & -           & -                         & -            & 26     \\ \hline
FM (QSO) &$\Lambda$CDM               & 24               & 0.014         & 2.21                     & 0.9      & -           & -                         & -            & 25      \\ \hline 
FM (all) &$\Lambda$CDM           & 24               & 0.045         & 6.40                     & 8.0      & -           & -                         & -            & 32     \\ \hline \hline
SF (LRG) &$w$CDM            & 87               & 0.45          & 1.20                     & 15     & 0.010         & 2.40                    & 0.67            & 103      \\ \hline
SF (ELG)  &$w$CDM              & 87               & 0.45          & 1.92                     & 24     & 0.010         & 2.40                    & 0.67            & 112      \\ \hline
SF (QSO)  &$w$CDM              & 87               & 0.45          & 1.92                     & 24     & 0.010         & 2.40                    & 0.67            & 112      \\ \hline
SF (all)  &$w$CDM          & 87              & 0.45          & 5.04                     & 63     & 0.015         & 2.40                     & 1.00            & 151     \\ \hline
FM (LRG) &$w$CDM             & 220               & 0.017         & 1.68                     & 0.8      & -           & -                         & -            & 221      \\ \hline
FM (ELG) &$w$CDM              & 220               & 0.017         & 2.52                     & 1.2      & -           & -                         & -            & 221      \\ \hline
FM (QSO) &$w$CDM               & 220               & 0.017         & 2.52                     & 1.2      & -           & -                         & -            & 221     \\ \hline
FM (all) &$w$CDM           & 220              & 0.056         & 5.72                     & 8.9      & -           & -                         & -            & 229   \\ \hline
\end{tabular}
\caption{The total run time (\(T\)) for \textit{ShapeFit} and \textit{Full-Modelling} analysis with the Taylor expansion emulators with the \(\Lambda\)CDM model (top half) and the \(w\)CDM model (bottom half). We use the same notation as Table~\ref{tab:time_LCDM_no}. Additionally, it takes \textit{ShapeFit} around 5.3 seconds and \textit{Full-Modelling} around 13.4 seconds to generate a single grid point. We use nine grid points centred around the truth for each parameter. We have four cosmological parameters in \(\Lambda\)CDM, so its total number of grid points is \(9^4 = 6561\). For \(w\)CDM, it will be \(9^5 = 59049\). The number of grid points is the same for \textit{ShapeFit} and \textit{Full-Modelling}. Lastly, we use \(T_G\) to denote the total time it takes to calculate the grid, \(L_M\) to denote the time it takes to run each iteration of MCMC for \textit{ShapeFit} and \textit{Full-Modelling}, and \(L_C\) to denote the time it takes to convert the \textit{ShapeFit} parameters to cosmological parameters for each iteration.}  
\label{tab:time_LCDM_yes}
\end{table}

\section{Correlation function}
\label{sec:corr_func}

Having fully validated and explored different modelling choices when applying the \textsc{PyBird} algorithm to our mock power spectrum measurements, we now demonstrate that these results also apply to the case of a correlation function analysis. \textsc{PyBird} does this by computing the spherical Bessel transform of the linear, 1-loop, and counter-term power spectrum during the instantiation with \textsc{FFTLog} \citep{Minas_2021}.\footnote{The spherical Bessel transforms of the stochastic terms are extremely small, so they are omitted when fitting the correlation function.} These correlation function terms, like the power spectrum terms, are added together to compute the theoretical correlation function. Our aim here is to present that the model can also be quickly and robustly applied to correlation function measurements --- given the implementation in \textsc{PyBird}, we expect our findings for the power spectrum to propagate through to the correlation function. All that remains is to demonstrate that an excellent fit to the data can be obtained for a reasonable choice of fitting scales and that the constraints from the power spectrum and correlation function are consistent. 

\begin{figure}
    \centering
    \includegraphics[width=1.0\textwidth]{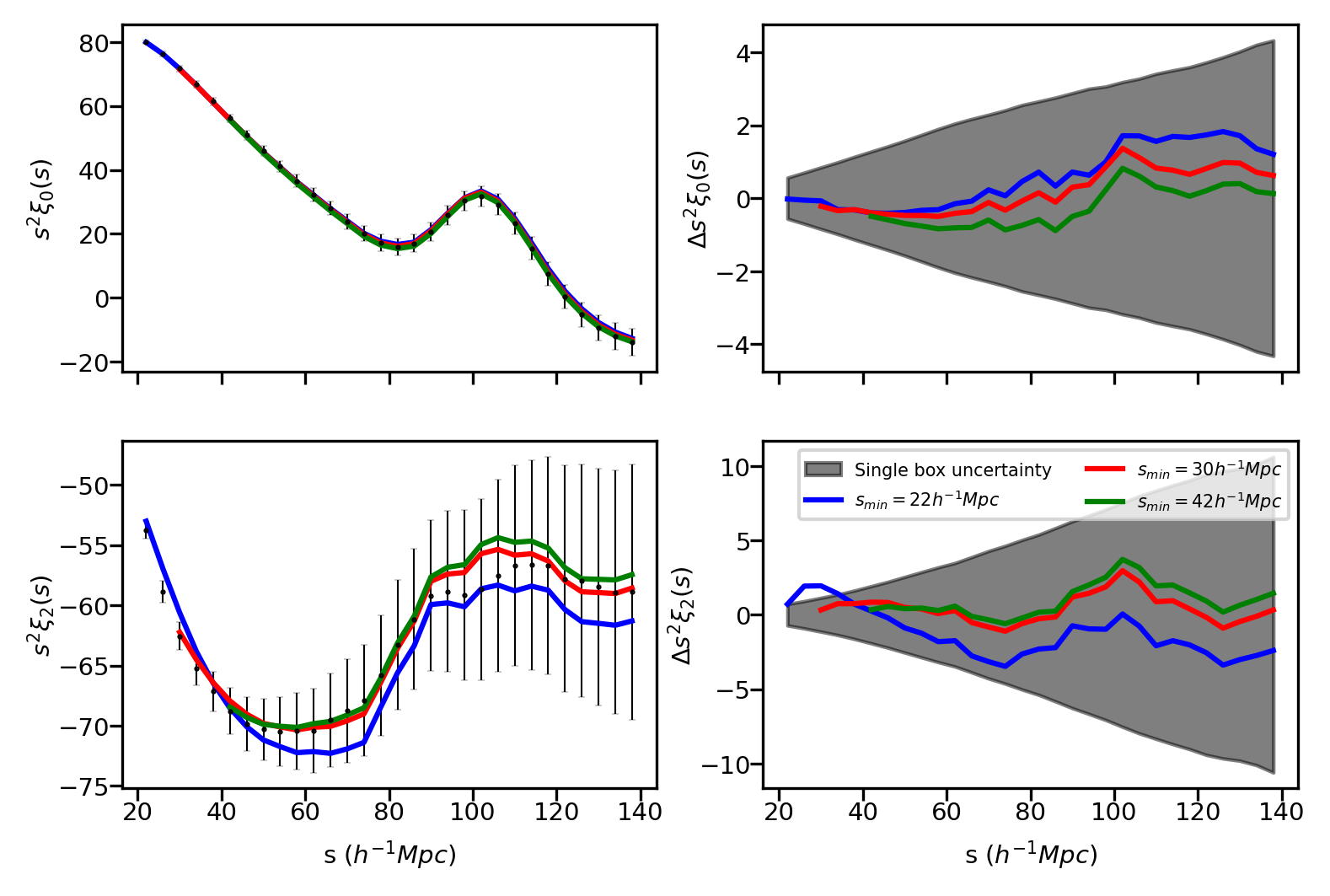}
    \caption{This plot shows the best-fit correlation functions against the data. The single-box covariance matrix gives the error bars. The grey band on the right corresponds to the error bars on the left. The best-fit correlation functions are consistent with each other for \(s_{\mathrm{min}} = 30 h^{-1} \mathrm{Mpc}\) and \(s_{\mathrm{min}} = 42 h^{-1} \mathrm{Mpc}\). However, the best-fit for \(s_{\mathrm{min}} = 22 h^{-1} \mathrm{Mpc}\) is consistently larger than the other two best-fits for the monopole and smaller than the other two best-fits for the quadrupole, arising from the fit being driven by the small-scale data which has smaller error bars. }
    \label{fig:CF_bestfit}
\end{figure}

To start, Fig.~\ref{fig:CF_bestfit} shows the best-fit correlation function and the data correlation function for different choices of minimum fitting scale $s_{\mathrm{min}}$ using the \textit{Full-Modelling} approach. The best-fit correlation functions are consistent with each other for the monopole for \(s_{\mathrm{min}} = 30 h^{-1} \mathrm{Mpc}\) and \(s_{\mathrm{min}} = 42 h^{-1} \mathrm{Mpc}\). However, the best-fit for \(s_{\mathrm{min}} = 22 h^{-1} \mathrm{Mpc}\) is constantly larger than the other two best-fits for the monopole and smaller than the other two best-fits for the quadrupole. This result is because the small-scale data has smaller error bars, so our model prioritizes fitting for the small-scale where the model is likely to break down.  

\begin{figure}
    \centering
    \includegraphics[width=1\linewidth]{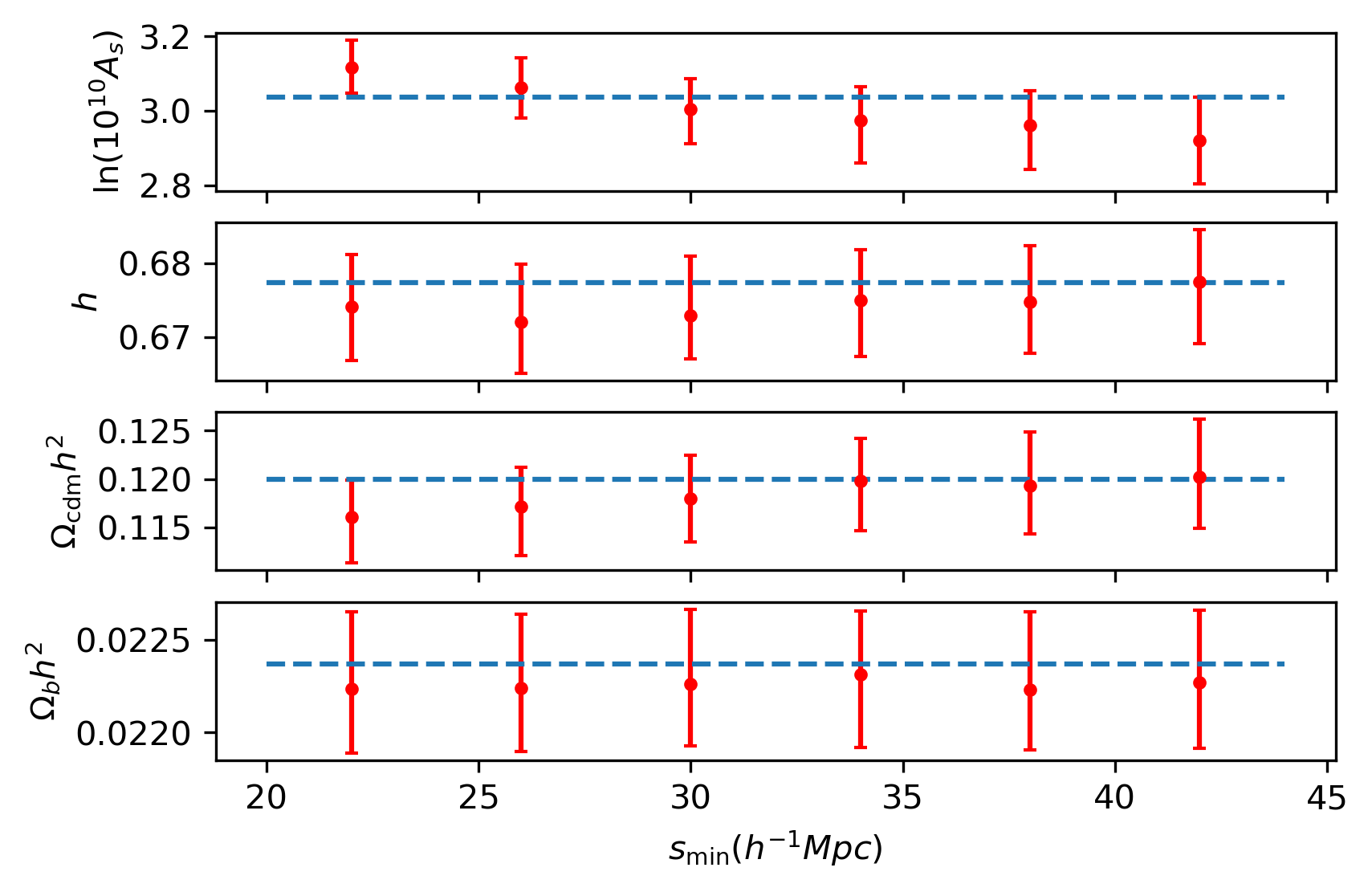}
    \caption{The effect of \(s_{\mathrm{min}}\) of the correlation function on the constraints of cosmological parameters within the \(\Lambda\)CDM cosmological model. Except for the prior dominated \(\omega_b\), increasing \(s_{\mathrm{min}}\) generally reduces the constraining power because the information on small scales is lost. The mean of the posterior for \(\ln({10^{10}A_s})\) decreases when increasing \(s_{\mathrm{min}}\). This result is similar to Fig.~\ref{fig:FS_kmax_nohex}, likely due to the model failure at the small scale. We find the most accurate model is obtained when $s_{\mathrm{min}} = 30\,h^{-1}\mathrm{Mpc}$.}
    \label{fig:CF_kmax}
\end{figure}

To explore this further, Fig.~\ref{fig:CF_kmax} and the corresponding Table~\ref{tab:CF_params} illustrate that increasing \(s_{\mathrm{min}}\) reduces the constraining power because we are losing information. Different from the power spectrum, the systematic error for \(\ln({10^{10}A_s})\) does not seem to always reduce when we remove small-scale information – there is a minimum beyond which the deviation of \(\ln({10^{10}A_s})\) from the truth seems to increase again. The source of this is unclear, but it is interesting to note that the best-fit residuals of the monopole around the BAO are smaller with larger \(s_{\mathrm{min}}\). This could indicate the mock data from the small-scale prefers a larger \(\ln{(10^{10}A_s)}\) while the large-scale mock data prefers a smaller \(\ln{(10^{10}A_s)}\). This could cause the trend in Fig.~\ref{fig:CF_kmax}. Furthermore, the best-fit residuals, shown in Fig~\ref{fig:CF_bestfit}, have a bump around the BAO scale for the quadrupole, which persists regardless of the value of \(s_{\mathrm{min}}\) used in the fits. Although this feature is smaller than the error bar, similar features are also seen in Ref.~\citep{Zhang_2022} with \textsc{PyBird}, indicating it is probably not a noise feature of our mock. One potential origin could be due to terms proportional to powers of \(k^2 Y_1\) in Ref.~\citep{Lewandowski_2018}. In \textsc{PyBird}, these terms are neglected beyond the first order. These neglected terms could contribute more significantly to the BAO peak in higher multipoles. Nonetheless, we stress that this feature is much smaller than the error bar for all \(s_{\mathrm{min}}\) tested, indicating the modelling of BAO in the EFTofLSS remains highly accurate.

\begin{table}
    \centering
    \renewcommand{\arraystretch}{1.4} 
    \caption{This table summarizes the deviation of the best fits from the truth values for the correlation function with different \(s_{\mathrm{min}}\). The deviations are generally less than \(1\sigma\) for all the tested scales.}
    \label{tab:CF_params}
    \begin{tabular}{c|c|c|c|c}
        \hline
		Model & $\Delta \ln(10^{10} A_s) \%$ & $\Delta h \%$ & $\Delta \Omega_{\mathrm{cdm}} h^2 \%$ & $\Delta \Omega_bh^2 \%$ \\ 
		\hline \hline
		$s_{\mathrm{min}} = 22 h^{-1} \mathrm{Mpc}$ & $2.6^{+2.4}_{-2.3}$ & $0.1^{+1.0}_{-1.1}$ & $-3.2^{+3.1}_{-4.0}$ & $-0.6^{+1.8}_{-1.6}$ \\ \hline
		$s_{\mathrm{min}} = 26 h^{-1} \mathrm{Mpc}$ & $0.9^{+2.6}_{-2.7}$ & $-0.2^{+1.2}_{-1.0}$ & $-2.4^{+3.4}_{-4.2}$ & $-0.6^{+1.8}_{-1.5}$ \\ \hline
		$s_{\mathrm{min}} = 30 h^{-1} \mathrm{Mpc}$ & $-1.0^{+2.6}_{-3.1}$ & $-0.11^{+1.20}_{-0.87}$ & $-1.7\pm 3.7$ & $-0.5^{+1.8}_{-1.5}$ \\ \hline
		$s_{\mathrm{min}} = 34 h^{-1} \mathrm{Mpc}$ & $-2.0^{+2.9}_{-3.8}$ & $0.2^{+1.0}_{-1.1}$ & $-0.1^{+3.7}_{-4.3}$ & $-0.3^{+1.5}_{-1.8}$ \\ \hline 
		$s_{\mathrm{min}} = 38 h^{-1} \mathrm{Mpc}$ & $-2.5^{+3.1}_{-3.9}$ & $0.2^{+1.1}_{-1.0}$ & $-0.6^{+4.6}_{-4.1}$ & $-0.6^{+1.9}_{-1.5}$ \\ \hline
		$s_{\mathrm{min}} = 42 h^{-1} \mathrm{Mpc}$ & $-3.8\pm 3.8$ & $0.6^{+1.0}_{-1.2}$ & $0.2^{+5.0}_{-4.4}$ & $-0.4^{+1.7}_{-1.6}$ \\ \hline
		\hline
    \end{tabular}
\end{table}

\begin{figure}
    \centering
    \includegraphics[width=0.495\textwidth]{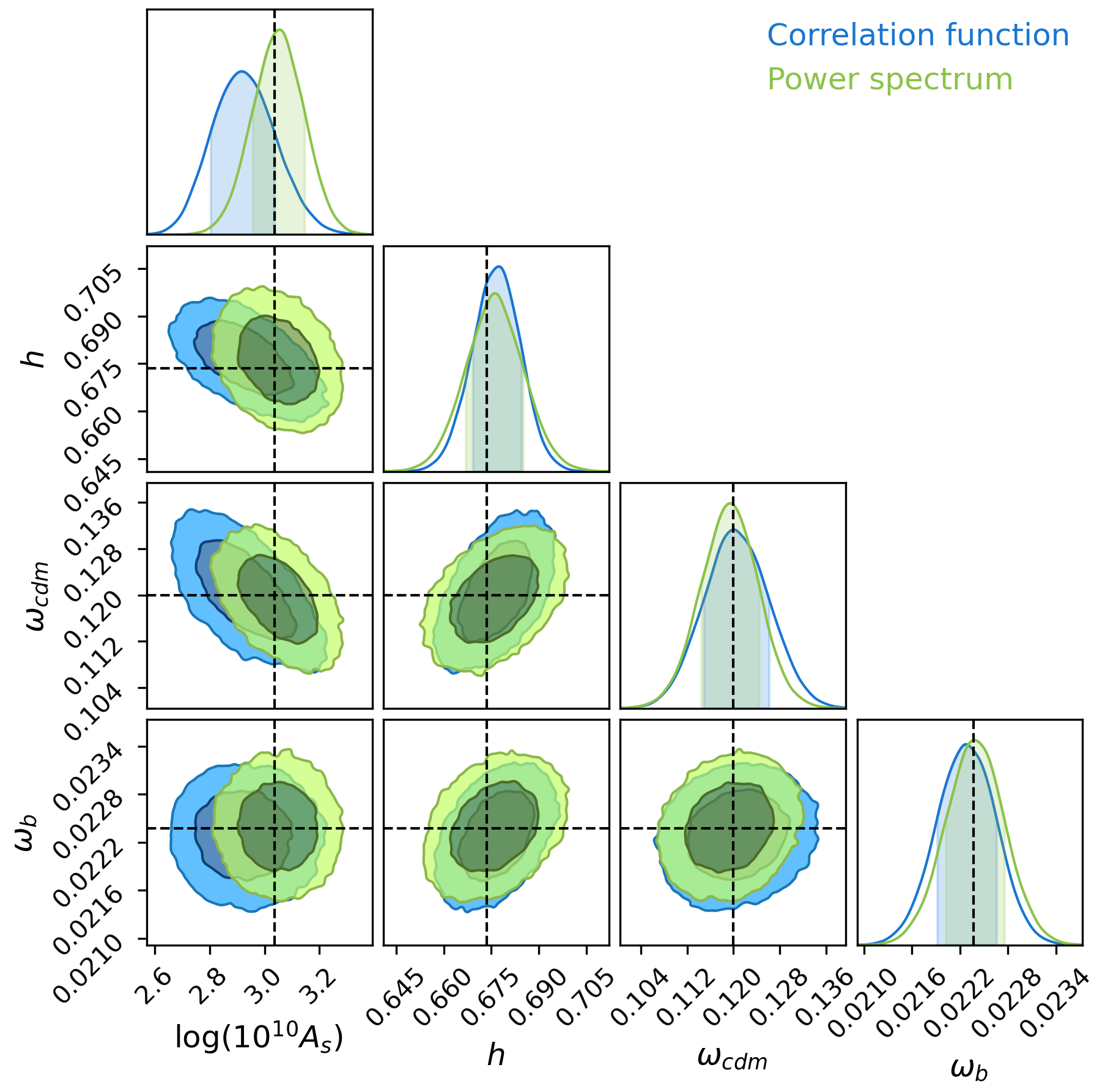}
    \includegraphics[width=0.495\textwidth]{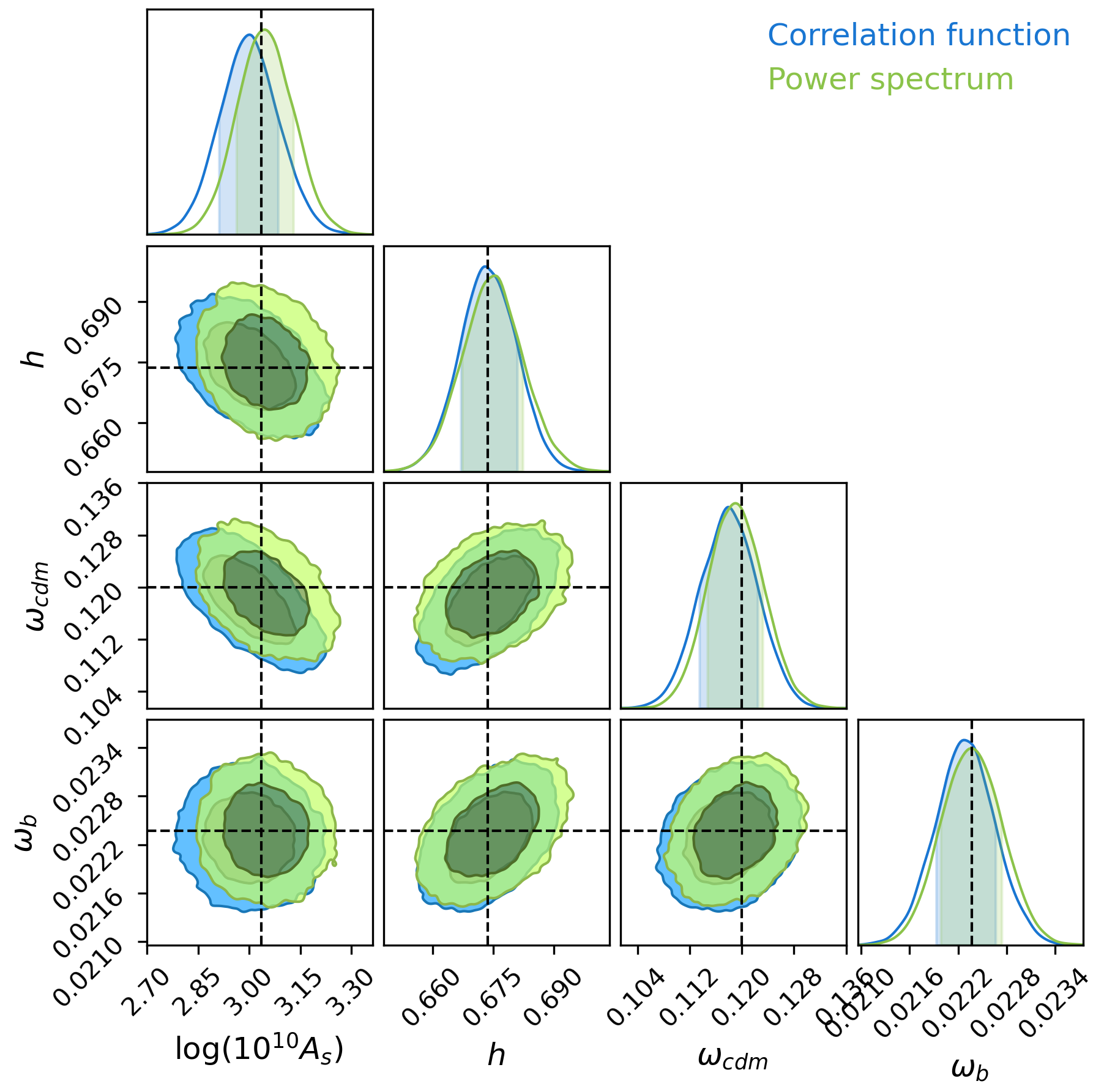}
    \includegraphics[width=0.495\textwidth]{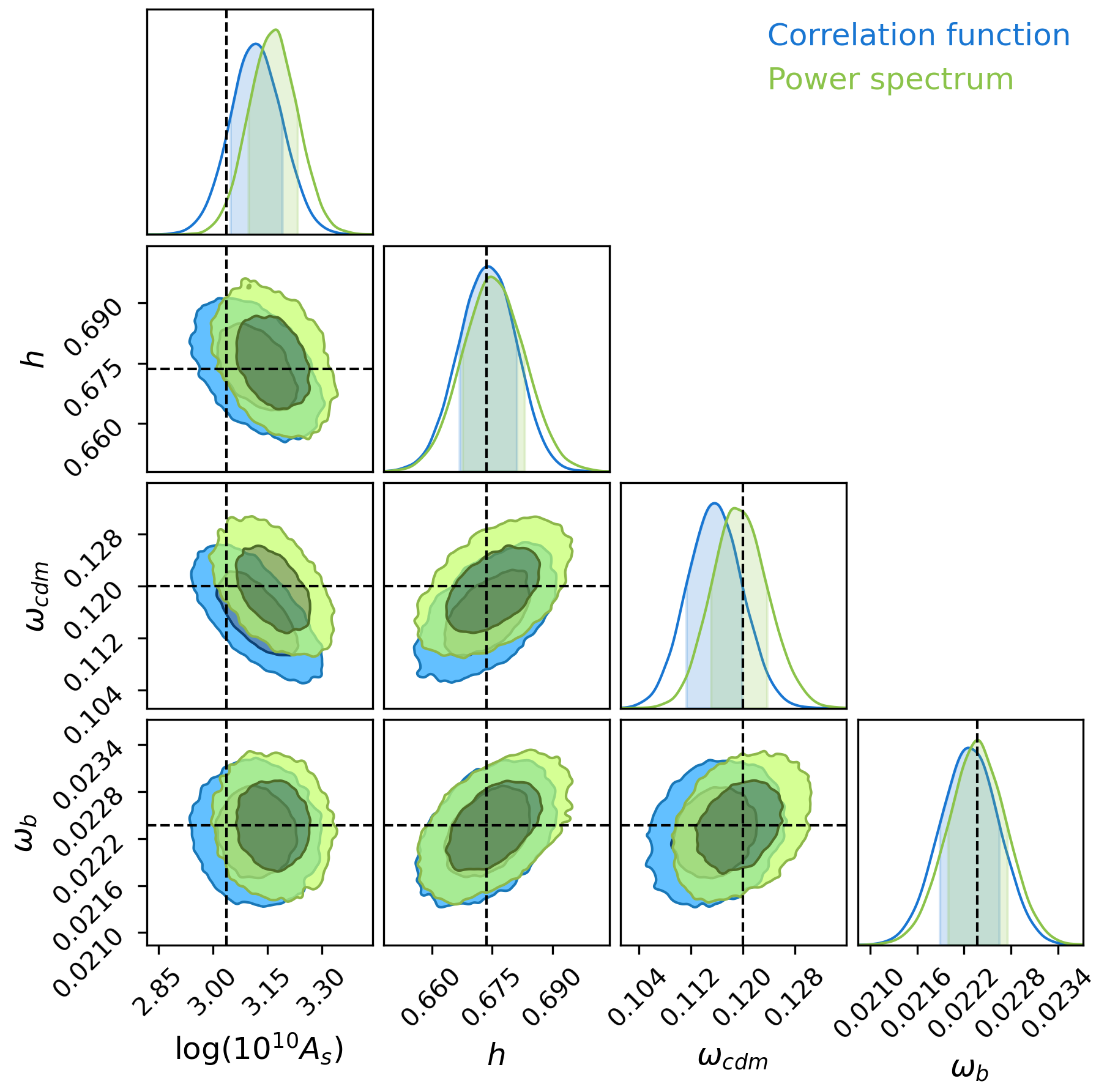}
    \caption{This plot compares the constraints from the correlation function and power spectrum. They both use the ``BOSS MaxF" prior and without the hexadecapole. For all three figures, \(s_{\mathrm{max}} = 138h^{-1} \mathrm{Mpc}\) for the correlation functions and \(k_{\mathrm{min}} = 0.02 h \mathrm{Mpc}^{-1}\) for the power spectra. On the top left, we use \(k_{\mathrm{max}} = 0.16 h \mathrm{Mpc}^{-1}\) for the power spectrum and \(s_{\mathrm{min}} = 42 h^{-1} \mathrm{Mpc}\) for the correlation function. On the top right, we use \(k_{\mathrm{max}} = 0.20 h \mathrm{Mpc}^{-1}\) for the power spectrum and \(s_{\mathrm{min}} = 30 h^{-1} \mathrm{Mpc}\) for the correlation function. On the bottom, we use \(k_{\mathrm{max}} = 0.28 h \mathrm{Mpc}^{-1}\) for the power spectrum and \(s_{\mathrm{min}} = 22 h^{-1} \mathrm{Mpc}\) for the correlation function. For all three plots, results from \(k_{\mathrm{max}} = 0.20 h \mathrm{Mpc}^{-1}\) and \(s_{\mathrm{min}} = 30 h^{-1} \mathrm{Mpc}\) are largely consistent with at most $0.5\sigma$ shifts in the constraints. We also find that the power spectrum becomes biased more quickly when including small scales, and the correlation function loses constraining power more quickly, fitting only larger scales.}
    \label{fig:CF_vs_FS_kmax}
\end{figure}

Finally, in Fig.~\ref{fig:CF_vs_FS_kmax}, we compare the constraints from the correlation function to the power spectrum with a similar range of scales. For the correlation function (power spectrum), we fix \(s_{\mathrm{max}} = 138 h^{-1} \mathrm{Mpc}\) (\(k_{\mathrm{min}} = 0.02 h \mathrm{Mpc}^{-1}\)). For the top left plot, we set \(s_{\mathrm{min}} = 42 h^{-1} \mathrm{Mpc}\) (\(k_{\mathrm{max}} = 0.16 h \mathrm{Mpc}^{-1}\)). For the top right plot, we set \(s_{\mathrm{min}} = 30 h^{-1} \mathrm{Mpc}\) (\(k_{\mathrm{max}} = 0.20 h \mathrm{Mpc}^{-1}\)). For the bottom plot, we set \(s_{\mathrm{min}} = 22 h^{-1} \mathrm{Mpc}\) (\(k_{\mathrm{max}} = 0.28 h \mathrm{Mpc}^{-1}\)). The \(s_{\mathrm{min}}\) and \(k_{\mathrm{max}}\) are chosen such that \(s_{\mathrm{min}} \approx \frac{2\pi}{k_{\mathrm{max}}}\). This figure illustrates that the constraints of cosmological parameters within the \(\Lambda\)CDM from the correlation function are consistent with the constraints from the power spectrum. For our optimum choice of scales (the top right panel), the constraining power is similar, and there is only a small shift ($<0.5\sigma$) in the constraints primarily in the $h$ and \(\ln{(10^{10}A_s)}\) parameters. As one goes to smaller scales, the correlation function stays robust longer than the power spectrum. Conversely, the correlation function loses constraining power more quickly as we cut to larger scales.

In summary, we expect consistent results from applying \textsc{PyBird} to the DESI data using the correlation function or power spectrum. Although we have only tested the correlation function method with the LRG mocks, we expect similar consistency for other tracers and the combined probes since, in our implementation, the correlation function is treated purely as the Fourier counterpart of the power spectrum.

\section{Conclusion}
\label{sec:conclusion}
This work has thoroughly tested and validated that the \textsc{PyBird} EFTofLSS model and algorithm can produce unbiased tight constraints for the cosmological parameters for the DESI survey and at a level of accuracy and precision sufficient for upcoming Year 1 analyses using the combination of LRG, ELG and QSO galaxy types that DESI is observing. Using this model, we have implemented two ways of extracting cosmological constraints: \textit{Full-Modelling} fitting of the data given a set of cosmological parameters and the \textit{ShapeFit} compression method. We find the constraints on cosmological parameters from both methods are remarkably consistent within the \(\Lambda\)CDM, $w$CDM, \(o\)CDM cosmological models, with $\sim 0.5\sigma$ systematic shifts on the best-fit cosmological parameters between \textit{Full-Modelling} and \textit{ShapeFit} analyses for \(\Lambda\)CDM, \(\leq 0.7 \sigma\) for \(w\)CDM, and \(\leq 0.5\sigma\) for \(o\)CDM. This difference is small compared to the statistical error, and we will leave it for future work to investigate its origin further. 

In more detail, for both the \textit{Full-Modelling} and \textit{ShapeFit} methods we find that using a maximum fitting scale of \(k_{\mathrm{max}} = 0.20 h \mathrm{Mpc}^{-1}\) gives accurate constraints with the single box covariance (equivalent to DESI Y5 volume) --- going beyond this does not substantially improve the constraining power of the model, and slightly increases the systematic errors. We also find that including the hexadecapole does not enhance the constraints of the \(\Lambda\)CDM model but gives a notable improvement on the constraints for the $w$CDM and the \(o\)CDM models. We also find a reduction in systematic bias when including the hexadecapole. As such, we recommend including this when fitting DESI or other next-generation data for extended cosmological models. 

We also test different choices of prior configurations and show that the ``BOSS MinF" case, where we use a minimal set of nuisance parameters and the local Lagrangian relation to fix the values of the non-linear galaxy bias parameters to functions of the linear galaxy bias, can slightly reduce systematic errors, and slightly improve the statistical precision. Furthermore, we find that without emulators, \textit{ShapeFit} is faster than \textit{Full-Modelling} for \(\Lambda\)CDM and \(w\)CDM cosmology. With the Taylor expansion emulator, \textit{ShapeFit} remains faster if the majority of computational time is in calculating the grid of models for the Taylor expansion (e.g. \(w\)CDM or \(o\)CDM). Otherwise, \textit{Full-Modelling} is faster (e.g. \(\Lambda\)CDM). However, this result depends on the setting of the Boltzmann code, \textsc{PyBird} and possibly emulators. Lastly, we also constrain the cosmological parameters with the correlation function. We find that the constraints are consistent with the results from the power spectrum. Although the correlation function tends to remain more robust as we include smaller-scale information, it loses constraining power more quickly when restricted to large fitting scales. 

This work forms one in a series validating different cosmological fitting pipelines for the DESI survey. A comparison between \textsc{PyBird} and other pipelines in DESI is summarized in the companion paper \citep{KP5s1-Maus}, where we find that the results in this paper are generally reflected in the other pipelines and that \textsc{PyBird} is in good agreement with other methods and the level required for DESI Year 1 analyses. Future work will apply \textsc{PyBird} and the configurations investigated herein to the data from DESI.

\appendix
\section{Co-evolution of \textsc{PyBird} bias parameters}
\label{sec:bias_transform}

Bias parameters in different models of Large-Scale Structure can be transformed to one another using the ``Monkey bias" parameters \citep{Fujita_2020}. For other codes tested as part of the DESI validation program (such as \textsc{Velocileptor} and \textsc{FOLPS\(\nu\)}), local Lagrangian relations \citep{Desjacques_2018} between their bias parameters are used to provide tighter constraints on cosmological parameters. However, to the best of the authors' knowledge, no such relation is known for bias expansion used in \textsc{PyBird}. To derive such relations, we here use the co-evolution relation in \textsc{Velocileptor} and transform it to \textsc{PyBird} bias parameters using the ``Monkey bias" relations. The conversion is summarized in Table \ref{tab: bias_convert}. 
 
 \begin{table}[t!]
\centering      
\renewcommand{\arraystretch}{1.6} 
\begin{tabular}{c|c|c}        
``Monkey bias'' parameter & \texttt{PyBird} & \texttt{Velocileptor} 
\\ \hline \hline
$a_1^m$ &  $b_1$ & $b_1^{v}+1$   \\
\hline
$b_3^m$ & $\frac{2}{7}b_2$ & $\frac{2}{7} + b_s^v $  \\
\hline
$f_1^m$ &  $\frac{2}{21}b_3$ & $\frac{42 - 145b_1^{v}-21b_3^{v}+630b_s^v}{441}$   \\
\hline
$b_1^m$ & $\frac{5}{7}b_2+b_4$ & $\frac{5}{7} + b_1^{v} + \frac{b_2^{v}}{2}-\frac{b_s^v}{3}$  \\
\end{tabular}
\caption{This table shows the conversion of bias parameters in \textsc{Velocileptor} and \textsc{PyBird} to the ``Monkey bias" parameters. The superscript \(v\) indicates the bias parameters are from the \textsc{Velocileptor}.}    
\label{tab: bias_convert} 
\end{table}

For the local Lagrangian relation in \textsc{Velocileptor}, \(b_3^v\) and \(b_s^v\) are set to zero. Therefore, 
\begin{align}
    b_1 &= b_1^v + 1 \nonumber \\
    b_2 &= \frac{7}{2}(\frac{2}{7} + b_s^v) = 1 \nonumber \\
    b_3 &= \frac{21(42 - 145b_1^v-21b_3^v+630b_s^v)}{882} = \frac{882-3045(b_1 - 1)}{882} \nonumber \\
    b_4 &= (b_1-1)+\frac{1}{2}b_2^v 
    \label{eq:conversion}
\end{align}
After the conversion, the only two free parameters are \(b_1\) and \(b_2^v\). To simplify the equations further, instead of putting priors on \(b_2^v\) and then computing the corresponding \(b_4\) during each iteration, we instead decide to put a flat prior on \(b_4\) directly. The range of this flat prior is determined by the prior on \(b_1\) and \(b_2^v\) used in the \textsc{Velocileptor} companion paper \citep{KP5s2-Maus}.

\section{\textit{ShapeFit} conversion nuances}
\label{sec:conversion}
\begin{figure}
    \centering
    \includegraphics[width=0.495\textwidth]{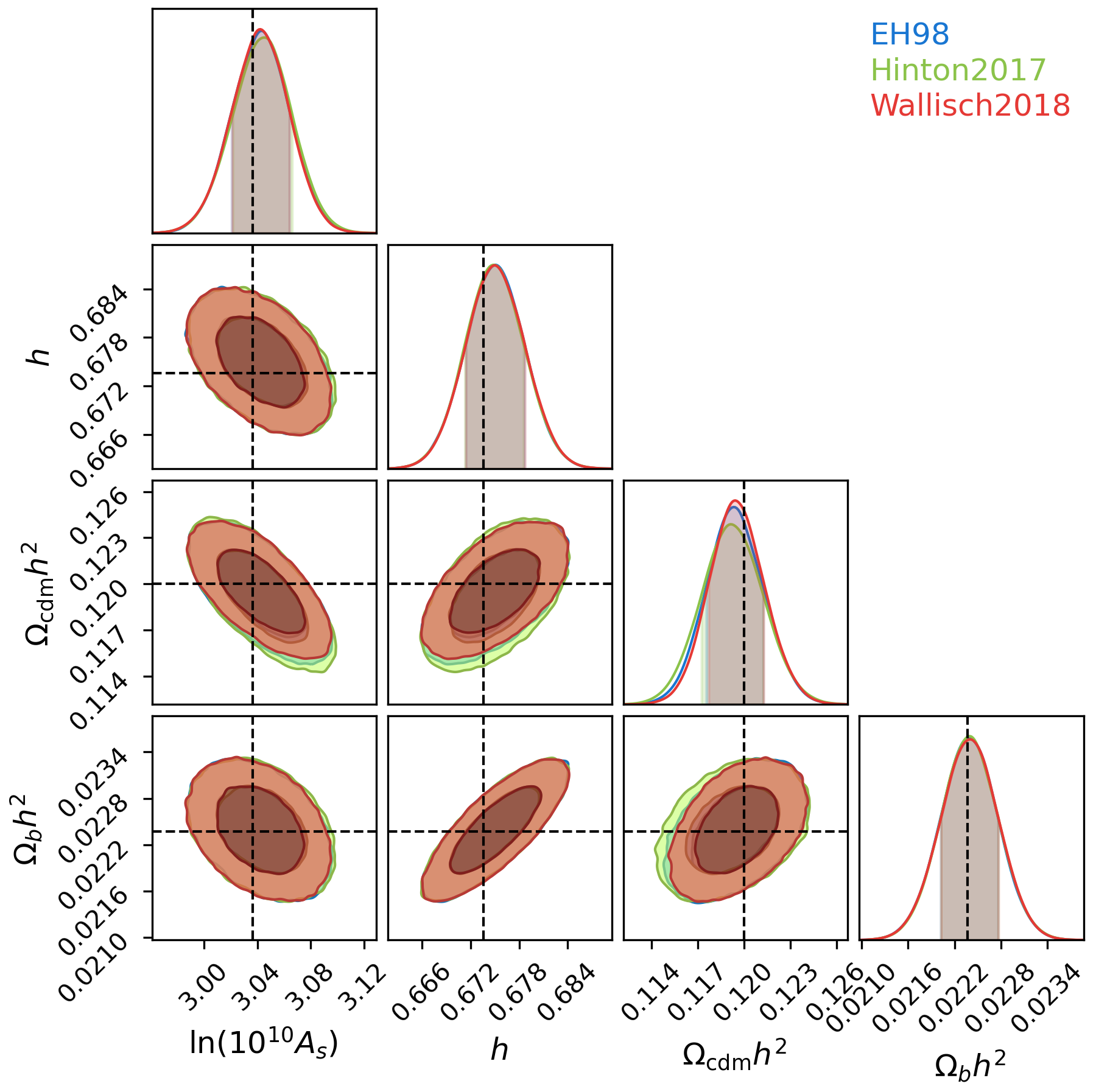}
    \includegraphics[width=0.495\textwidth]{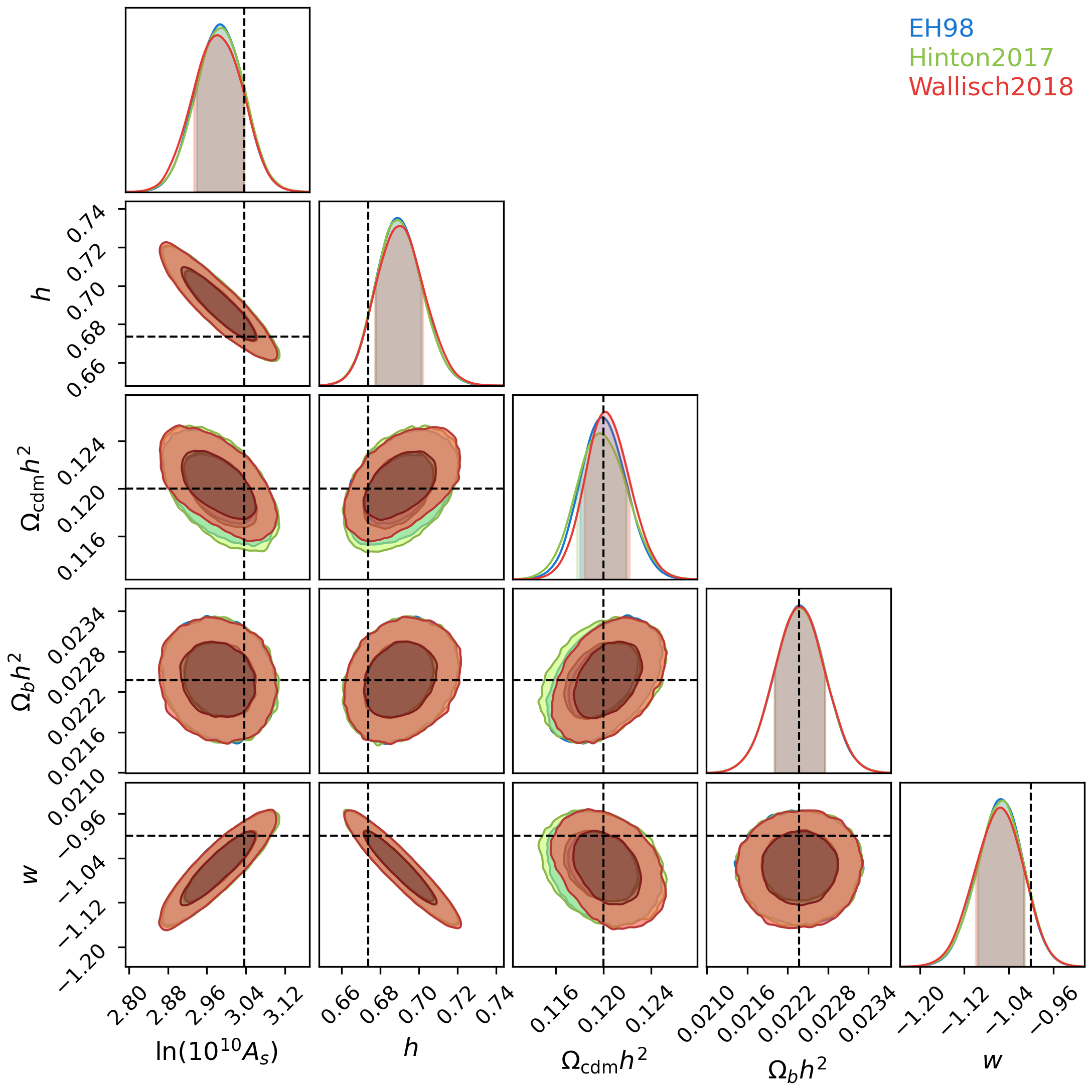}
    \includegraphics[width=0.495\textwidth]{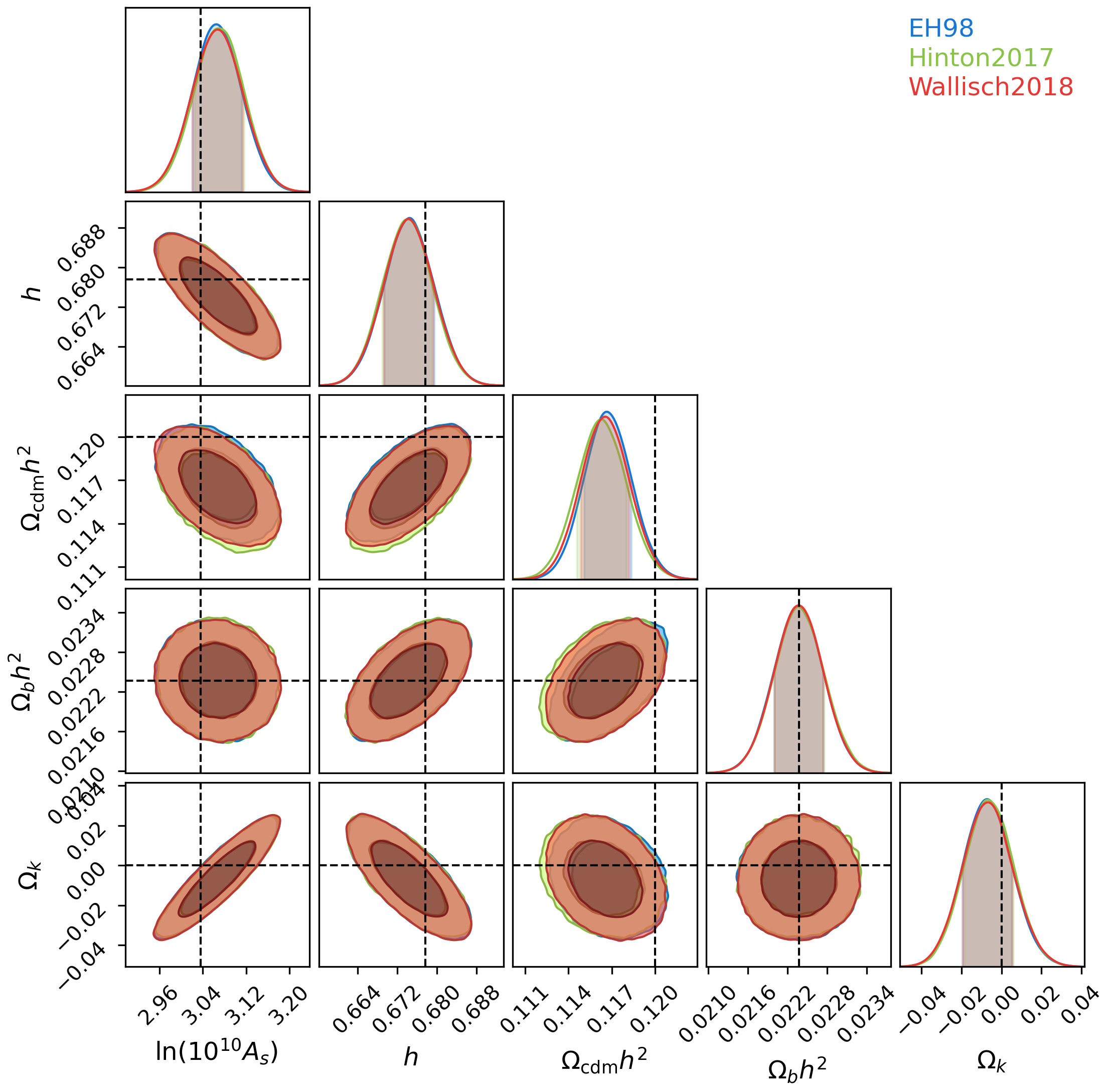}
    \caption{Different methods of calculating the no-wiggle linear power spectrum when converting from \textit{ShapeFit} to cosmological parameters applied to the mean of the ELG mocks assuming the \(\Lambda\)CDM (top left) and the $w$CDM models (top right), and to the mean of the LRG mocks for the \(o\)CDM models (bottom). These fits are done with a $25\times$ reduced covariance matrix to test whether different de-wiggle algorithms will introduce bias into the constraints of cosmological parameters. We find excellent agreements among all three methods for the three cosmological models.} 
    \label{fig:conversion_LRG_LCDM_nohex}
\end{figure}
When converting \textit{ShapeFit} constraints to cosmological parameters, we use equation (\ref{eq: alpha_perp}), (\ref{eq: alpha_par}), (\ref{eq: fsigma8}), and (\ref{eq: m_convert}). The conversion from these expressions is straightforward for the \(\alpha\) parameters and \(f\sigma_8\). However, for the slope parameter \(m\), there are different ways one can calculate the ``de-wiggled'' transfer function (i.e., the transfer function of baryons and dark matter, but without the presence of BAO), so we want to make sure the final constraints on the cosmological parameters do not depend on how this is obtained.

Ref.~\citep{Brieden_2021} suggested using the analytical Eisenstein-Hu transfer function \citep{Eisenstein_1999} (hereon referred to as EH98). However, these fitting functions are not necessarily accurate enough for modern cosmological inference, and numerical algorithms have been developed since then, particularly within the scope of BAO fitting. In this work, we also tested the polynomial algorithm from Ref.~\citep{Hinton_2017} (hereon referred to as Hinton2017) and spectral decomposition algorithm from Ref.~\citep{Wallisch_2018} based on Ref.~\citep{Hamann_2010} (hereon referred to as Wallisch2018). 

Fig. \ref{fig:conversion_LRG_LCDM_nohex} compares the cosmological constraints obtained using these algorithms within the context of the \(\Lambda\)CDM, the $w$CDM, and the \(o\)CDM models. We use these three algorithms to convert the \textit{ShapeFit} parameters to cosmological parameters. For \(\Lambda\)CDM and \(w\)CDM, we use the reduced covariance matrix with the mean of the 25 ELG mocks. For \(o\)CDM, we use the reduced covariance matrix with the mean of the 25 LRG mocks. All three algorithms give consistent constraints on the cosmological parameters. The Hinton2017 algorithm gives a slightly larger error bar for \(\Omega_{\mathrm{cdm}}h^2\), but it is not significant. Therefore, we conclude that our constraints on the cosmological parameters are independent of the de-wiggle algorithm.

\section{Performance of Taylor expansion}
\label{sec:Taylor_grid}
\begin{figure}
    \centering
	\includegraphics[width=0.75\textwidth]{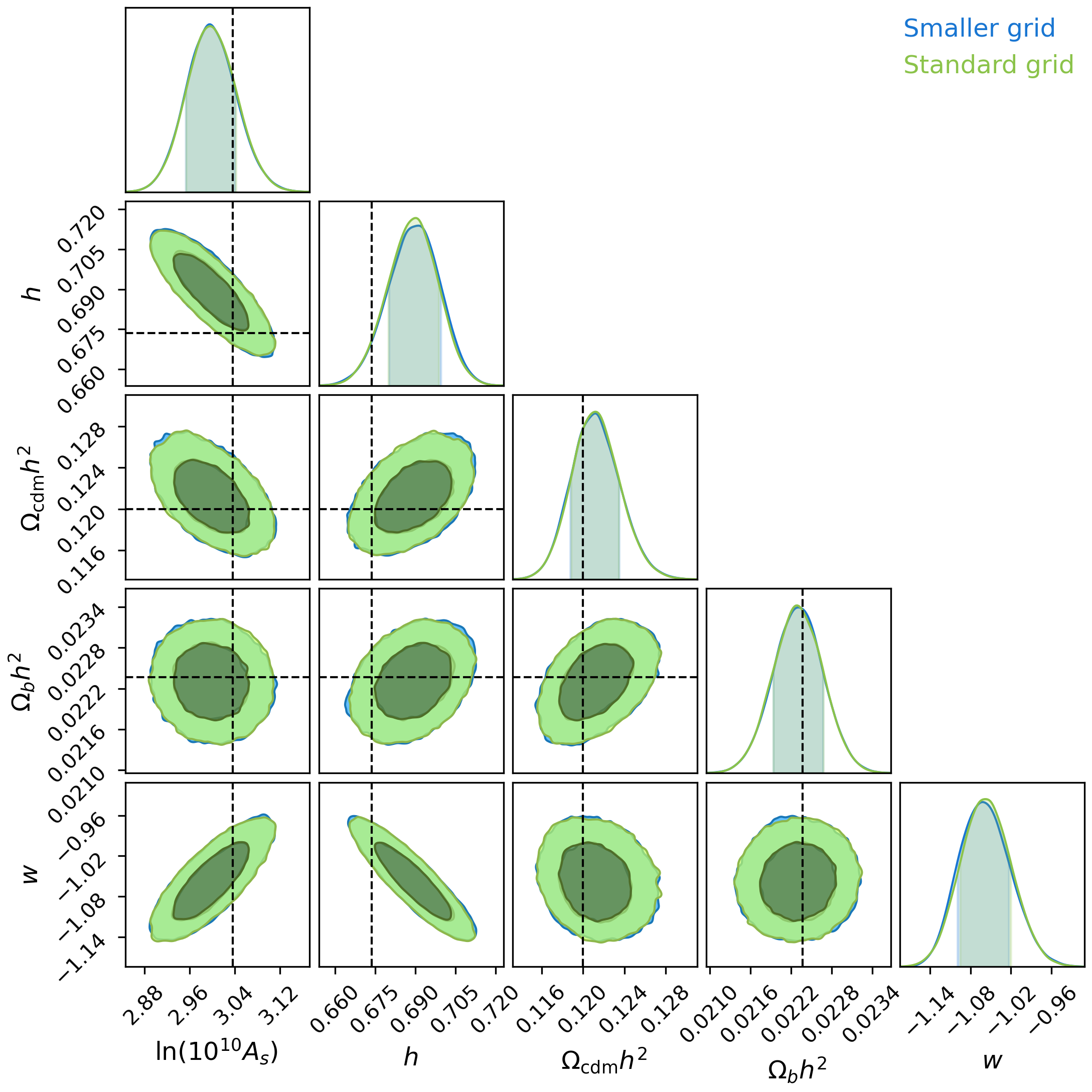}
    \caption{This plot shows the constraints on the $w$CDM cosmological parameters fitting to the mean of the LRG mocks with a $25\times$ reduced covariance matrix for two very different choices of the Taylor expansion grid resolution. Compared to the standard grid size, the smaller grid size shrinks \(\Delta \ln{(10^{10} A_s)}\) from \(0.25\) to \(0.05\), \(\Delta h\) from \(0.03\) to \(0.015\), \(\Delta \omega_{cdm}\) from \(0.01\) to \(0.0025\), \(\Delta \omega_b\) from 0.001 to 0.0004, and \(\Delta w\) from 0.075 to 0.05. Overall, we see that changing the size of the grid for Taylor expansion has no impact on the constraints of the cosmological parameters, indicating for \(w\)CDM models that the Taylor expansion does not introduce bias into our final constraints.}
    \label{fig:Taylor_DE}
\end{figure}

\begin{figure}
    \centering
	\includegraphics[width=0.75\textwidth]{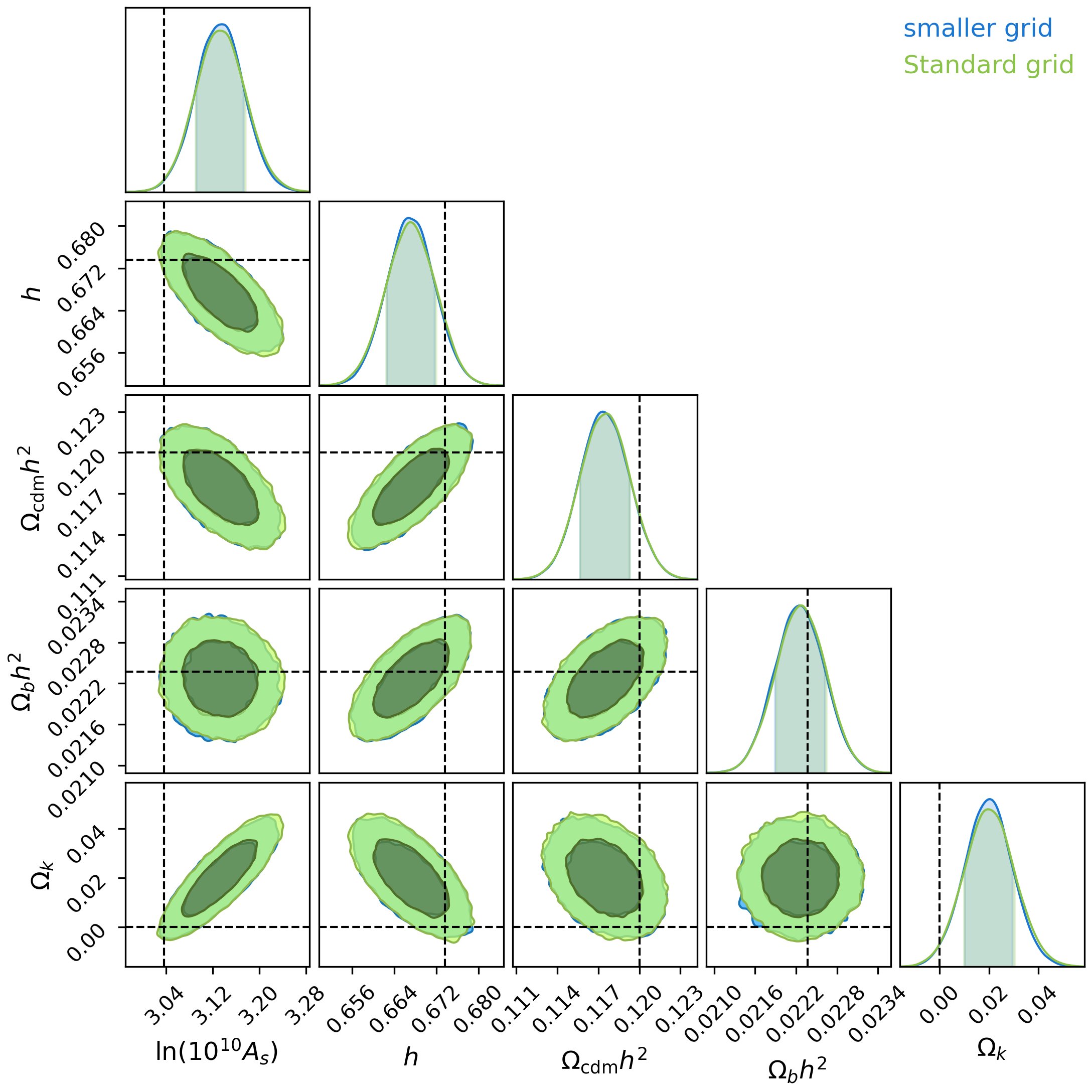}
    \caption{This plot shows the constraints on the $o$CDM cosmological parameters fitting to the mean of the LRG mocks with a $25\times$ reduced covariance matrix for two very different choices of the Taylor expansion grid resolution. Compared to the standard grid size, the smaller grid size shrinks \(\Delta \ln{(10^{10} A_s)}\) from \(0.25\) to \(0.075\), \(\Delta h\) from \(0.03\) to \(0.015\), \(\Delta \omega_{cdm}\) from \(0.01\) to \(0.0025\), \(\Delta \omega_b\) from 0.001 to 0.0004, and \(\Delta \Omega_k\) from 0.0625 to 0.025. Similar to Fig.~\ref{fig:Taylor_DE}, changing the grid size has a negligible impact on the constraints of the cosmological parameters. Therefore, for \(o\)CDM models, the Taylor expansion does not introduce bias into our final constraints.}
    \label{fig:Taylor_omegak}
\end{figure}

To speed up our fitting, we use a third-order Taylor expansion around a set of grid points to evaluate the model power spectra and the \textit{ShapeFit} parameters given a set of cosmological parameters. For the \(\Lambda\)CDM case, Ref. \citep{Colas_2020} has done extensive tests on the Taylor expansion emulator with \textsc{PyBird}. Therefore, in this work, we focus on testing the implementation of the Taylor expansion for the \(w\)CDM and \(o\)CDM models. We fit the $25\times$ reduced covariance matrix to the mean of the LRG mocks to provide a much more rigorous test, Fig.~\ref{fig:Taylor_DE} and Fig.~\ref{fig:Taylor_omegak} illustrate that the grid size of the Taylor expansion does not affect the constraints on the cosmological parameters in the $w$CDM model and the \(o\)CDM model. 

\section{Marginalized vs unmarginalized constraints}
\label{sec:marg}
\begin{figure}
    \centering
    \includegraphics[width=1\linewidth]{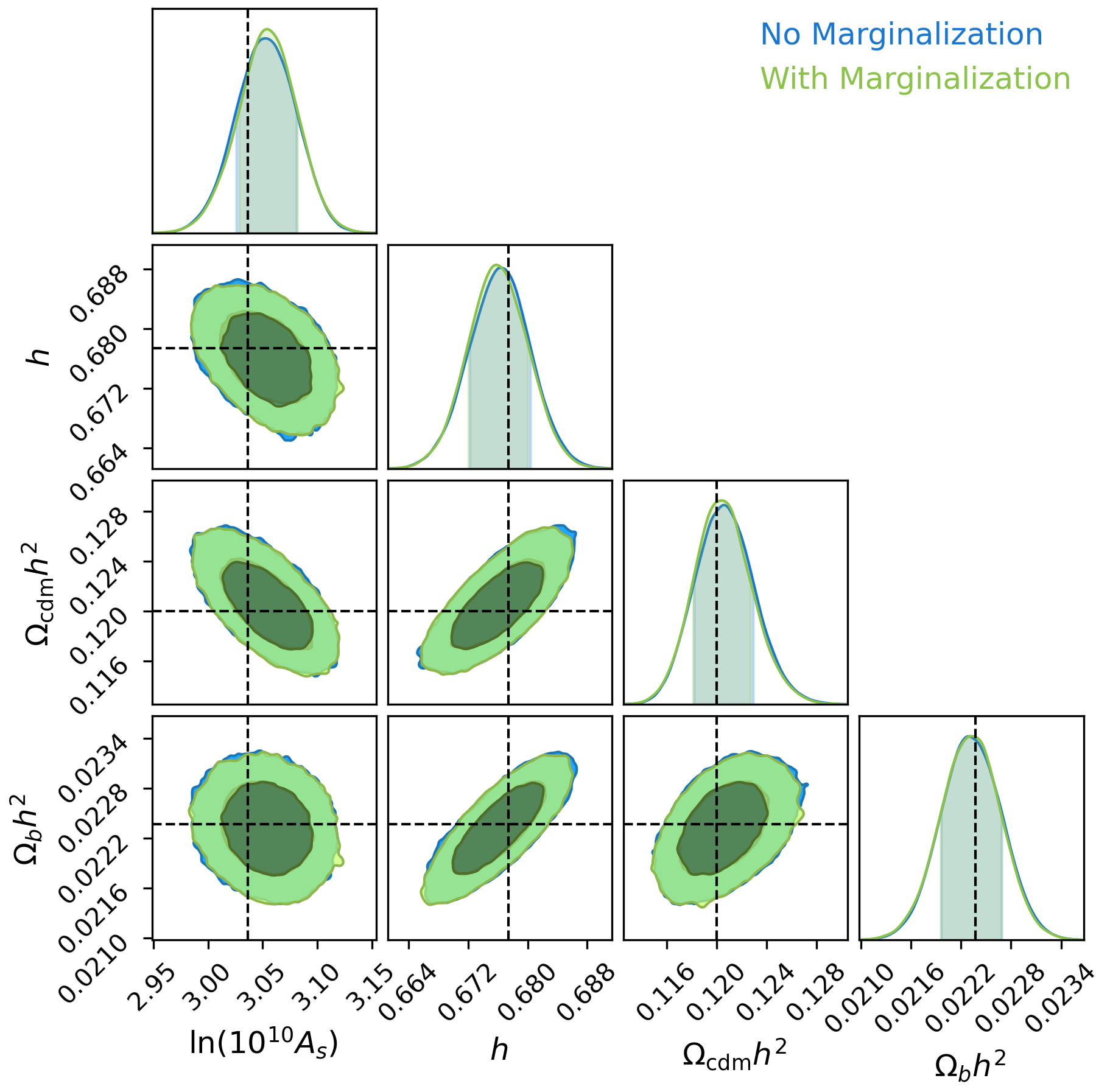}
    \caption{Parameter constraints with the mean of the LRG mocks and a $25\times$ reduced covariance matrix using the analytically marginalized and unmarginalized likelihood functions. The constraints on the cosmological parameters from both methods are the same.}
    \label{fig:Marg_vs_no_marg}
\end{figure}
Our results in this work have used analytic marginalization over linear order nuisance parameters to speed up our fitting. By fitting the mean of the LRG mocks with a $25\times$ reduced covariance matrix, Fig.~\ref{fig:Marg_vs_no_marg} illustrates that the constraints on cosmological parameters with and without analytic marginalization are consistent. This result validates our use of the analytically marginalized likelihood to speed up the analysis. 

\section{Effect of different \textsc{PyBird} versions.}
\label{sec:version}
\begin{figure}
    \centering
	\includegraphics[width=0.495\textwidth]{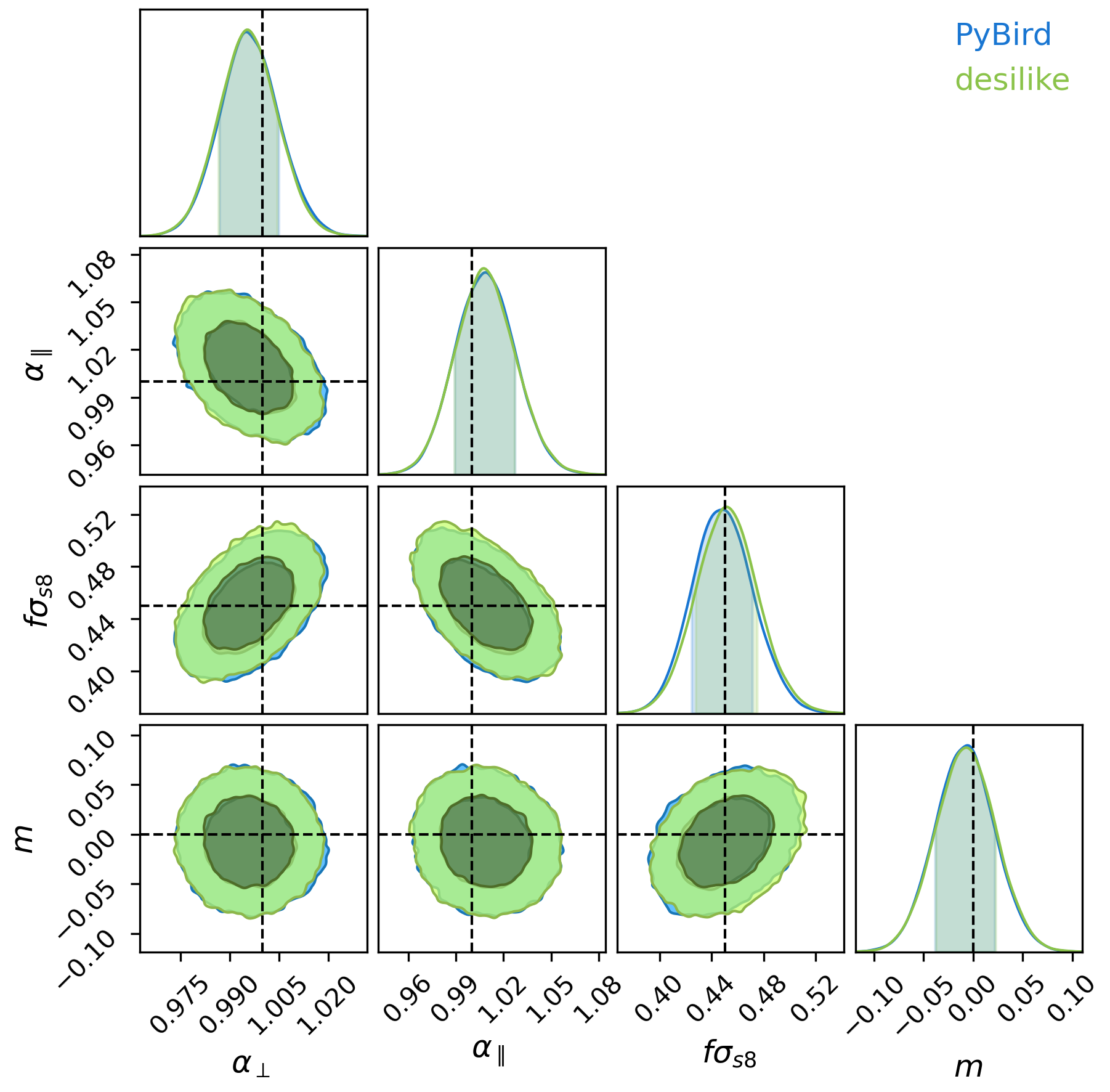}
    \includegraphics[width=0.495\textwidth]{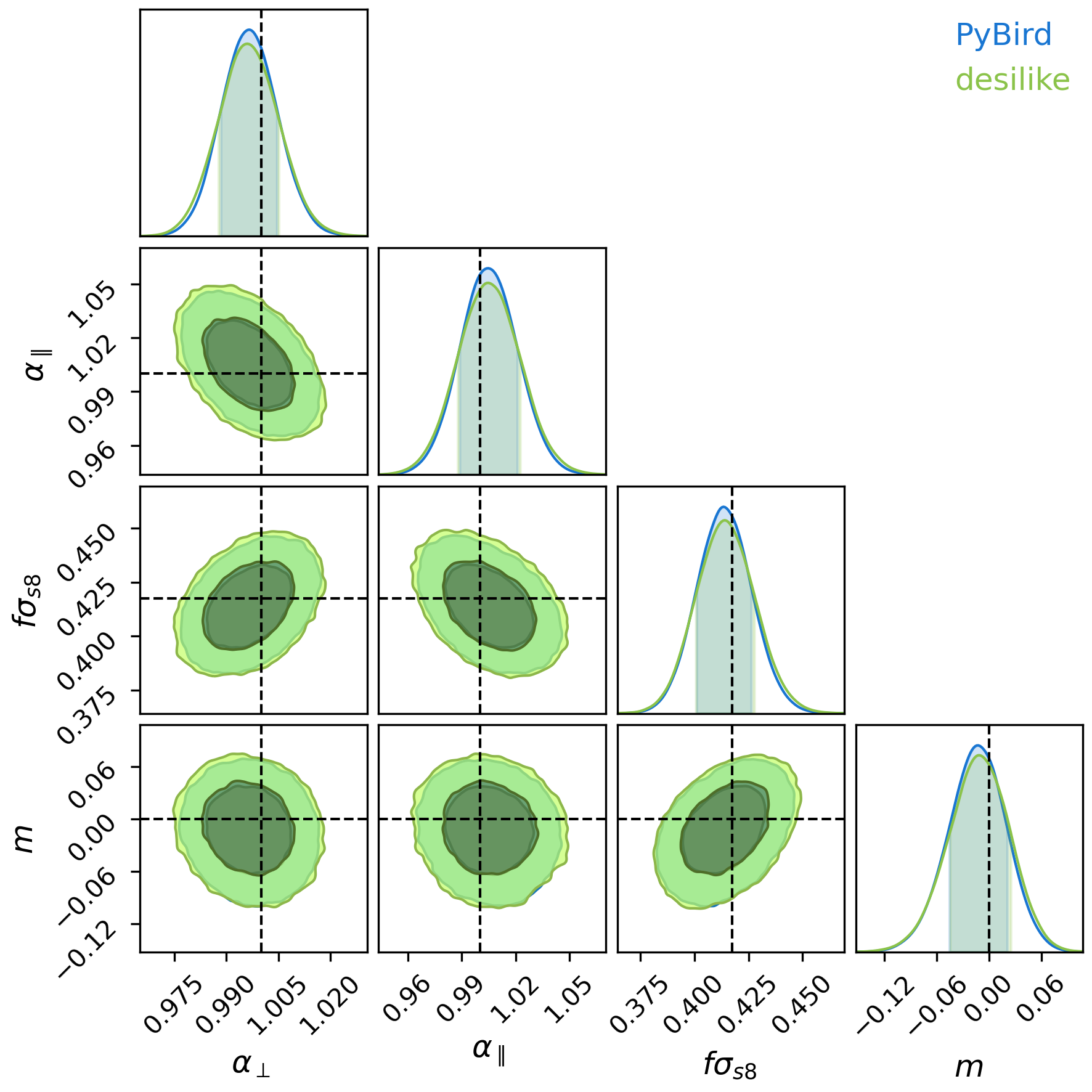}
    \includegraphics[width=0.495\textwidth]{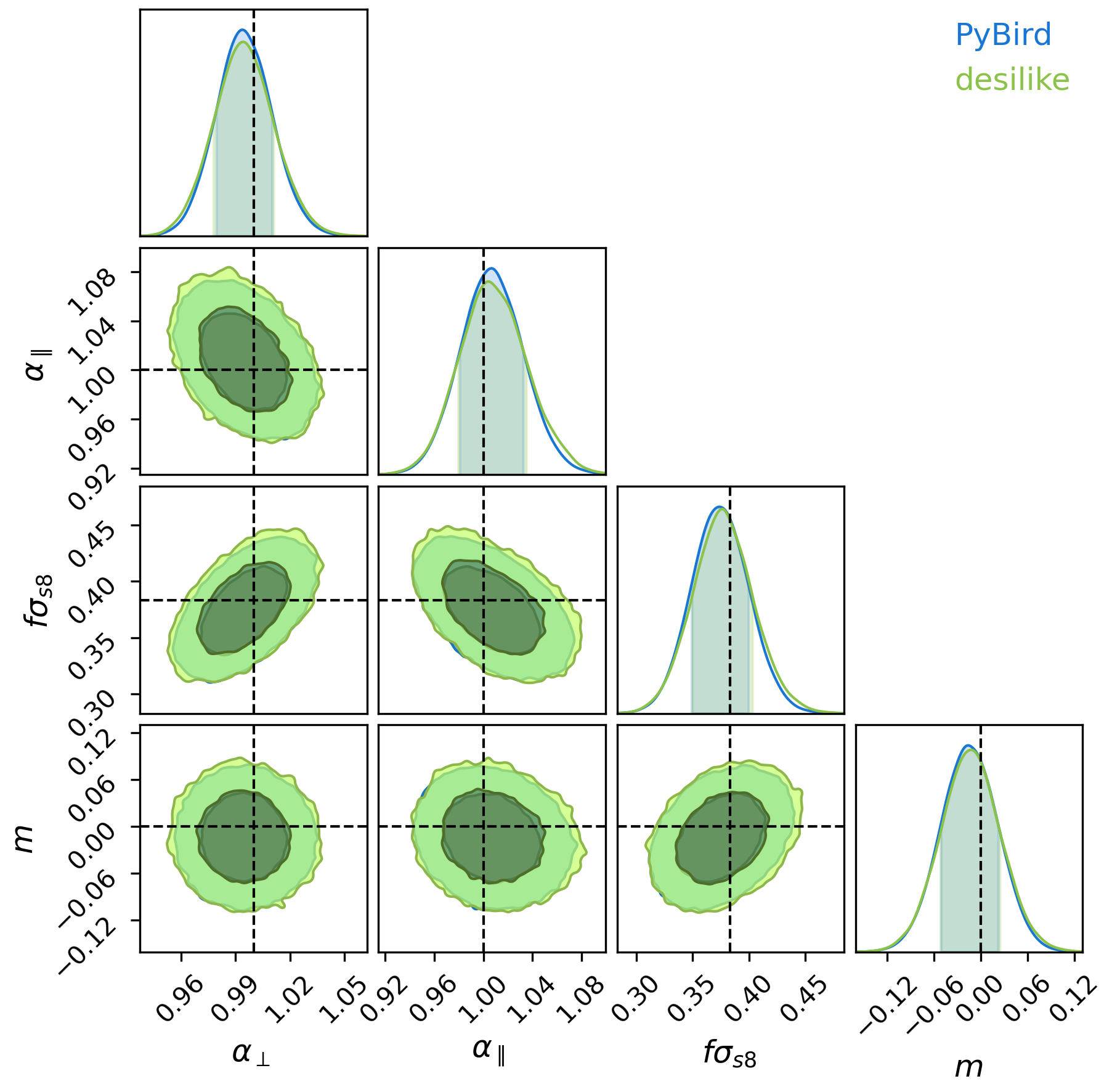}
    
    \caption{This table compares constraints from the version of \textsc{PyBird} in this work to the \textsc{desilike} version for LRG (top left), ELG (top right), and QSO (bottom). They are consistent with each other. The minor difference is due to \(k_r\) in \textsc{desilike} is much smaller, so the prior on the counter terms are larger in \textsc{desilike}, which weakens the constraints.}
    \label{fig:version}
\end{figure}

The version of \textsc{PyBird} we used is slightly different from the one in \textsc{desilike}, which is used for DESI fitting. To understand whether other versions of \textsc{PyBird} will affect the cosmological constraints we compare the constraints on the Shapefit parameters using two different versions of \textsc{PyBird} for LRG (top left), ELG (top right), and QSO (bottom). We find that the constraints are mostly consistent with each other. The \textsc{desilike} version gives \(\approx 5\%\) larger error bars for ELG and QSO. This difference is probably because \(k_r\) in \textsc{desilike} is set to \(0.25 h \mathrm{Mpc}^{-1}\), which is around three times smaller than the one in this paper (\(0.70 h \mathrm{Mpc}^{-1}\)). This setting means the prior on the counter terms in \textsc{desilike} is larger than the ones in this paper, so we see slightly weaker constraints. 

\section{The derived parameters in \(\Lambda\)CDM}
\label{sec:derived}
\begin{figure}
    \centering
    \includegraphics[width=1\linewidth]{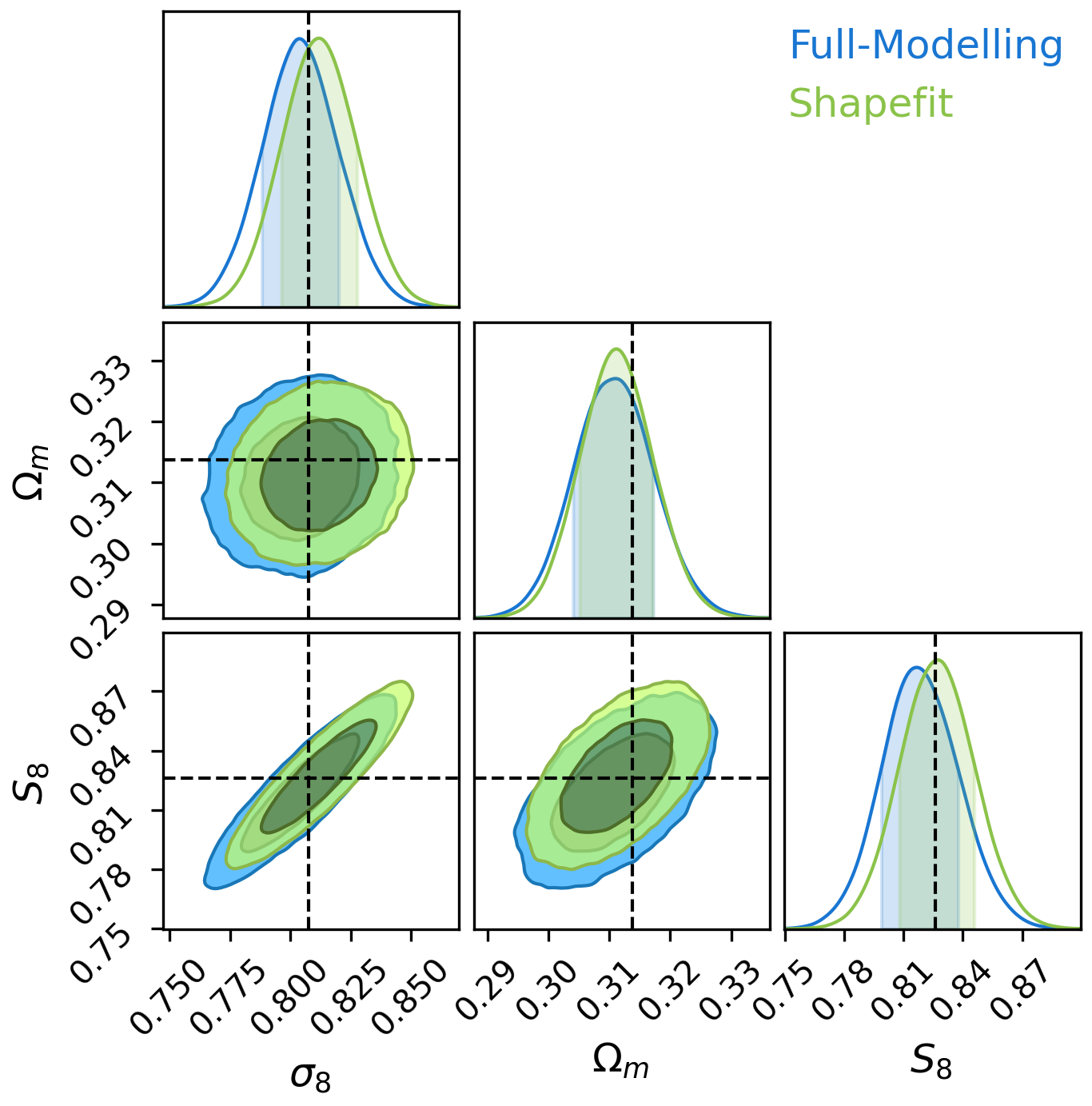}
    \caption{The derived parameters \(\sigma_8\), \(\Omega_m\), and \(S_8\) with the \(\Lambda\)CDM model. These parameters are generated with the posteriors from Fig.~\ref{fig:FS_vs_SF_bestfit}. We use the second fitting configuration with a joint fit to the mean of the LRG, ELG, and QSO mocks. Table~\ref{tab:derived} shows the constraints on these three parameters. The constraints on these three parameters from \textit{ShapeFit} and \textit{Full-Modelling} are both within \(0.5\sigma\) from the truth. Furthermore, the constraints of these three derived parameters between \textit{ShapeFit} and \textit{Full-Modelling} are all consistent within \(\sim 0.5\sigma\).}
    \label{fig:derived}
\end{figure}

\begin{table}[]
\centering
\begin{tabular}{c|c|c|c|} 
Model     & $\sigma_8$      & $\Omega_m$    & $S_8$          \\ \hline \hline
FM        & $0.803_{-0.015}^{+0.017} (0.806)$ & $0.311_{-0.007}^{+0.006} (0.311) $ & $0.817_{-0.019}^{+0.020} (0.821)$ \\ \hline
SF & $0.812_{-0.016}^{+0.016} (0.813)$ & $0.311_{-0.006}^{+0.006} (0.311)$ & $0.827_{-0.019}^{+0.019} (0.828)$ \\ \hline
$\Delta_{\mathrm{FM}}$   & $0.28\sigma (0.10\sigma)$ & $0.40\sigma (0.44\sigma)$ & $0.47\sigma (0.27\sigma)$ \\ \hline
$\Delta_{\mathrm{SF}}$   & $0.28\sigma (0.31\sigma)$ & $0.42\sigma (0.39\sigma)$ & $0.06\sigma (0.10\sigma)$ \\ \hline
$\Delta_{\mathrm{FM-SF}}$ & $0.56\sigma (0.41\sigma)$ & $0.00\sigma (0.00\sigma)$ & $0.53\sigma (0.37\sigma)$
\end{tabular}
\caption{The constraints on the derived parameters \(\sigma_8\), \(\Omega_m\), and \(S_8\) for \textit{Full-Modelling} (FM) and \textit{ShapeFit} (SF). The best-fit parameters are in the brackets. Here, \(\Delta_{\mathrm{SF}}\) and \(\Delta_{\mathrm{FM}}\) denote the deviation from the truth with \textit{ShapeFit} and \textit{Full-Modelling} respectively. Furthermore, \(\Delta_{\mathrm{FM-SF}}\) denotes the difference in constraints between \textit{ShapeFit} and \textit{Full-Modelling}. Both \textit{ShapeFit} and \textit{Full-Modelling} are consistent with the truth values to within \(0.5\sigma\). Lastly, the constraints from \textit{ShapeFit} and \textit{Full-Modelling} are also consistent with each to within \(\sim 0.5\sigma\) level.}
\label{tab:derived}
\end{table}

Despite fitting the ensemble of current data extremely well, several tensions arise within the \(\Lambda\)CDM model. One of the most famous tensions is the \(\sigma_8\) or \(S_8\) tension \citep{Kuijken_2015, Hamana_2020, Abbott_2022, Planck_2020}. This section investigates whether \textit{ShapeFit} can be used to investigate this tension. Fig. \ref{fig:derived} and Table~\ref{tab:derived} demonstrate the constraints on \(\sigma_8\), \(\Omega_m\), and \(S_8 = \sigma_8\sqrt{\frac{\Omega_m}{0.3}}\) using the posteriors from Fig.~\ref{fig:FS_vs_SF_bestfit}. All three parameters are interpolated back to redshift zero. Generally, both \textit{Full-Modelling} and \textit{ShapeFit} constraints are consistent with the truth to within \(0.5\sigma\). Furthermore, the constraints from \textit{ShapeFit} and \textit{Full-Modelling} are also consistent with each other within \(\sim 0.5\sigma\). The remaining small difference between the two is consistent with the discrepancies between the constraints from these two methods shown in Fig.~\ref{fig:FS_vs_SF_bestfit}. 

\section{Data Availability}
The data used in this analysis is available at \url{https://doi.org/10.5281/zenodo.10846264}. This paper's version of the \textsc{PyBird} code is available at \url{https://github.com/pierrexyz/pybird/tree/desi}.

\acknowledgments

We thank Arnaud de Mattia for implementing \textsc{PyBird} into the DESI pipelines and members of the Galaxy and Quasar Clustering working group within DESI for helpful discussions. We also thank Pierre Zhang for helpful advice on \textsc{PyBird}.

YL and CH acknowledge support from the Australian Government through the Australian Research Council’s Laureate Fellowship (project FL180100168) and Discovery Project (project DP20220101395) funding schemes. YL is also supported by an Australian Government Research Training Program Scholarship. MM is supported by the DOE. HN is supported by Ciencia de Frontera grant No. 319359. SR and HN are supported by the Investigacion in Ciencia Basica CONAHCYT grant No. A1-S-13051 and PAPIIT IN108321 and IN116024, and Proyecto PIFF. HGM acknowledges support through the program Ramón y Cajal (RYC-2021-034104) of the Spanish Ministry of Science and Innovation. This research has made use of NASA's Astrophysics Data System Bibliographic Services and the \texttt{astro-ph} pre-print archive at \url{https://arxiv.org/}, the {\sc matplotlib} plotting library \citep{Hunter:2007}, and the {\sc chainconsumer} and {\sc emcee} packages \citep{Hinton2016, Foreman_Mackey_2013}. This research was done using the Getafix supercomputer at the University of Queensland and NERSC.

This material is based upon work supported by the U.S. Department of Energy (DOE), Office of Science, Office of High-Energy Physics, under Contract No. DE–AC02–05CH11231, and by the National Energy Research Scientific Computing Center, a DOE Office of Science User Facility under the same contract. Additional support for DESI was provided by the U.S. National Science Foundation (NSF), Division of Astronomical Sciences under Contract No. AST-0950945 to the NSF’s National Optical-Infrared Astronomy Research Laboratory; the Science and Technology Facilities Council of the United Kingdom; the Gordon and Betty Moore Foundation; the Heising-Simons Foundation; the French Alternative Energies and Atomic Energy Commission (CEA); the National Council of Humanities, Science and Technology of Mexico (CONAHCYT); the Ministry of Science and Innovation of Spain (MICINN), and by the DESI Member Institutions: \url{https://www.desi.lbl.gov/collaborating-institutions}. Any opinions, findings, and conclusions or recommendations expressed in this material are those of the author(s) and do not necessarily reflect the views of the U. S. National Science Foundation, the U. S. Department of Energy, or any of the listed funding agencies.

The authors are honored to be permitted to conduct scientific research on Iolkam Du’ag (Kitt Peak), a mountain with particular significance to the Tohono O’odham Nation.

\section{Author affiliations}
\label{sec:affiliations}

\begin{hangparas}{.5cm}{1}

$^{1}${School of Mathematics and Physics, The University of Queensland, QLD 4072, Australia.}

$^{2}${University of California, Berkeley, 110 Sproul Hall \#5800 Berkeley, CA 94720, USA}

$^{3}${Institut de Ciències del Cosmos (ICCUB), Universitat de Barcelona, Martí i Franquès, 1, E08028 Barcelona, Spain.}

$^{4}${Instituto de Ciencias F\'isicas, Universidad Nacional
Autónoma de México,  62210, Cuernavaca, Morelos.}

$^{5}${Instituto de F\'{\i}sica, Universidad Nacional Aut\'{o}noma de M\'{e}xico,  Cd. de M\'{e}xico  C.P. 04510,  M\'{e}xico}

$^{6}${Sorbonne Université, Université Paris Diderot, Sorbonne Paris Cité, CNRS, Laboratoire de Physique Nucléaire et de Hautes Energies (LPNHE), 4 place Jussieu, F-75252, Paris Cedex 5, France}

$^{7}${Lawrence Berkeley National Laboratory, 1 Cyclotron Road, Berkeley, CA 94720, USA}

$^{8}${Physics Dept., Boston University, 590 Commonwealth Avenue, Boston, MA 02215, USA}

$^{9}${University of Michigan, Ann Arbor, MI 48109, USA}

$^{10}${Instituto Avanzado de Cosmolog\'{\i}a A.~C., San Marcos 11 - Atenas 202. Magdalena Contreras, 10720. Ciudad de M\'{e}xico, M\'{e}xico}

$^{11}${Department of Physics \& Astronomy, University College London, Gower Street, London, WC1E 6BT, UK}

$^{12}${Institute for Advanced Study, 1 Einstein Drive, Princeton, NJ 08540, USA}

$^{13}${Department of Physics and Astronomy, The University of Utah, 115 South 1400 East, Salt Lake City, UT 84112, USA}

$^{14}${Departamento de F\'isica, Universidad de los Andes, Cra. 1 No. 18A-10, Edificio Ip, CP 111711, Bogot\'a, Colombia}

$^{15}${Observatorio Astron\'omico, Universidad de los Andes, Cra. 1 No. 18A-10, Edificio H, CP 111711 Bogot\'a, Colombia}

$^{16}${Institut d'Estudis Espacials de Catalunya (IEEC), 08034 Barcelona, Spain}

$^{17}${Institute of Cosmology and Gravitation, University of Portsmouth, Dennis Sciama Building, Portsmouth, PO1 3FX, UK}

$^{18}${Institute of Space Sciences, ICE-CSIC, Campus UAB, Carrer de Can Magrans s/n, 08913 Bellaterra, Barcelona, Spain}

$^{19}${Center for Cosmology and AstroParticle Physics, The Ohio State University, 191 West Woodruff Avenue, Columbus, OH 43210, USA}

$^{20}${Department of Physics, The Ohio State University, 191 West Woodruff Avenue, Columbus, OH 43210, USA}

$^{21}${The Ohio State University, Columbus, 43210 OH, USA}

$^{22}${NSF's NOIRLab, 950 N. Cherry Ave., Tucson, AZ 85719, USA}

$^{23}${Departament de F\'{i}sica, Serra H\'{u}nter, Universitat Aut\`{o}noma de Barcelona, 08193 Bellaterra (Barcelona), Spain}

$^{24}${Institut de F\'{i}sica d'Altes Energies (IFAE), The Barcelona Institute of Science and Technology, Campus UAB, 08193 Bellaterra Barcelona, Spain}

$^{25}${Instituci\'{o} Catalana de Recerca i Estudis Avan\c{c}ats, Passeig de Llu\'{\i}s Companys, 23, 08010 Barcelona, Spain}

$^{26}${Department of Physics and Astronomy, University of Sussex, Brighton BN1 9QH, U.K}

$^{27}${Departamento de F\'{i}sica, Universidad de Guanajuato - DCI, C.P. 37150, Leon, Guanajuato, M\'{e}xico}

$^{28}${Instituto Avanzado de Cosmolog\'{\i}a A.~C., San Marcos 11 - Atenas 202. Magdalena Contreras, 10720. Ciudad de M\'{e}xico, M\'{e}xico}

$^{29}${IRFU, CEA, Universit\'{e} Paris-Saclay, F-91191 Gif-sur-Yvette, France}

$^{30}${Department of Physics and Astronomy, University of Waterloo, 200 University Ave W, Waterloo, ON N2L 3G1, Canada}

$^{31}${Perimeter Institute for Theoretical Physics, 31 Caroline St. North, Waterloo, ON N2L 2Y5, Canada}

$^{32}${Waterloo Centre for Astrophysics, University of Waterloo, 200 University Ave W, Waterloo, ON N2L 3G1, Canada}

$^{33}${Space Sciences Laboratory, University of California, Berkeley, 7 Gauss Way, Berkeley, CA  94720, USA}

$^{34}${Department of Physics, Kansas State University, 116 Cardwell Hall, Manhattan, KS 66506, USA}

$^{35}${Department of Physics and Astronomy, Sejong University, Seoul, 143-747, Korea}

$^{36}${CIEMAT, Avenida Complutense 40, E-28040 Madrid, Spain}

$^{37}${Department of Physics, University of Michigan, Ann Arbor, MI 48109, USA}

$^{38}${SLAC National Accelerator Laboratory, Menlo Park, CA 94305, USA}

$^{39}${National Astronomical Observatories, Chinese Academy of Sciences, A20 Datun Rd., Chaoyang District, Beijing, 100012, P.R. China}

\end{hangparas}




\bibliographystyle{JHEP}
\bibliography{biblio.bib}






\end{document}